\documentclass[12pt]{article}
\usepackage[sort]{natbib}
\usepackage[margin=1in]{geometry}
\usepackage{times}
\usepackage[italic, noexclam, nopunctuation, noplus, nominus, noplusnominus, noequal, noparenthesis, nospecials, defaultnormal]{mathastext}
\usepackage{amsmath,amssymb,amsfonts, mathtools}
\usepackage{amsthm}
\usepackage{setspace}
\usepackage[labelfont=bf]{caption}
\usepackage{caption}
\usepackage{subcaption}
\usepackage{hyperref}
\usepackage{multirow}
\usepackage{longtable}
\usepackage[table]{xcolor}
\definecolor{maroon}{cmyk}{0,0.87,0.68,0.32}
\usepackage{hhline}
\usepackage{csquotes}
\usepackage{fancyhdr}
\usepackage{color}

\usepackage{tikz}
\usetikzlibrary{decorations.pathreplacing,angles,quotes}

\usepackage{newtxtext}

\usepackage{amsmath}

\usepackage{pgfplots}
\pgfplotsset{compat = newest}

\newcommand{\R}{\mathbb{R}}

\newcommand{\sgn}{\mathrm{sgn}}
\newcommand{\cL}{\mathcal{L}}

\newtheorem{definition}{Definition}
\newtheorem{axiom}{Axiom}
\newtheorem{theorem}{Theorem}

\newtheorem{lemma}{Lemma}
\newtheorem{example}{Example}
\newtheorem{remark}{Remark}

\begin{document}

\author{Ganesh Karapakula\footnote{The Janeway Institute, Faculty of Economics, The University of Cambridge. Email: vgk22@cam.ac.uk. I first and foremost thank Nathan Hendren and Ben Sprung-Keyser for providing extremely helpful feedback on this paper, for explaining the historical development of this research area, for kindly sharing their resources (including their fully replicable code and data), and for being generous with their time to help me build on their highly influential paper. I thank Don Andrews, Zheng Fang, Hiro Kaido, and Jessie Li for feedback on uniformly valid statistical inference methods. I am very grateful to Oliver Linton and Melvyn Weeks for providing immensely beneficial constructive critiques and suggestions that have vastly improved this paper and my other related research. I also thank Michael Ashby, Debopam Bhattacharya, Matt Elliott, Eric French, Alexei Onatski, Debraj Ray, Ariel Rubinstein, Mikhail Safronov, Liyang Sun, and Wei Xiong for their insightful questions and comments that helped me refine and expand my framework. In addition, I thank Laura Araújo de Freitas and Shane Mahen for useful comments. I gratefully acknowledge financial support from the Janeway Institute for my research. This paper previously circulated last year under an abstruse title: ``A Double Staircase Index for Welfare Analysis.'' I previously gave the name ``Double Staircase Index'' (DSI) to my cost--benefit measure, the ``Relative Policy Value'' (RPV), because the RPV bears strong visual resemblance to an interesting type of architectural design called the double staircase. However, I have since decided to leave architecture out of this paper for now!}}
\title{An Axiomatic Framework for Cost--Benefit Analysis}
\date{\normalsize}

\maketitle

\vspace{-10mm}
\begin{center}
    First Version: 31 May 2021 \\
    Current Version: 28 April 2022
\end{center}

\vspace{-2mm}
\begin{abstract}
\noindent 
In recent years, the Marginal Value of Public Funds (MVPF) has become a popular tool for conducting cost--benefit analysis; the MVPF relies on the ratio of willingness-to-pay for a policy divided by its net fiscal cost. The MVPF gives policymakers important information about the equity--efficiency trade-off that is not necessarily conveyed by absolute welfare measures. However, I show in this paper that the usefulness of MVPF for comparative welfare analysis is limited, because it suffers from several empirically important economic paradoxes and statistical irregularities. There are also several practical issues in using the MVPF to aggregate welfare across policies or across population subgroups. To address these problems, I develop a new axiomatic framework to construct a measure that quantifies the equity--efficiency trade-off in a better way. I do so without compromising on the core features of the MVPF: its unit-free property, and the main preference orderings underlying it. My axiomatic framework delivers a unique (econo)metric that I call the Relative Policy Value (RPV), which can be weighted to conduct both comparative and absolute welfare analyses (or a hybrid combination thereof) and to intuitively aggregate welfare (without encountering the issues in MVPF-based aggregation). I also propose computationally convenient methods to make uniformly valid statistical inferences on welfare measures. After reanalyzing several government policies using my new econometric methods, I conclude that there is substantial economic and statistical uncertainty about welfare of some policies that were previously reported to have very high or even ``precisely estimated infinite'' MVPF values.
\end{abstract}

\clearpage

\onehalfspacing

\section{Introduction}
\label{sec:introduction}

At its core, cost--benefit analysis involves quantifying the equity--efficiency trade-off. There are two main challenges in analyzing the welfare impacts of various policies. First, their equity and efficiency aspects must be operationalized and measured. Second, the resulting measurements need to be combined in an economically coherent manner for welfare calculations. The first challenge has been successfully tackled in various contexts by \cite{mayshar1990measures}, \cite{hendren2016policy}, \cite{hendren2020unified}, \cite{finkelstein2020welfare}, \cite{bhattacharya2021incorporating}, \cite{bhattacharya2022empirical}, and others, but I argue that the second crucial challenge has not been properly addressed in the existing economic literature. Thus, I propose new welfare econometrics for systematically completing the cost--benefit analysis of a collection of public policies. 

\begin{figure}[ht]
\begin{center}
\caption{Willingness-To-Pay, Net Fiscal Cost, and the Equity--Efficiency Trade-off}
\label{figure:eetradeoff}
\begin{tikzpicture}[scale=0.95]
\begin{axis}[xlabel={$c$}, ylabel={$p$}, ymin=-2,ymax=2,xmax=2,xmin=-2, axis lines = middle]
\node[rotate=90] at (0.15,0.85) {\scriptsize willingness-to-pay};
\node[rotate=90] at (-0.35,1) {\scriptsize effectiveness or equity};
    \addplot[
        scatter/classes={a={white}, b={red}},
        scatter, mark=, only marks, 
        scatter src=explicit symbolic,
        nodes near coords*={\Label},
        visualization depends on={value \thisrow{label} \as \Label} 
    ] table[meta=class, row sep=\\]
    {
        x y class label \\
        1 0 a {\scriptsize --- net fiscal cost $\longrightarrow$} \\
        -1.25 0 a {\scriptsize $\longleftarrow$ efficiency ------} \\
        -1.5 -1 a {\footnotesize Tax enforcement policies} \\
        -1.5 -1.5 a {\footnotesize Spending cuts} \\
        -1.55 1 a {\footnotesize Efficient and equitable} \\
        -1.55 0.7 a {\footnotesize reforms and investments} \\
        1.5 1.25 a {\footnotesize Public projects} \\
        1.5 0.75 a {\footnotesize Tax reduction policies} \\
        1.5 -1 a {\footnotesize Counterproductive policies} \\
        1.5 -1.5 a {\footnotesize Boondoggles} \\
    };
\end{axis}
\end{tikzpicture} \begin{tikzpicture}[scale=0.95]
\begin{axis}[xlabel={$c$}, ylabel={$p$}, ymin=-2,ymax=2,xmax=2,xmin=-2, axis lines = middle]
\draw (-2,-2) to (2,2);
\draw (-2,2) to (2,-2);
\node[rotate=90] at (0.15,0.85) {\scriptsize -- more equitable $\to$};
\node[rotate=90] at (0.15,-1) {\scriptsize $\longleftarrow$ less equitable ---};
    \addplot[
        scatter/classes={a={white}, b={red}},
        scatter, mark=, only marks, 
        scatter src=explicit symbolic,
        nodes near coords*={\Label},
        visualization depends on={value \thisrow{label} \as \Label} 
    ] table[meta=class, row sep=\\]
    {
        x y class label \\
        1 0 a {\scriptsize --- less efficient $\longrightarrow$} \\
         -1 0 a {\scriptsize $\longleftarrow$ more efficient ---} \\
        1.5 -1 a {\footnotesize IV-B} \\
        0.75 -1.75 a {\footnotesize IV-A} \\
        -0.75 -1.75 a {\footnotesize III-B} \\
        -1.5 -1 a {\footnotesize III-A} \\
        -1.5 0.75 a {\footnotesize II-B} \\
        -0.75 1.5 a {\footnotesize II-A} \\
        1.5 0.75 a {\footnotesize I-A} \\
        0.75 1.5 a {\footnotesize I-B} \\
    };
\end{axis}
\end{tikzpicture}
\end{center}
\footnotesize \textit{Note}: The left figure shows examples of policies with different values of the net fiscal cost $c$ and willingness-to-pay $p$. The right figure shows the sub-quadrants (I-A, I-B, II-A, II-B, III-A, III-B, IV-A, and IV-B) of the $(c,p)$-plane that are associated with different equity--efficiency trade-offs, along with diagonal ($p = c$) and antidiagonal ($p = -c$) axes.
\end{figure}
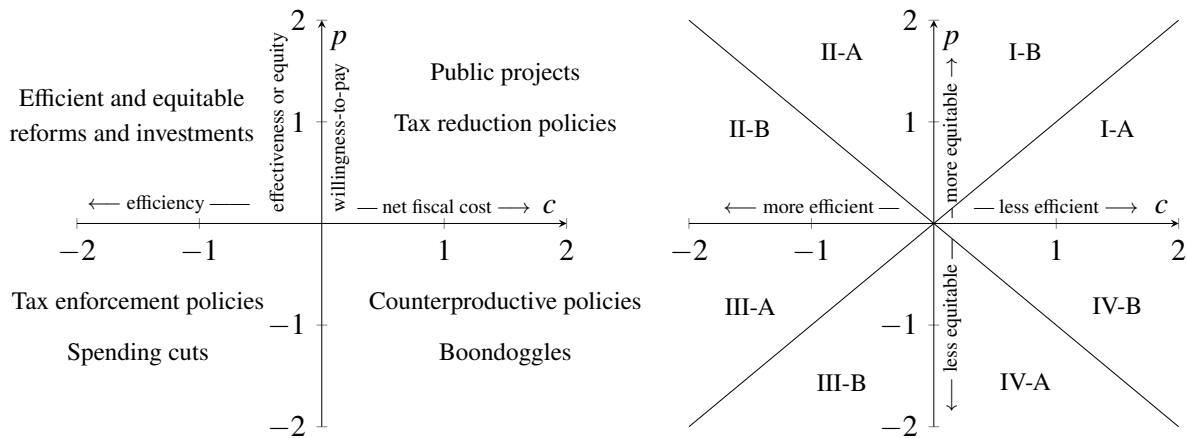

There is a growing consensus among economists, especially those in the field of public finance, that there are two fundamental building blocks of empirical welfare analysis: $(c, p) \in \R^2$, where $c$ is the net fiscal cost of a policy (so that $-c$ represents the net fiscal revenue), and $p$ is the impacted population's willingness to pay for the policy \citep{finkelstein2020welfare}. The latter quantity $p$ represents the policy's equity aspect, and $-c$ represents the policy's efficiency aspect.  Figure \ref{figure:eetradeoff} provides some examples of policies that fall on the four quadrants of the $(c,p)$-plane, in addition to showing the eight sub-quadrants with different equity--efficiency trade-offs.

To make comparisons across policies, $c$ and $p$ need to be defined appropriately. Specifically, for a public policy of interest, let $p = \eta_p\,\tilde{p} \in \R$ be the associated normalized measure of the impacted population's willingness-to-pay (WTP), where $\tilde{p}$ is a raw money metric that measures the WTP for the policy in a way that is not necessarily comparable across policies, and $\eta_p > 0$ is a policymaker's subjective scaling factor that enables cross-policy comparisons. In some contexts, $\eta_p$ has a theoretical interpretation as the ``average social marginal utility'' \citep{hendren2020unified, kleven2006marginal}. In addition, let $c = \eta_c\,\tilde{c} \in \R$ be the policy's net fiscal cost, where $\tilde{c}$ is a raw measure of the present discounted value of the fiscal expenditure after accounting for any fiscal externalities, and $\eta_c > 0$ is a subjective or objective scaling factor that enables cross-policy comparisons. Negative values of $c$ indicate positive net fiscal revenue.

It is typically the case that $c > 0$ and $p > 0$ for a tax reduction policy or a public project. Efficient and equitable reforms and investments usually have $c < 0$ and $p > 0$. Tax enforcement policies and spending cuts generally have $c < 0$ and $p < 0$. Counterproductive policies and boondoggles not only have $c > 0$ but also $p < 0$. As shown in Figure \ref{figure:eetradeoff}, the four quadrants (I, II, III, IV) of the $(c,p)$-plane can be further divided into eight sub-quadrants so that quadrant I is the union of the sub-quadrants I-A and I-B, i.e, I $ = $
I-A $\cup$ I-B, and the same applies to other quadrants as well, i.e., II $ = $ II-A $\cup$ II-B, III $ = $ III-A $\cup$ III-B, and IV $ = $ IV-A $\cup$ IV-B.

Based on \citeauthor{hendren2020unified}'s (\citeyear{hendren2020unified}) estimates, some examples of policies in the various sub-quadrants as follows. For a tax deduction policy (implemented in 2006) affecting a subgroup of single filers with post-secondary tuition expenses, the estimated value of $(c, p)$ is $(\hat{c}, \hat{p}) = (-1.13, 1.00)$, which falls on II-B. For the Kemp--Roth tax cut (a crucial part of the Economic Recovery Tax Act of 1981), $(\hat{c}, \hat{p}) = (-0.51, 1.00)$, which falls on II-A. For the Affordable Care Act-related tax changes implemented in 2013, $(\hat{c}, \hat{p}) = (0.86, 1.00)$, which falls on I-B. For the unemployment insurance benefit changes in New York (implemented in 1989), $(\hat{c}, \hat{p}) = (1.31, 1.17)$, which falls on I-A. For the youth component of the Job Training Partnership Act, $(\hat{c}, \hat{p}) = (0.91, -0.21)$, which falls on IV-B. For the Hope Tax Credit for a subgroup of single filers, $(\hat{c}, \hat{p}) = (1.16, -1.91)$, which falls on IV-A. If a standard labor supply model is used to estimate $(c,p)$ for marginal tax rate increases in the two extreme tax bracket deciles in the UK, then the associated estimates of $(c, p)$ are $(-1.09, -1.00)$ and $(-0.37, -1.00)$, which respectively fall on III-A and III-B, according to the estimates reported by \cite{kleven2006marginal}, although they show that the estimates change substantially when a more realistic model is used.

At this point, it is important to clarify the scope of the welfare econometrics in this paper. As mentioned earlier, there exist reasonable approaches to measuring $(c, p)$ for policies \citep[see][]{finkelstein2020welfare, bhattacharya2022empirical}. This paper does not get into these issues and instead takes measurement of $(c, p)$ as given. For this reason, I focus on welfare calculations after the intermediate measurements. Thus, this paper requires empirical researchers to operationalize $(c, p)$ for the policies of interest. Given information on $(c, p)$, I propose new approaches to:~measuring welfare so as to enable comparisons across policies; aggregating welfare in different ways; and making statistical inferences about welfare of policy collections. Subsections \ref{subsec:overview_problem} and \ref{subsec:overview_solution} respectively provide brief overviews of the motivating problem and my proposed solutions.

\subsection{A Brief Overview of the Paper: The Problem}
\label{subsec:overview_problem}

Given a value of $(c,p)$, the current practice is to measure welfare using the difference $p - c$, which is based on an absolute notion of welfare, or the ratio $p/c$, which is instead based on a comparative notion. The difference $p - c$ is called the social surplus\footnote{See \cite{bhattacharya2021incorporating} and \cite{kamat2020estimating} for examples of empirical versions of the social surplus for non-marginal policy changes.} and is the opposite of the excess burden $c - p$. In the case of marginal policy changes \citep{mayshar1990measures}, $p - c$ is called the marginal social surplus (MSS). Its opposite $c - p = (-p) - (-c)$ is known as the marginal excess burden (MEB), and the ratio $p/c = (-p)/(-c)$ is known as the marginal cost of funds (MCF).\footnote{In the literature, the phrases ``marginal excess burden'' (MEB) and ``marginal cost of funds'' (MCF) have been used variously and in confusing ways. \cite{hendren2020unified} and \cite{finkelstein2020welfare} discuss the etymology of these terms. \cite{hendren2016policy} also discusses how \citeauthor{mayshar1990measures}'s (\citeyear{mayshar1990measures}) definitions of MEB and MCF differ from the traditional connotations of the terms. Different name choices for the concepts would have avoided the issue. Nevertheless, this paper uses the terms MEB and MCF as defined in equations (8b) and (9) on page 267 of \cite{mayshar1990measures}, respectively, to avoid confusion for readers who may refer to \cite{mayshar1990measures} while reading this paper. \cite{mayshar1990measures} defines the MEB as $MS - MR$, where $MS = -p$ is the (minus of) marginal (private) surplus and $MR = -c$ is the marginal (net) revenue. Thus, in the definition of MEB, ``the [marginal net] revenue change [is] deducted from the [marginal] change in consumer surplus'' \citep{mayshar1990measures}. \cite{mayshar1990measures} defines the MCF as $MS/MR$, where $MS = -p$ is the (minus of) marginal (private) surplus and $MR = -c$ is the marginal (net) revenue, when $MR > 0$.}

In recent years, empirical researchers have been increasingly measuring welfare based on a comparative notion called the marginal value of public funds (MVPF), which is an extension of MCF.\footnote{This paper uses the term ``marginal value of public funds'' (MVPF) in the same way \cite{hendren2016policy}, \cite{hendren2020unified}, and \cite{finkelstein2020welfare} use it after accounting for $\eta_p$ (i.e., the $\eta_p$-adjusted MVPF).} The MVPF equals $\infty$ if $p \geq 0$ and $c \leq 0$ and equals $p/c$ if $c > 0$. As detailed in Subsection \ref{subsec:economic_paradoxes}, the MVPF is difficult to extend to the third quadrant (containing policies such as the usual tax increases and spending cuts) where $c < 0$ and $p < 0$, but some researchers augment the MVPF by setting it equal to $p/c$ in this case, giving rise to the paradoxes discussed in the next paragraph. Many analysts and policymakers prefer to use the MVPF instead of the MSS because the latter is an unstandardized measure that does not necessarily distinguish between policies with different equity--efficiency trade-offs in the various quandrants of Figure \ref{figure:eetradeoff}. For example, policies satisfying the equation $p - c = -2$ have the same MSS regardless of whether $p$ is positive or negative, whereas the MVPF is negative if $p < 0$ and positive if $p > 0$, assuming that $c > 0$.

Even though the MVPF, which standardizes $p$ by $c$, is useful because of its unit-free property, it is economically paradoxical as a welfare measure. For example, a policy with $(c,p) = (-0.001,-1000)$ has an (augmented) MVPF equal to a million, whereas a policy with $(c, p) = (0.001, 1)$ has an MVPF of 1000; this would imply a contradictory conclusion that policies with negligible net fiscal revenue but a highly negative willingness-to-pay can have higher welfare than policies with negligible net fiscal cost but a relatively higher positive willingness-to-pay. Two additional examples illustrate how the (augmented) MVPF is puzzling: a policy with $(c,p) = (-0.01, -10)$ has an MVPF of $1000$, but a minuscule change in $c$ while keeping $p$ unchanged, such as changing the value of $(c,p)$ to $(0.01, -10)$, causes the MVPF to drop to $-1000$; in addition, a policy with $(c,p) = (-10,0.01)$ has an infinite MVPF, but a small change in $p$ while keeping $c$ unchanged, such as changing the value of $(c,p)$ to $(-10,-0.01)$, causes the MVPF to drop to $0.001$. There are also several empirically relevant issues in using the MVPF to aggregate welfare across policies. For example, if a policy has an infinite MVPF and another policy has a negative MVPF, then MVPF-based welfare aggregation methods can be problematic, as detailed in Subsection \ref{subsec:limitations_existing}. In addition, since the MVPF is a ratio that is not fully differentiable, the conventional methods (such as the delta method-based confidence intervals or the percentile bootstrap-based confidence intervals or their bias-corrected variants) fail to provide valid statistical inferences in general. Thus, the seemingly convenient and simple-looking ratio $p/c$ can be econometrically intractable.

Instead of using the aforementioned ad hoc measures of welfare, this paper develops a new ``econo''-metric from the ground up, after taking stock of useful properties of both the MVPF and the MSS. The MVPF is unit-free, i.e., homogeneous of degree zero, but the MSS is not so. In addition, the MVPF (minus one), i.e., $p/c - 1 = (p-c)/c$, can be interpreted intuitively as shortfall in WTP relative to the net fiscal cost when $c > 0$ and $p \in [-c, c]$, but the MSS is an unstandardized measure that does not have such an interpretation. However, the MSS satisfies two useful properties that the MVPF does not. Consider two symmetrically opposite policies such as $(c,p) = (2,1)$ and $(c',p') = (1,2)$ that lie on different sides of the break-even line $p = c$. Their combination has the same total MSS (i.e., the sum of MSS values) as the MSS of $(c^*,p^*) = (2+1,1+2) = (3,3)$. In other words, for these policies, aggregate welfare is the same regardless of the aggregation method (i.e., summing their MSS values versus computing the MSS of their sum). In addition, the combination of opposite policies such as $(c,p) = (1,2)$ and $(c',p') = (-1,-2)$ has the same total MSS as the MSS of $(c^*,p^*) = (1-1,2-2) = (0,0)$. Again, for these policies, aggregate welfare is the same regardless of the aggregation method (i.e., summing their MSS values versus computing the MSS of their sum). However, these properties do not hold for the MVPF. A natural question arises: Is there a measure satisfying all four desirable properties? If so, is it unique? As outlined in Subsection \ref{subsec:overview_solution}, my answers are ``yes'' and ``yes'' to both of these questions!

\subsection{A Brief Overview of the Paper: The Solution}
\label{subsec:overview_solution}

After taking stock of the useful properties of both the MVPF and the MSS, I develop an axiomatic framework incorporating the four aforementioned properties, and I then prove that there is only one metric that satisfies them all: the ``Relative Policy Value'' (RPV). The RPV is zero at the origin but otherwise has the formula $\phi(c,p) = (p -c)/\mathrm{max}\{|\,p\,|,|\,c\,|\}$. Its contour plot is shown in Figure \ref{fig:square_dots_rpv} below, in addition to a square with vertices $\{-1, 1\} \times \{-1, 1\}$, on which the maximum norm equals one. The RPV looks slightly odd at first glance, but it has an intuitive description and explanation. It is based on a preference relation between $(c,p)$ values satisfied by both the MVPF and the MSS: $(-1,1) \succ (-1,-1) \sim (1,1) \succ (1,-1)$, where $\succ$ denotes strict preference and $\sim$ represents indifference. The RPV satifies these preference relations by using the MSS $(p - c)$ to measure welfare on the square formed by those four points, i.e., $\phi(c,p) = p - c$ when $\mathrm{max}\{|\,p\,|,|\,c\,|\} = 1$. Hence, imposing the unit-free property (i.e., degree-zero homogeneity) on the welfare measure leads to the RPV, which standardizes the MSS $(p - c)$ by using the maximum norm $\mathrm{max}\{|\,p\,|,|\,c\,|\}$. The reciprocal $1/\mathrm{max}\{|\,p\,|,|\,c\,|\}$ serves as a multiplicative factor that makes the RPV scale-free. In addition, a useful feature of the RPV is that $\phi(c, p) \times ||(c, p)||_\infty = p - c$. In other words, the RPV and the maximum norm are sufficient statistics for both comparative and absolute welfare analyses!

\begin{figure}
\begin{center}
\caption{Contour Plot of the Relative Policy Value (RPV)}
\vspace{-3mm}
\label{fig:square_dots_rpv}
\includegraphics[scale=1]{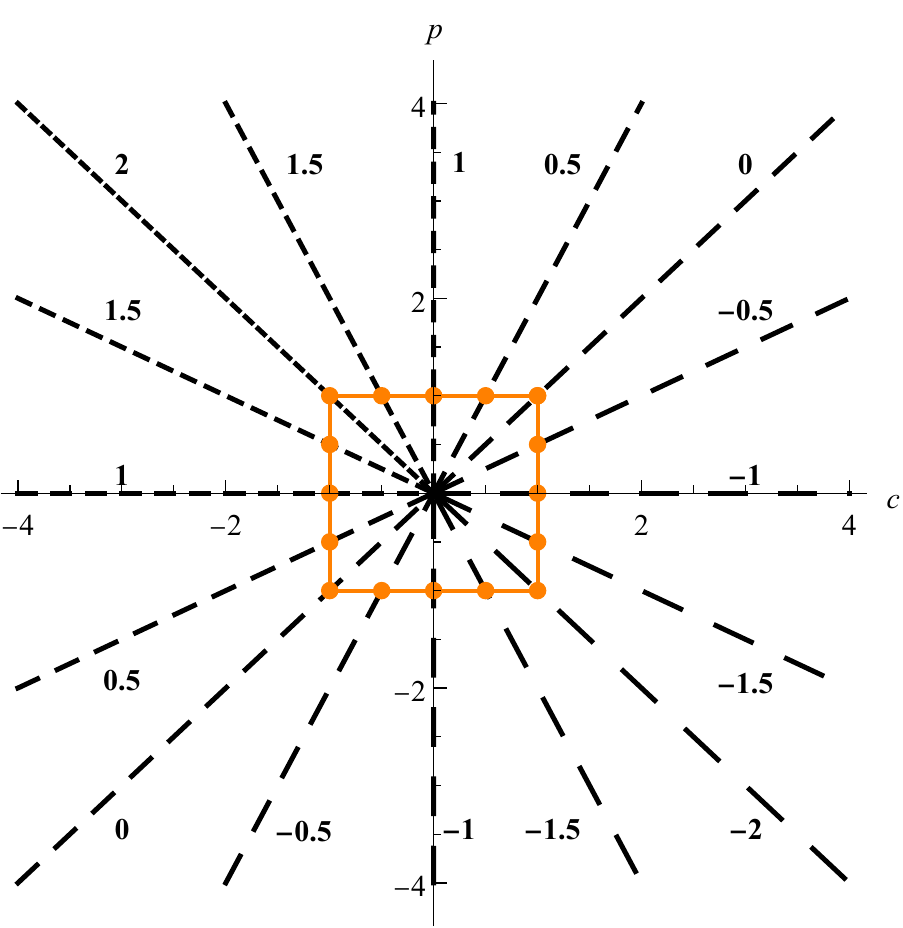}
\end{center}
\footnotesize\vspace{-2mm} \textit{Note}: In the above graph, the thick dashed lines represent contours of the Relative Policy Value (RPV), which normalizes the social surplus $(p - c)$ by the maximum norm $||(c,p)||_\infty = \mathrm{max}\{|\,p\,|,|\,c\,|\}$. The values of the contours are indicated near them in boldface. In addition, the square with vertices $\{-1, 1\} \times \{-1, 1\}$ is superimposed on the RPV contours.
\end{figure}

Having developed an axiomatic foundation for my new measure, I explore how to conduct valid statistical inferences on welfare measures. Because the MVPF and the RPV are not regular functions, we could obtain very misleading statistical inferences about welfare of policies if we use standard methods such as the delta method and percentile bootstrap confidence intervals. In addition to not being fully differentiable, the MVPF and the RPV are neither convex nor Lipschitz continuous. Some recent methods in econometrics rely on the latter two properties to generate uniformly valid inferences for functions that are only directionally differentiable, and so these recent methods are also not generally applicable in the case of the MVPF and the RPV. Hence, I develop a computationally convenient two-step simulation-based procedure for conducting uniform inferences on the welfare of policies.

A uniformly valid $(1-\alpha)$-confidence interval $\hat{C}^\alpha_n$ (based on a sample of $n$ observations) for the RPV has the following property: the worst-case probability that $\hat{C}^\alpha_n$ contains $\phi(c,p)$, the true value of RPV, is at least $1-\alpha$ asymptotically. More technically, the interval $\hat{C}^\alpha_n$ is a uniformly valid $(1-\alpha)$-confidence interval for the RPV if $\mathrm{lim\,inf}_{n \to \infty} \,\mathrm{inf}_{F \in \mathbb{F}}\,\mathbb{P}_F\{\phi(c,p) \in \hat{C}^\alpha_n\} \geq 1 - \alpha$, where $\mathbb{F}$ is a class of distributions that satisfy standardized uniform integrability. My inference procedure has two steps: the first step uses a resampling method and appropriate test statistics to construct a joint confidence set $\hat{S}^\alpha_n$ for $(c,p)$; and the simulation-based second step approximately projects the boundary of $\hat{S}^\alpha_n$ (rather than the full region $\hat{S}^\alpha_n$) using the RPV function $\phi$ to obtain the confidence interval $\hat{C}^\alpha_n$. Since the uniform confidence interval $\hat{C}^\alpha_n$ may be a bit conservative, I also construct a less conservative (but also less uniformly valid) version: a minimalist confidence interval $\tilde{C}^\alpha_n$ that only projects the original resampled points within $\hat{S}^\alpha_n$ using the RPV function $\phi$.

This paper also conducts some simulation exercises to demonstrate that the conventional procedures can have very poor coverage. Thus, for example, the usual procedures may lead a policymaker to believe that a certain policy has an infinite MVPF even though the data do not warrant that conclusion. In contrast, my new statistical procedures perform very well in simulation exercises, as expected based on the theory. I apply my statistical procedures to reanalyze over a hundred government policies that were implemented in the United States over the past half-century using the dataset assembled by \cite{hendren2020unified}. My empirical results show that there is substantial economic and statistical uncertainty about the welfare of many important policies, especially some policies that were previously reported to have infinite MVPF values.

The axiomatic framework-based RPV is a new addition to the cost--benefit analysis toolkit and has several advantages over the existing welfare measures. My measure can also be used to intuitively aggregate welfare across policies or across population subgroups in different ways without running into the issues that arise in MVPF-based welfare aggregation. My statistical procedures can be used to make uniformly valid statistical inferences on the RPV to better inform risk-averse policymakers. Overall, my research shows that different types of policy preferences and econometric frameworks can lead to very different conclusive or inconclusive empirical statements about policies, even when the same data and same resamples are utilized.

This paper is organized as follows. Section \ref{sec:motivation} motivates my new welfare measurement framework by discussing the limitations of the existing methods for comparative welfare analysis. Section \ref{sec:axiomatic_construction} uses an axiomatic framework to develop the Relative Policy Value (RPV) for cost--benefit analysis and also provides intuitive interpretations of the RPV. Section \ref{sec:welfare_agg} discusses welfare aggregation using the RPV. Section \ref{sec:stat_inference} discusses uniformly valid statistical inference. Section \ref{section:reanalysis} presents empirical reanalysis of a selected set of government policies using the new methods. Section \ref{section:conclusion} concludes.

\section{Motivation for a New Welfare Measurement Framework}
\label{sec:motivation}

In this section, I critically evaluate the existing welfare measures and motivate the need for a new framework for cost--benefit analysis. Recall that $p = \eta_p\,\tilde{p}$ is the impacted population's willingness-to-pay (WTP) for a policy and $c = \eta_c\,\tilde{c}$ is the policy's net fiscal cost, where $\eta_p$ and $\eta_c$ are the policymaker's subjective scaling factors that enable cross-policy comparisons. In public finance, the excess burden $c - p = (-p) - (-c)$ and its opposite $p - c$, the social surplus, are some of the most commonly used metrics to measure social welfare in an absolute sense. In the context of marginal policy changes, \cite{mayshar1990measures} calls $(-p) - (-c)$ the marginal excess burden (MEB). Its opposite is the marginal social surplus (MSS), which is a function $\beta: \R^2 \to \R$ given by
$$\beta(c, p) = p - c.$$
The conventional benefit-to-cost ratio (BCR) can be expressed as a simple affine transformation of a specialized version of the MSS. For example, using \citeauthor{hendren2020unified}'s (\citeyear{hendren2020unified}) operationalization of $(c, p)$, the BCR reduces to $\beta(c, p) + 1$, which is just the MSS plus one.\footnote{To see this, let $\tilde{a} \in \R_{> 0}$ represent the initial upfront government spending on a public policy. Then, the unscaled net fiscal cost is $\tilde{c} = \tilde{a} + (\tilde{c} - \tilde{a})$, where $(\tilde{c} - \tilde{a})$ represents the fiscal externality. \cite{hendren2020unified} set $\eta_p = \eta_c = 1/\tilde{a}$. Then, $p = \eta_p\,\tilde{p} = \tilde{p}/\tilde{a}$ and $c = \eta_c\,\tilde{c} = \tilde{c}/\tilde{a}$. Then, the benefit-to-cost ratio (BCR) is given by $$\frac{\tilde{p} - (\tilde{c} - \tilde{a})}{(1 + r)\,\tilde{a}} = \frac{p - c + 1}{1 + r} = \frac{\beta(c, p) + 1}{1 + r},$$ where $r$ is some nonnegative discount rate representing the deadweight loss of raising government revenue. Note that $\beta(c, p) + 1$, which is the BCR when $r = 0$, is sufficient to calculate the BCR for any positive value of $r$. For this reason, \cite{hendren2020unified} report the BCR as $\beta(c, p) + 1$ (i.e., the MSS plus one), which is convenient.} Thus, I focus most of my subsequent discussion on the MSS, which is the main absolute welfare measure.

While the MSS $\beta(c, p) = p - c$ is defined everywhere on $\R^2$, the ratio $p/c$ is undefined when $c = 0$, and so researchers have historically used ratio-based measures on restricted domains. For example, \cite{mayshar1990measures} defines a ratio measure called the marginal cost of funds (MCF) $\underline{m}: \R_{< 0} \times \R \to \R$ for policies with positive net fiscal revenue, and \cite{slemrod2001integrating} define a ``symmetrical concept [called] the marginal benefit of public projects'' (MBP) $\overline{m}: \R_{> 0} \times \R \to \R$ for policies with positive net fiscal cost. They are given, respectively, by
$$\underline{m}(c,p) = (-p)/(-c) = p/c\,\, \text{ if }\,\, c < 0 \quad \text{ and } \quad \overline{m}(c,p) = p/c \,\,\text{ if }\,\, c > 0.$$
The MCF was originally developed to analyze policies with negative net fiscal costs, and so it equals $(-p)/(-c)$ for $c < 0$ and is undefined elsewhere. In the context of some tax hikes that have no fiscal externalities, the MCF represents the ``[marginal] loss in [consumer] surplus due to raising a marginal dollar of tax revenue'' \citep{mayshar1990measures}. Thus, a revenue-raising policy with an MCF below 1 would be preferable to that with an MCF above 1. Policies with a negative MCF, i.e., $\underline{m}(c, p) \in (-\infty, 0)$, are Pareto superior (i.e., equitable and efficient) because they not only have a positive willingness-to-pay (WTP) but also raise revenue.

To make comparisons between public projects with positive net fiscal costs, one can use ``the marginal benefit of public projects, or MBP, which indicates the value to individuals of the dollars spent'' \citep{slemrod2001integrating}. Thus, a public project with an MBP above 1 would be preferable to that with an MBP below 1. Public projects with a negative MBP, i.e., $\overline{m}(c, p) \in (-\infty, 0)$, are Pareto inferior (i.e., inequitable and inefficient) because they have a negative willingness-to-pay (WTP). On the other hand, the MBP tends to infinity, i.e., $\overline{m}(c, p) \to \infty$, as $c \to 0^+$ if $p > 0$. Since policies with $(c, p)$ in the second quadrant are Pareto superior (i.e., both efficient and equitable),  the MBP can be augmented by assigning a value of $\infty$ to such policies. \cite{hendren2020unified} use this notion to define an extension of the MBP.

\cite{hendren2020unified} define the marginal value of public funds (MVPF) on quadrants I, II, and IV of Figure \ref{figure:eetradeoff} as $m: \R^2 \setminus \R^2_{\leq 0} \to \R \cup \{\infty\}$ given by
$$m(c,p) = p/c\,\, \text{ if }\,\, c > 0, \quad m(c,p) = \infty \,\,\text{ if }\,\, c \leq 0 \text{ and } p \geq 0, \quad m(c,p) = \,\,? \,\,\text{ if }\,\, (c, p) \in \R^2_{< 0},$$ where $\,?\,$ denotes an undefined object. Whenever \cite{hendren2020unified} encounter either raw estimates or resampled estimates of $(c, p)$ that fall on $\R^2_{\leq 0}$, including many estimates that are far away from the origin $(0, 0)$, in their empirical analysis, the authors have missing values for the MVPF, consistent with the above definition. The MVPF is undefined on the third quadrant, where $(c, p) \in \R^2_{\leq 0}$ for policies such as some usual tax increases and spending cuts, because there are conceptual and mathematical difficulties in extending the MVPF to the third quadrant. Subsection \ref{subsec:economic_paradoxes} contains an in-depth discussion of this issue. Nevertheless, at least for the policies with $(c, p)$ on the restricted domain $\R^2 \setminus \R^2_{\leq 0}$, the MVPF can serve as a useful alternative to the MSS. Policies with a negative MVPF, i.e., $m(c, p) \in (-\infty, 0)$, are Pareto inferior; and policies with an infinite MVPF, i.e., $m(c, p) \in \{\infty\}$, are Pareto superior. This importance of this kind of relative welfare measurement is discussed next in Subsection \ref{subsec:importance_comparative}.

\subsection{The Importance of Comparative Welfare Analysis}
\label{subsec:importance_comparative}

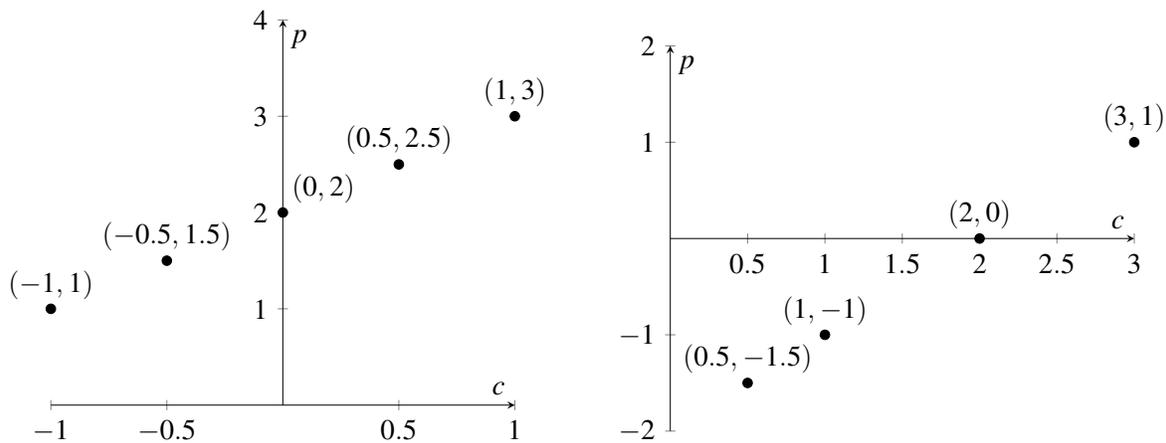
\begin{figure}[ht]
\caption{Limitations of Absolute Welfare Measures in Measuring the Equity--Efficiency Trade-off}
\label{fig:limitations_absolute}
\begin{center}
\begin{tikzpicture}[scale=0.9]
\begin{axis}[xlabel={$c$}, ylabel={$p$}, ymin=0,ymax=4,xmax=1,xmin=-1, axis lines = middle]
    \addplot[
        scatter/classes={a={black}, b={white}},
        scatter, mark=*, only marks, 
        scatter src=explicit symbolic,
        nodes near coords*={\Label},
        visualization depends on={value \thisrow{label} \as \Label} 
    ] table[meta=class, row sep=\\]
    {
        x y class label \\
        -1 1 a {$(-1,1)$} \\
        -0.5 1.5 a {$(-0.5,1.5)$} \\
        0 2 a {} \\
        0.175 2 b {$(0, 2)$} \\
        0.5 2.5 a {$(0.5, 2.5)$} \\
        1 3 a {$(1,3)$} \\
    };
\end{axis}
\end{tikzpicture}
\hspace{5mm}
\begin{tikzpicture}[scale=0.9]
\begin{axis}[xlabel={$c$}, ylabel={$p$}, ymin=-2,ymax=2,xmax=3,xmin=0, axis lines = middle]
    \addplot[
        scatter/classes={a={black}, b={red}},
        scatter, mark=*, only marks, 
        scatter src=explicit symbolic,
        nodes near coords*={\Label},
        visualization depends on={value \thisrow{label} \as \Label} 
    ] table[meta=class, row sep=\\]
    {
        x y class label \\
        0.5 -1.5 a {$(0.5,-1.5)$} \\
        1 -1 a {$(1,-1)$} \\
        2 0 a {$(2,0)$} \\
        3 1 a {$(3,1)$} \\
    };
\end{axis}
\end{tikzpicture}
\end{center}
\footnotesize\vspace{-2mm} \textit{Note}: The left figure shows points which lie on different quadrants of the $(c,p)$-plane but with a constant social surplus of two, i.e., $p - c = 2$. The right figure shows points which lie on different quadrants of the $(c,p)$-plane but with a constant excess burden of two, i.e., $p - c = -2$. Thus, the social surplus measure $p - c$ does not necessarily distinguish between points that lie on separate quadrants with different equity--efficiency trade-offs.
\end{figure}

Comparative welfare measures such as the MVPF, MBP, and MCF can provide useful information on the equity--efficiency trade-off that the absolute measures such as the MSS do not always reveal. For example, the MVPF orders the various $(c, p)$ points shown on the right panel of Figure \ref{fig:limitations_absolute} as follows: $(3,1) \succ (2, 0) \succ (1, -1) \succ (0.5, -1.5)$. More generally, if we fix $p - c = -2 \implies p/c = 1 - \frac{2}{c}$, then $p/c \to 1$ as $c \to \infty$, but $p/c < 0$ when $c < 2$, and $\lim_{c \to 0^+} p/c = - \infty$ when $p - c = -2$. However, the MSS is the same, i.e., $\beta(c, p) = p - c = -2$, for all of these points. Specifically, consider the points $(1, -1)$ and $(3, 1)$, and suppose they represent a counterproductive policy and a disability insurance policy, respectively. The point $(1, -1)$ representing the counterproductive policy is both inefficient and inequitable. The point $(3, 1)$ representing the disability insurance policy has a positive equity aspect even though it is more inefficient because it has higher net fiscal cost. However, the MSS does not distinguish between these points, and so the MSS in this case is not useful for policymakers who wish to fund disability insurance programs to improve equity while eliminating counterproductive policies. On the other hand, the MVPF can be very useful to those policymakers because it quantifies equity relative to inefficiency for the points $(1, -1)$ and $(3, 1)$. Similarly, the MVPF orders the various $(c, p)$ points shown on the left panel of Figure \ref{fig:limitations_absolute} as follows: $(1, 3) \prec (0.5, 2.5) \prec (0, 2) \preceq (-0.5, 1.5) \preceq (-1, 1)$. However, the MSS is the same, i.e., $\beta(c, p) = p - c = 2$, for all of these points. The MVPF again recognizes that the point $(-1, 1)$ is more efficient and equitable on a relative basis compared to the point $(1, 3)$, but the MSS fails to recognize the difference between these two economically different points.

The above critique of the MSS has been voiced before in the literature. For example, \cite{mayshar1990measures} says that the MSS and the MEB, which are unstandardized absolute measures, suffer from the ``index-number problem of ranking invariance to the choice of numeraire.'' \cite{mayshar1990measures} then endorses the MCF as ``the cornerstone concept of applied tax analysis'' and as an\begin{quote} \small \singlespacing \vspace{-5mm}
``intuitive measure of the marginal loss of surplus per dollar of additional revenue [that] avoids the impasse between the compensated and equivalent measures of changed surplus. Because it is constructed as a ratio of the marginal surplus and revenue, it is a unit free, pure number and is immune to the index-number problem of ranking invariance to the choice of numeraire.''
\end{quote}

Since the MBP and the MVPF are also relative measures like the MCF, the above endorsement also applies to the MBP and the MVPF. In addition, since the benefit-to-cost ratio (BCR) is just a transformation of the MSS, the above critique of the MSS also applies to the BCR, as \cite{hendren2020unified} concisely explain:\begin{quote} \small \singlespacing \vspace{-5mm}
``The difference between the MVPF and benefit--cost ratio in these cases reflects the fact that the benefit--cost ratio places all causal effects of the program in the numerator while the MVPF incorporates effects based on their incidence. In particular, the numerator of the MVPF captures the effects on beneficiaries while the denominator captures all effects on the government budget.''
\end{quote} 
In other words, the MVPF measures equity relative to inefficiency and does not combine them, whereas the MSS (and BCR) does not distinguish them very well because it is the sum of equity and efficiency. The MVPF and the MSS are useful in different ways, but the above examples clearly demonstrate that comparative welfare measures are economically important and provide useful information to policymakers. However, as I argue next, the existing relative welfare measures, such as the MVPF, have several limitations and give rise to many undesirable economic paradoxes.

\subsection{Economic Paradoxes in Inflexible Comparative Notions}
\label{subsec:economic_paradoxes}

As argued above, economically coherent comparative welfare measures are very useful to policymakers. However, as I show in this subsection, inflexible ways of comparing equity and efficiency can give rise to many undesirable economic paradoxes. Recognizing the issues with the raw ratio $p/c$, \cite{hendren2016policy} develops the MVPF to overcome some of the problems with it. However, I show that the MVPF also has several practical issues. To my knowledge, this paper is the first to clarify the paradoxes underlying the MVPF as well.

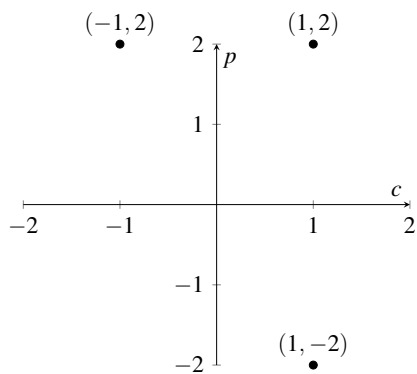
\begin{figure}[ht]
\caption{Limitations of the Raw Ratio of Equity to Inefficiency in Ordering Policies}
\label{fig:limitations_ratio}
\begin{center}
\begin{tikzpicture}[scale=0.75]
\begin{axis}[xlabel={$c$}, ylabel={$p$}, ymin=-2,ymax=2,xmax=2,xmin=-2, axis lines = middle]
    \addplot[
        scatter/classes={a={black}, b={red}},
        scatter, mark=*, only marks, 
        scatter src=explicit symbolic,
        nodes near coords*={\Label},
        visualization depends on={value \thisrow{label} \as \Label} 
    ] table[meta=class, row sep=\\]
    {
        x y class label \\
        -1 2 a {$(-1,2)$} \\
        1 2 a {$(1,2)$} \\
        1 -2 a {$(1,-2)$} \\
    };
\end{axis}
\end{tikzpicture}
\end{center}
\footnotesize\vspace{-2mm} \textit{Note}: The above figure shows three points on the $(c, p)$-plane that cannot be ordered sensibly using the raw ratio $p/c$.
\end{figure}

The issues with the raw ratio $p/c$ are obvious. While the ratio $p/c$ may be useful for comparing policies on only either the right half-plane or the left half-plane, it is problematic generally. For example, consider the three points $(-1, 2)$, $(1, 2)$, and $(1, -2)$ on different quadrants of the $(c, p)$-plane, as shown in Figure \ref{fig:limitations_ratio}. Since $(-2)/1 = 2/(-1)$, the raw ratio $p/c$ is indifferent between the points $(1, -2)$ and $(-1, 2)$, although common sense suggests that $(-1,2) \succ (1,2) \succ (1,-2)$. In addition, $\mathrm{lim}_{c \,\to\, 0^+} (2)/c = \infty$ while $\mathrm{lim}_{c \,\to\, 0^-} (2)/c = -\infty$, and so the raw ratio has undesirable paradoxical discontinuities.

Since there are Pareto superior (i.e., efficient and equitable) public projects for which $c < 0$ and $p > 0$, Hendren (2016) develops the MVPF, which replaces $p/c$ with $\infty$ on the second quadrant, to overcome the above issue with the raw ratio $p/c$. However, the MVPF is not defined on the third quadrant. For example, whenever \cite{hendren2020unified} encounter (raw or resampled) estimates of $(c, p)$ that fall on $\R^2_{\leq 0}$, including many resampled estimates that are far away from the origin $(0, 0)$, in their empirical analysis, the authors have missing values for the MVPF, consistent with the definition of the MVPF as $m: \R^2 \setminus \R^2_{\leq 0} \to \R \cup \{\infty\}$ given by
$$m(c,p) = p/c\,\, \text{ if }\,\, c > 0, \quad m(c,p) = \infty \,\,\text{ if }\,\, c \leq 0 \text{ and } p \geq 0, \quad m(c,p) = \,\,? \,\,\text{ if }\,\, (c, p) \in \R^2_{< 0},$$ where $\,?\,$ denotes an undefined object.

\begin{figure}[ht]
\caption{Difficulty with Extending the MVPF to the Third Quadrant Using the MCF}
\label{fig:third_quadrant_1}
\begin{center}
\includegraphics[scale=0.5]{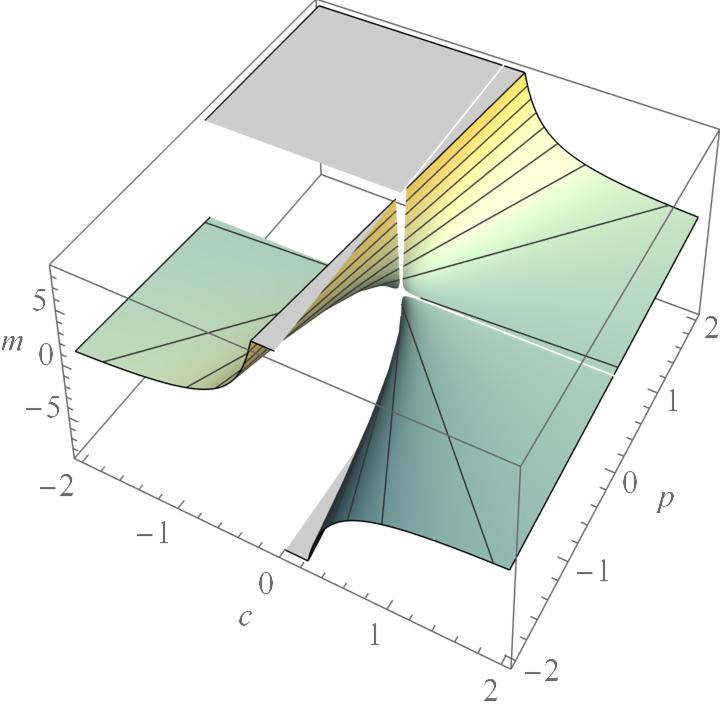}
\hspace{10mm}
\begin{tikzpicture}[scale=1] \scriptsize
\begin{axis}[xlabel={$c$}, ylabel={$p$}, ymin=-1000,ymax=1000,xmax=0.2,xmin=-0.2, axis lines = middle]
    \addplot[
        scatter/classes={a={black}, b={red}},
        scatter, mark=*, only marks, 
        scatter src=explicit symbolic,
        nodes near coords*={\Label},
        visualization depends on={value \thisrow{label} \as \Label} 
    ] table[meta=class, row sep=\\]
    {
        x y class label \\
        -0.1 -1000 a {$p/c = 10^4$} \\
        0.1 1000 a {$p/c = 10^4$} \\
        0.1 100 a {$p/c = 10^3$} \\
        -0.1 -400 a {$p/c = 4000$} \\
        0.1 -400 a {$p/c = -4000$} \\
    };
\end{axis}
\end{tikzpicture}
\end{center}
\footnotesize\vspace{-2mm} \textit{Note}: The left figure shows the surface plot of the MVPF $m(c, p)$ augmented with the MCF $\underline{m}(c,p) = (-p)/(-c) = p/c$ on the third quadrant of the $(c, p)$-plane. The resulting measure equals infinity on the second quadrant and $p/c$ elsewhere. Since this function has an infinite range, the above figure clips the extreme portions of the function (colored in gray). The right figure shows a few points on the $(c, p)$-plane that cannot be ordered sensibly using the augmented MVPF.
\end{figure}
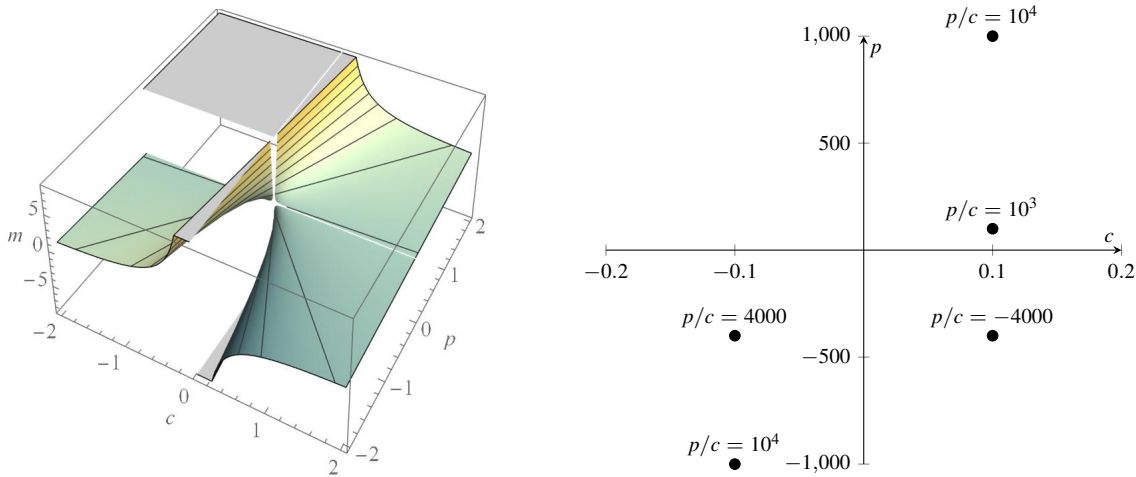

Conceptual and mathematical difficulties arise if one uses the MCF $\underline{m}(c,p) = p/c$ or its opposite $- \underline{m}(c,p) = -p/c$ to define the MVPF on the third quadrant. For example, if one sets $m(c, p) = \underline{m}(c,p) = p/c$ for all $(c, p) \in \R^2_{<0}$, then this would imply the following preference ordering, which is illustrated in Figure \ref{fig:third_quadrant_1}: $(-0.1,-1000) \sim (0.1, 1000) \succ (0.1, 100) \implies (-0.1, -1000) \succ (0.1, 100)$, since the $p/c$ ratio for both $ (-0.1, -1000)$ and $ (0.1, 1000)$ is $-1000/(-0.1) = 1000/0.1 = 10^4$, which is much higher than $10^3 = 100/0.1$, which is the $p/c$ ratio for $(0.1, 100)$. A similar reasoning would imply that $(-1, -10000) \succ (1, 1000)$ and also that $(-10, -10^6) \succ (10, 10^5)$ if the MCF is used to extend the MVPF to the third quadrant, resulting in a contradictory conclusion that harmful policies (with highly negative WTP) that generate negligible government revenue (when compared with the magnitude of the WTP) can have higher welfare than socially beneficial policies that have a negligible net fiscal cost to the government. This is in contrast to the following preference ordering based on common sense, even if we are interested only in measuring welfare using a comparative notion: $(0.1, 1000) \succ (0.1, 100) \succ (-0.1, -1000)$.

\begin{figure}[ht]
\caption{Difficulty with Extending the MVPF to the Third Quadrant Using the Inverted MCF}
\label{fig:third_quadrant_2}
\begin{center}
\includegraphics[scale=0.5]{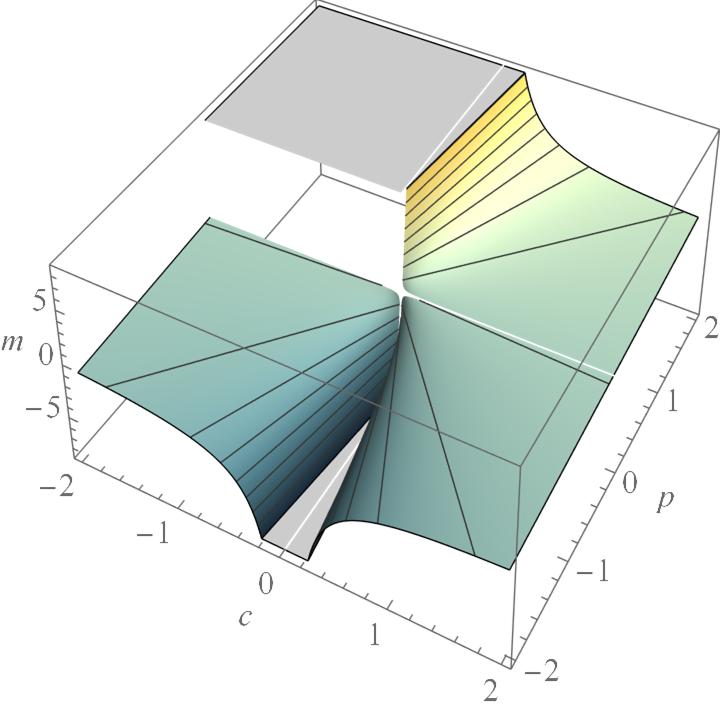}
\hspace{10mm}
\begin{tikzpicture}[scale=1] \scriptsize
\begin{axis}[xlabel={$c$}, ylabel={$p$}, ymin=-2,ymax=2,xmax=2,xmin=-2, axis lines = middle]
    \addplot[
        scatter/classes={a={black}, b={red}},
        scatter, mark=*, only marks, 
        scatter src=explicit symbolic,
        nodes near coords*={\Label},
        visualization depends on={value \thisrow{label} \as \Label} 
    ] table[meta=class, row sep=\\]
    {
        x y class label \\
        1 -2 a {$p/c = -2$} \\
        -1 -2 a {$-p/c = -2$} \\
        -1 -1 a {$-p/c = -1$} \\
        1 1 a {$p/c = 1$} \\
    };
\end{axis}
\end{tikzpicture}
\end{center}
\footnotesize\vspace{-2mm} \textit{Note}: The left figure shows the surface plot of the MVPF $m(c, p)$ augmented with inverted MCF $- \underline{m}(c,p) = -p/c$ on the third quadrant of the $(c, p)$-plane. The resulting measure equals infinity on the second quadrant and $p/|c|$ elsewhere. Since this function has an infinite range, the above figure clips the extreme portions of the function (colored in gray). The right figure shows a few points on the $(c, p)$-plane that cannot be ordered sensibly using the augmented MVPF.
\end{figure}
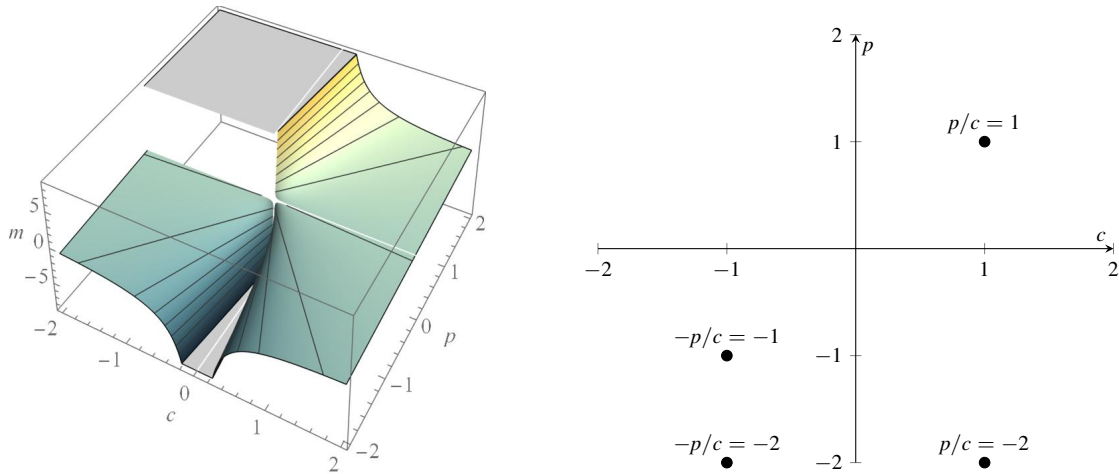

Another example (illustrated in the right panel of Figure \ref{fig:third_quadrant_1}) shows how the MVPF is puzzling: a policy with $(c,p) = (0.1, -400)$ has an MVPF of $-4000$, but a minuscule change in $c$ while keeping $p$ unchanged, such as changing the value of $(c,p)$ to $(-0.1, -400)$, causes the (augmented) MVPF to jump unreasonably to $4000$. In addition, as illustrated in the left panel of Figure \ref{fig:third_quadrant_1}, there is a huge discontinuity in the (augmented) MVPF at $p = 0$ when $c < 0$. For example, the MVPF of $(c, p) = (-10, 0.001)$ is $m(-10, 0.001) = \infty$, but the MCF of $(c, p) = (-10, -0.001)$ is $-0.001/(-10) = 0.0001$. Thus, moving from $(c, p) = (-10, 0.001)$ to $(c, p) = (-10, -0.001)$ causes the (augmented) MVPF to unreasonably change from $\infty$ to $0.0001$. Different paradoxes appear if we use $-$MCF to augment the MVPF, as shown in Figure \ref{fig:third_quadrant_2}.

Using the opposite of MCF to augment the MVPF on the third quadrant does not solve the problem. Specifically, attempting to set $m(c, p) = -\underline{m}(c,p) = -p/c < 0$ for all $(c, p) \in \R^2_{<0}$ would counterintuitively imply that all of the policies on the third quadrant are Pareto inferior, just like the policies on the fourth quadrant that have a negative MVPF. As shown in Figure \ref{fig:third_quadrant_2}, the augmented MVPF (using $-$MCF on third quadrant) would result in the preference ordering $(1, 1) \succ (-1, -1)$, implying that a tax enforcement policy that is neutral (i.e., without fiscal externalities) is always Pareto inferior (in contrast to a neutral tax cut). This would lead to the recommendation that the government should always prefer neutral tax cuts over neutral tax increases, which is in contrast with the common sense perspective that $(-1, -1) \sim (1, 1)$. In addition, as shown in the right panel of Figure \ref{fig:third_quadrant_2}, the augmented MVPF is different between $(-1, -2)$ and $(1, -2)$, because $-[(-2)/(-1)] = -2 = (-2)/(1)$. This is in contrast with the common sense perspective that $(-1, -2) \succ (1, -2)$, since $p = -2$ for both of these points but $(c, p) = (-1, -2)$ is efficient whereas $(c, p) = (1, -2)$ is inefficient.

Because of the issues explained above, several practically relevant economic paradoxes arise if the MVPF is augmented with either the MCF or $-$MCF on the third quadrant of the $(c, p)$-plane. As will be clear later in Section \ref{sec:axiomatic_construction}, these paradoxes can be eliminated if we use a measure that has a more flexible denominator than the MVPF (or its augmented version), which puts $p$ in the numerator and $c$ in the denominator. For comparative welfare analysis, the signs and relative magnitudes of $c$ and $p$ are the most important considerations. The choice of whether to put $c$ or $p$ in the denominator is arbitrary in a sense, because $p/c = 1/(c/p)$ and so $p/c$ is just the inverse of $c/p$. As discussed in Section \ref{sec:axiomatic_construction}, conveniently incorporating both $p/c$ and $c/p$ in a welfare measure can avoid the paradoxes that plague the augmented versions of the MVPF, which has an inflexible denominator. Thus, the MVPF can be meaningfully used to only make comparisons between policies not on the third quadrant (i.e., $(c, p) \not\in \R^2_{\leq 0}$), and direct numerical comparisons are only possible between policies with $c > 0$. Thus, it is difficult to create a complete preference ordering with the MVPF. In addition, the non-numerical MVPF value (i.e., $\infty$) on the second quadrant and the unbounded nature of MVPF make it difficult to aggregate welfare, as discussed in Subsection \ref{subsec:limitations_existing}.

\subsection{Limitations of Existing Comparative Measures for Welfare Aggregation}
\label{subsec:limitations_existing}

Suppose $(c_j, p_j)_{j = 1}^J$ are the values of net fiscal cost and willingness-to-pay for either $J$ separate policies or for a single policy by $J$ population subgroups. Since the subscripts $j$ are agnostic as to whether $(c_j, p_j)_{j = 1}^J$ relate to either $J$ different policies or a single policy for $J$ population subgroups, the discussion in this subsection applies to both contexts. How should we aggregate welfare across policies or across population subgroups using a unit-free welfare measure $\gamma(c,p)$, such as the MVPF $m(c,p)$? \cite{hendren2020unified} suggest calculating welfare $\gamma(C(\lambda), P(\lambda))$ of a ``category average'' $(C(\lambda), P(\lambda)) = \sum_{j = 1}^J \lambda_j\times(c_j, p_j)$ where $\lambda = (\lambda_1, \dots, \lambda_J) \in \R^J_+$ are subjective scaling factors. However, welfare of scaling-factor-weighted (i.e., $\lambda$-weighted) policy average is difficult to interpret unless, e.g., $\{(c_j, p_j)\}_{j = 1}^J$ are measured per-capita for a federal policy implemented across $J$ states and $\lambda$ contains the population weights. However, if alternatively $w = (w_1, \dots, w_J) \in \R^J_+$ represent subjective importance weights a policymaker attaches to welfare from different policies, then it is straightforward to interpret the importance-weighted welfare sum $\sum_{j = 1}^J w_j\,\gamma(c_j, p_j)$ as the policymaker's total utility.

Consider a hypothetical ``policy category'' (in \citeauthor{hendren2020unified}'s (\citeyear{hendren2020unified}) terminology) with two policies $(c^\prime_1,p^\prime_1) = (-4,0.5)$ and $(c^\prime_2,p^\prime_2) = (2,-1)$. If the policymaker attaches the subjective scaling weights $\lambda'_1 = \lambda'_2 = 0.5$ to these policies, then \cite{hendren2020unified} define the MVPF of the ``category average'' as $m(C(\lambda'), P(\lambda'))$ where $(C(\lambda'), P(\lambda')) = \lambda'_1 (c^\prime_1,p^\prime_1) + \lambda'_1 (c^\prime_2, p^\prime_2) = [(-4,0.5) + (2,-1)]/2 = (-1,-0.25)$. Although both $m(c^\prime_1, p^\prime_1) = \infty$ and $m(c^\prime_2, p^\prime_2) = -0.5$ are defined, the MVPF of the category average is undefined, i.e., $m(C, P) = m(-1, -0.25) = \,?$. In addition, the average MVPF with importance weights $w^\prime_1 = w^\prime_2 = 0.5$, i.e., $w^\prime_1\,m(c^\prime_1, p^\prime_1) +w^\prime_2\, m(c^\prime_2, p^\prime_2) = (\infty - 0.5)/2$, is also ambiguous. Even if we consider non-additive types of aggregation such as multiplicative aggregation using a geometric average, this issue of ambiguity persists, because it is unclear how to multiply infinite MVPF values and negative MVPF values. In addition, geometric averages are not very meaningful in this context because MVPF can take negative values, unlike the non-negative ratios of prices. Therefore, even if the MVPFs of individual policies are defined, the MVPF of ``cateogory average'' and the average MVPF may be undefined or ambiguous.

The above concerns are not merely theoretical. For example, the above examples, $(c^\prime_1,p^\prime_1)$ and $(c^\prime_2,p^\prime_2)$, can be replaced with estimates for two policies considered by \cite{hendren2020unified}: ``Tax Deduction for Postsecondary Tuition, Single Filers at Phase Start'' (Tuition Deduc. (SS)) and ``Hope and Lifetime Learners Tax Credits'' (HOPE/LLC). \cite{hendren2020unified} estimate $(c^\prime_1,p^\prime_1) = (-5.10, 5.38)$ for the former policy and estimate $(c^\prime_2,p^\prime_2) = (4.86, -42.82)$ for the latter policy. In this case, the ``category average'' is given by $(C, P) = \lambda'_1 (c^\prime_1,p^\prime_1) + \lambda'_1 (c^\prime_2, p^\prime_2) = [(-5.10, 5.38) + (4.86, -42.82)]/2 = (-0.24,-37.44)$. Again, both $m(c^\prime_1, p^\prime_1) = m(-5.10, 5.38) = \infty$ and $m(c^\prime_2, p^\prime_2) = m(4.86, -42.82) = -8.81$ are defined, but the MVPF of the category average is undefined, i.e., $m(C, P) = m(-0.24,-37.44) = \,?$, and the average MVPF, i.e., $w^\prime_1\,m(c^\prime_1, p^\prime_1) +w^\prime_2\, m(c^\prime_2, p^\prime_2) = (\infty - 8.81)/2$, is also ambiguous in this case.

\begin{figure}[ht]
\caption{Example of MVPF's Limitations in Aggregating Welfare of ``College Adult'' Policies}
\vspace{-5mm}
\label{fig:welfare_agg_example}
\begin{center}
\fbox{
\begin{tikzpicture}[scale=1.2] \tiny
\begin{axis}[xlabel={$c$}, ylabel={$p$}, axis lines = middle]
    \addplot[
        scatter/classes={a={black}, b={blue}},
        scatter, mark=*, only marks, 
        scatter src=explicit symbolic,
        nodes near coords*={\Label},
        visualization depends on={value \thisrow{label} \as \Label} 
    ] table[meta=class, row sep=\\]
    { 
        x y class label \\
        1.57 3.42 a 2.18 \\
        0.53 5.36 a {} \\
        0.42 5.27 a 12.58 \\
        4.86 -42.82 a $-8.81$ \\
        1.29 1 a 0.77 \\
        1.38 -0.03 a {} \\
        -1.13 1 a $\infty$ \\
        -5.1 5.38 a  $\infty$ \\
        0.48 -2.68 b $-5.59$ \\
    };
\end{axis}
\end{tikzpicture}
}
\fbox{ \footnotesize \,\,\,
\begin{tabular}{c} 
 \hphantom{\quad\,\,} $m(\frac{1}{8}\sum_j c_j, \, \frac{1}{8}\sum_j p_j) = m(0.48, -2.68) = -5.59$ \hphantom{\quad\quad} \\ and $\frac{1}{8} \sum_j m(c_j, p_j) = \,\,?$
\end{tabular} 
} \vspace{5mm}

\fbox{
\begin{tikzpicture}[scale=1.2]  \scriptsize
\begin{axis}[xlabel={$c$}, ylabel={$p$}, axis lines = middle] 
    \addplot[
        scatter/classes={a={black}, b={blue}},
        scatter, mark=*, only marks, 
        scatter src=explicit symbolic,
        nodes near coords*={\Label},
        visualization depends on={value \thisrow{label} \as \Label} 
    ] table[meta=class, row sep=\\]
    { 
        x y class label \\
        1.57 3.42 a 2.18 \\
        0.53 5.36 a {} \\
        0.42 5.27 a 12.58 \\
        0.486 -4.282 a $-8.81$ \\
        1.29 1 a 0.77 \\
        1.38 -0.03 a {} \\
        -1.13 1 a $\infty$ \\
        -5.1 5.38 a  $\infty$ \\
        -0.07 2.14 b $\infty$ \\
    };
\end{axis}
\end{tikzpicture}
}
\fbox{ \footnotesize
\begin{tabular}{c} 
 \hphantom{\quad\quad-;} $m(\frac{1}{8}\sum_j c_j, \, \frac{1}{8}\sum_j p_j) = m(-0.07, 2.14) = \infty$ \hphantom{\quad\quad- -} \\ and $\frac{1}{8} \sum_j m(c_j, p_j) = \,\,?$
\end{tabular}
}
\end{center}
\footnotesize \textit{Note}: The upper figure shows a scatter plot of eight points (using black markers) indicating the $(c, p)$ values of policies in the ``College Adult'' category (as defined by \cite{hendren2020unified}), along with labels indicating the MVPF values of these eight policies. The MVPF of the ``category average'' for these policies is  $m(\frac{1}{8}\sum_j c_j, \, \frac{1}{8}\sum_j p_j) = m(0.48, -2.68) = -5.59$ (indicated above the blue marker), but their average MVPF is ambiguous, i.e., $[\sum_j m(c_j, p_j)]/8 = \,\,?$. The lower figure shows the same scatter plot but with only one modification: for the policy with MVPF equal to $-8.81$, its $(c, p)$ value is rescaled by a factor of 0.1 without changing its MVPF ($-8.81$). With this modification, the MVPF of the ``category average'' jumps to  $m(\frac{1}{8}\sum_j c_j, \, \frac{1}{8}\sum_j p_j) = m(-0.07, 2.14) = \infty$.
\end{figure}
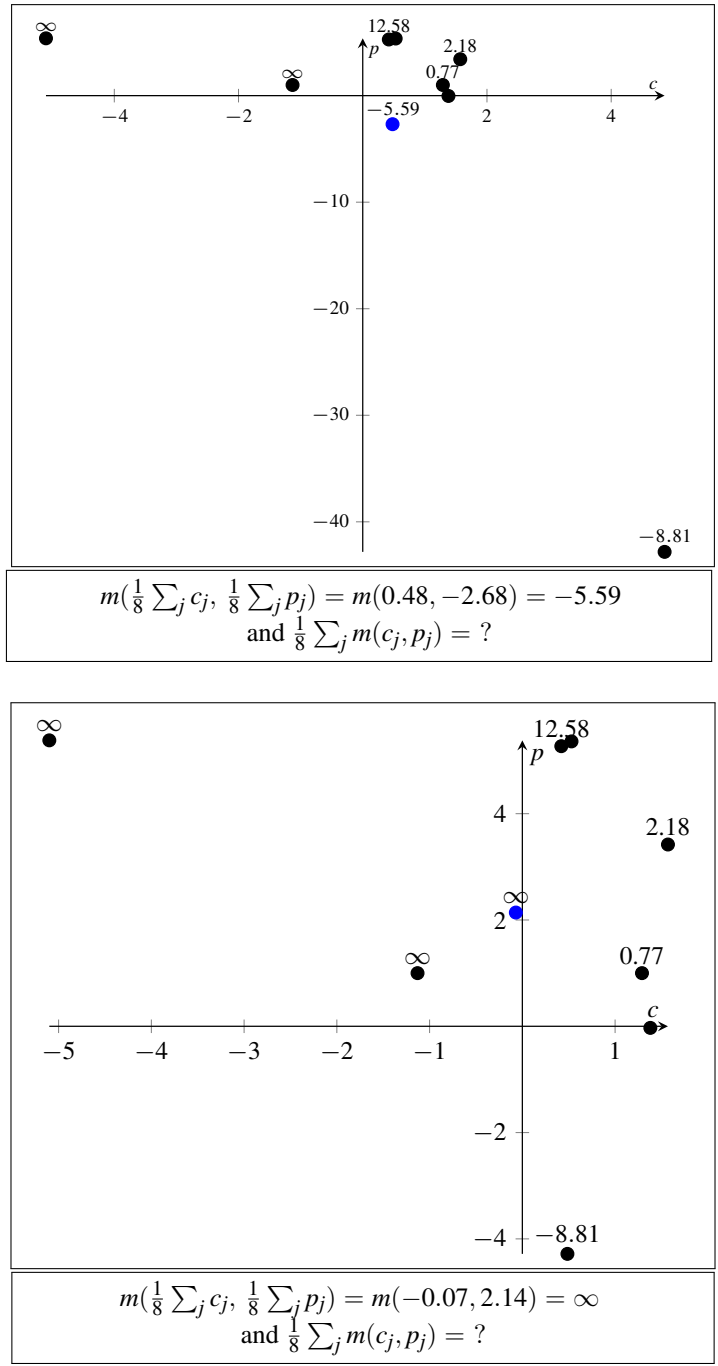

\cite{hendren2020unified} analyze a ``policy category'' consisting of eight policies that they label ``College Adult'' policies; see their paper for descriptions of these policies. The upper panel of Figure \ref{fig:welfare_agg_example} shows a scatter plot of eight points indicating their $(c, p)$ values along with labels indicating their MVPF values. Using this ``College Adult'' policy category, I further demonstrate the MVPF's limitations in aggregating welfare across policies (or across population subgroups).

\clearpage

As shown in the upper panel of Figure \ref{fig:welfare_agg_example}, the MVPF of the ``category average'' of the ``College Adult'' policies is  $m(\frac{1}{8}\sum_j c_j, \, \frac{1}{8}\sum_j p_j) = m(0.48, -2.68) = -5.59$, but their average MVPF is ambiguous, i.e., $[\sum_j m(c_j, p_j)]/8 = \,\,?$. The lower panel of Figure \ref{fig:welfare_agg_example} shows the same scatter plot as the upper panel but with only one modification: for the HOPE/LLC policy with MVPF equal to $-8.81$, its $(c, p)$ value is rescaled by a factor of 0.1 without changing its MVPF ($-8.81$). With this modification, the MVPF of the ``category average'' jumps to  $m(\frac{1}{8}\sum_j c_j, \, \frac{1}{8}\sum_j p_j) = m(-0.07, 2.14) = \infty$, but the new average MVPF is still ambiguous, i.e., $[\sum_j m(c_j, p_j)]/8 = \,\,?$ because the associated MVPF values are unchanged.

Mathematically, the jump in the MVPF of the ``category average'' from $m(0.48, -2.68) = -5.59$ to $m(-0.07, 2.14) = \infty$ is understandable because $(0.48, -2.68)$ and $(-0.07, 2.14)$ represent two hypothetical policies with very different equity--efficiency trade-offs; specifically, the former is Pareto inferior and the latter is Pareto superior. However, it is very difficult to interpret the MVPF of the ``category average'' $(C(\lambda), P(\lambda)) = \sum_{j = 1}^J \lambda_j\times(c_j, p_j)$ in this context (i.e., for the ``College Adult'' policy category) because it represents a hypothetical policy with arbitrary scaling that was never implemented. Such a ``category average'' would have been easier to interpret if the eight points in Figure \ref{fig:welfare_agg_example} represented per-capita measurements for the same policy implemented in different states and if the scaling factors were population weights.

On the other hand, even if $(c_j, p_j)_{j = 1}^J$ are the net fiscal costs and the WTP values associated with different policies implemented in different years and settings, it is possible to intuitively interpret the weighted average of a unit-free welfare measure $\gamma(c, p)$. Specifically, if $w = (w_1, \dots, w_J) \in \R^J_+$ represent subjective importance weights a policymaker attaches to welfare from different policies, then it is straightforward to interpret the importance-weighted welfare sum $\sum_{j = 1}^J w_j\,\gamma(c_j, p_j)$ as the policymaker's total utility. Unlike $\gamma(C(\lambda), P(\lambda))$, where $(C(\lambda), P(\lambda))$ need not always represent a policy that was actually implemented, the weighted sum $\sum_{j = 1}^J w_j\,\gamma(c_j, p_j)$ is a weighted sum of welfare values associated with policies that were actually implemented. It would be desirable to compute both $\sum_{j = 1}^J w_j\,\gamma(c_j, p_j)$ and $\gamma(C(\lambda), P(\lambda))$ for an arbitrary set of policies and weightings or scaling factors, but the MVPF-based welfare aggregation is problematic on both fronts, as demonstrated in this subsection.

\subsection{Desirable Qualities of a Comparative System for Welfare Calculations}
\label{subsec:desirable_qualities}

Even though $c$ and $p$ can be measured using well-developed microeconomic foundations \citep{bhattacharya2022empirical, finkelstein2020welfare}, empirical researchers combine $c$ and $p$ on an ad hoc basis to measure welfare based on the seemingly convenient ratio $p/c$ or the difference $p - c$. To my knowledge, the existing public finance literature does not have a sound micro-theoretic framework that first specifies a set of preference orderings (for the equity--efficiency trade-off) and then derives the appropriate metric for those preference orderings to measure welfare on a relative basis. While the MVPF $m(c,p)$ and the MSS $\beta(c,p)$ both have desirable properties, they have different implications for policy-making preferences. Rather than thinking backwards about the implications of a given ad hoc welfare measure, we could instead build a welfare measure from the ground up. Before moving on to Section \ref{sec:axiomatic_construction} that develops a new axiomatic framework for comparative welfare analysis, it is useful to take stock of useful properties of existing measures. 

If we would like to use a metric $\varphi(c, p)$ for comparative welfare analysis, what properties or qualities would be desirable to have in $\varphi(c, p)$? For comparative welfare analysis, the signs and relative magnitudes of $c$ and $p$ are the most important considerations. Thus, for $\varphi(c, p)$ to be even regarded as a comparative welfare measure, it should at the very least have the unit-free property that the MVPF has. Specifically, $\varphi(c, p)$ needs to be homogeneous of degree zero, i.e., $\varphi(\lambda\,c,\lambda\,p) = \varphi(c, p)$ for all $(c, p) \in \R^2$ and any $\lambda > 0$.

In addition, there will necessarily be a discontinuity at the origin $(0, 0)$ for any relative measure such as the MVPF that distinguishes Pareto superior points (in quadrant II), which are both efficient and equitable, from Pareto inferior points (in quadrant IV), which are both inefficient and inequitable. If we would like $\varphi(c, p)$ to have this property (like the MVPF), then we would require that $\varphi(c', p') > \varphi(c'', p'')$ for any $(c', p') \in \R_{< 0} \times \R_{> 0}$ and any $(c'', p'') \in \R_{> 0} \times \R_{< 0}$. Thus, $\varphi(c, p)$, like the MVPF, would be discontinuous at $(0, 0)$. However, this requirement does not impose discontinuities in other parts of the real plane. In fact, it would be desirable if $\varphi(c, p)$ is continuous everywhere else because infinitesimal changes in $(c, p)$ outside the origin should not drastically change welfare. (This desirable continuity property is of course not satisfied by the augmented MVPF, as shown in Subsection \ref{subsec:economic_paradoxes}.)

The MVPF (minus one) has a very useful interpretation as relative shortfall in willingness-to-pay when $c > 0$ and $p \in [-c, c]$. Note that $m(c, p) - 1 = p/c - 1 = p/c - c/c = (p - c)/c = \beta(c, p)/c$, which measures the extent to which $p$ falls short of $c$ on a relative basis when $c > 0$ and $p \in [-c, c]$. Thus, it would also be desirable for the comparative welfare measure $\varphi(c, p)$ to have this property, i.e., $\varphi(c, p) = p/c - 1$ whenever $c > 0$ and $p \in [-c, c]$. This would imply that $(3, 1) \succ (1, -1)$ from a comparative perspective using $\varphi(c, p)$ because the point $(1, -1)$ has a worse relative shortfall (since $\frac{-1 - 1}{1} = -2 < \frac{1 - 3}{3} = -2/3$) than the point $(3, 1)$, even though both points have the same social surplus, i.e., $(1) - (3) = (-1) - (1) = -2$. Thus, $\varphi(c, p)$ would, like the MVPF, provide information about the equity--efficiency trade-off that is not necessarily conveyed by the MSS.

In addition, both the MSS and the MVPF (augmented with the MCF) satisfy the preference ordering  $(-1, 1) \succ (-1, -1) \sim (1, 1) \succ (1, -1)$ for the four points shown in Figure \ref{fig:square_vertices} that are the vertices of a square centered at the origin. Thus, it would be desirable to have the measure $\varphi(c, p)$ satisfy the inequality $\varphi(-1, 1) > \varphi(-1, -1) = \varphi(1, 1) > \varphi(1, -1)$. Since $\varphi(c, p)$ is a desired comparative measure, $\varphi(-\lambda, \lambda) > \varphi(-\lambda, -\lambda) = \varphi(\lambda, \lambda) > \varphi(\lambda, -\lambda)$ for all $\lambda > 0$ as well.

\clearpage

\begin{figure}[ht]
\caption{Points of Agreement Between the MSS and the Augmented MVPF}
\label{fig:square_vertices}
\begin{center}
\begin{tikzpicture}[scale=0.6]
\begin{axis}[xlabel={$c$}, ylabel={$p$}, ymin=-1,ymax=1,xmax=1,xmin=-1, axis lines = middle]
    \addplot[
        scatter/classes={a={black}, b={red}},
        scatter, mark=*, only marks, 
        scatter src=explicit symbolic,
        nodes near coords*={\Label},
        visualization depends on={value \thisrow{label} \as \Label} 
    ] table[meta=class, row sep=\\]
    {
        x y class label \\
        -1 1 a {$(-1,1)$} \\
        1 1 a {$(1,1)$} \\
        1 -1 a {$(1,-1)$} \\
        -1 -1 a {$(-1,-1)$} \\
    };
\end{axis}
\end{tikzpicture}
\end{center}
\footnotesize\vspace{-2mm} \textit{Note}: The above figure shows four points that have the same order using both the MSS and the augmented MVPF.
\end{figure}
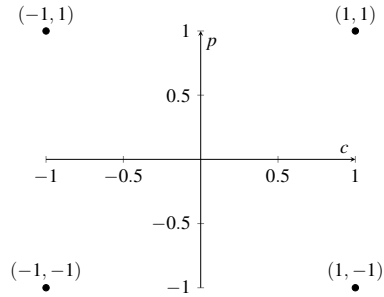

\begin{figure}[ht]
\begin{center}
\caption{Points With Symmetric and Asymmetric Equity--Efficiency Trade-offs}
\label{figure:symmetrical_points}
\begin{tikzpicture}[scale=0.9]
\begin{axis}[xlabel={$c$}, ylabel={$p$}, ymin=-2,ymax=2,xmax=2,xmin=-2, axis lines = middle]
\draw (-2,-2) to (2,2);
\draw (-2,2) to (2,-2);
\node[rotate=90] at (0.15,0.85) {\scriptsize -- more equitable $\to$};
\node[rotate=90] at (0.15,-1) {\scriptsize $\longleftarrow$ less equitable ---};
    \addplot[
        scatter/classes={a={white}, b={red}},
        scatter, mark=, only marks, 
        scatter src=explicit symbolic,
        nodes near coords*={\Label},
        visualization depends on={value \thisrow{label} \as \Label} 
    ] table[meta=class, row sep=\\]
    {
        x y class label \\
        1 0 a {\scriptsize --- less efficient $\longrightarrow$} \\
         -1 0 a {\scriptsize $\longleftarrow$ more efficient ---} \\
    };
    \addplot[
        scatter/classes={a={black}, b={red}},
        scatter, mark=*, only marks, 
        scatter src=explicit symbolic,
        nodes near coords*={\Label},
        visualization depends on={value \thisrow{label} \as \Label} 
    ] table[meta=class, row sep=\\]
    {
        x y class label \\
        2 1 a {\footnotesize $(2, 1)$} \\
        1 2 a {\footnotesize $(1, 2)$} \\
        -1 -2 a {\footnotesize $(-1, -2)$} \\
        -2 -1 a {\footnotesize $(-2, -1)$} \\
        -1 2 a {\footnotesize $(-1, 2)$} \\
        -2 1 a {\footnotesize $(-2, 1)$} \\
        1 -2 a {\footnotesize $(1, -2)$} \\
        2 -1 a {\footnotesize $(2, -1)$} \\
    }; 
\end{axis}
\end{tikzpicture}
\end{center}
\footnotesize \textit{Note}: The above figure shows examples of policies on eight different sub-quadrants of the $(c,p)$-plane that are associated with different equity--efficiency trade-offs, along with diagonal ($p = c$) and antidiagonal ($p = -c$) axes. Points that are reflections about the antidiagonal axis, such as $(1,2)$ and $(-2,-1)$, can be thought of as having symmetric welfare values (due to similar equity--efficiency trade-offs). Points that are reflections about the diagonal axis, such as $(1,2)$ and $(2,1)$, can be thought of as having asymmetric welfare values (due to dissimilar equity--efficiency trade-offs).
\end{figure}
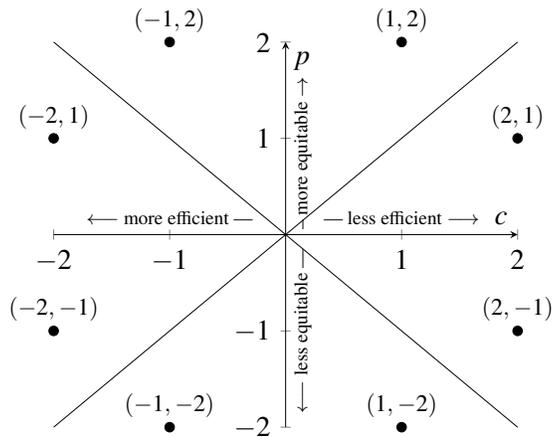

Finally, counterintuitively, and most importantly, I argue that comparative welfare analysis should actually take inspiration from absolute welfare measures in recognizing some important symmetries and asymmetries in equity--efficiency trade-offs on the $(c, p)$-plane. After all, there is no complete disconnect between the absolute and comparative approaches, although they obviously do differ in significant ways. Note that there is a symmetry between equity $p = \eta_p\,\tilde{p}$, which is the adjusted willingness-to-pay, and efficiency $-c = - \eta_c\,\tilde{c}$, the adjusted net fiscal revenue. The policymaker's subjective scaling factors $\eta_p$ and $\eta_c$ not only enable cross-policy comparisons but also enable substitutability between $p$ and $-c$ from an absolute welfare perspective. Because of the flexible scaling factors $\eta_p$ and $\eta_c$, the policymaker can weight $p$ and $-c$ equally without loss of generality, giving rise to the MSS $\beta(c, p) = p - c$ as a natural measure of absolute welfare.

I argue that the notion of equal substitutability between $p$ and $-c$ is also key to conducting sound comparative analysis. For example, as shown in Figure \ref{figure:symmetrical_points}, points that are reflections about the antidiagonal axis (i.e., the line $p = -c$), such as $(1,2)$ and $(-2,-1)$, can be thought of as having symmetric welfare values. They of course have the same surplus, e.g., $\beta(1, 2) = (2) - (1) = 1 = (-1) - (-2) = \beta(-2, -1)$, but it would also be desirable to have $(1, 2) \sim (-2, -1)$ on a comparative basis because these points are similar except for the symmetric switch in the labels; specifically, the point $(c, p) = (1, 2)$ is twice as ``equitable'' as it is ``inefficient,'' whereas the point $(c, p) = (-2, -1)$ is twice as ``efficient'' as it is ``inequitable.'' Similarly, it would be useful if the desired comparative metric $\varphi(c, p)$ emulated the MSS $\beta(c, p)$ in satisfying, e.g., the following indifference preference orderings: $(-2, 1) \sim (-1, 2)$, $(-1, -2) \sim (2, 1)$, and $(1, -2) \sim (2, -1)$. In other words, it would be desirable to have $\varphi(c', p') = \varphi(-p', -c')$ for all $(c', p') \in \R^2$.

In addition, as shown in Figure \ref{figure:symmetrical_points}, points that are reflections about the diagonal axis (i.e., about the break-even line $p = c$), such as $(1,2)$ and $(2,1)$, can be thought of as having asymmetric welfare values. They of course have opposite surpluses, e.g., $\beta(1, 2) = 2 - 1 = 1$, which is the opposite of $-1 = 1 - 2 = \beta(2, 1)$. In the same spirit, it would also be desirable to have $\varphi(2, 1) = -\varphi(1, 2)$ on a comparative basis because these points have an asymmetric switch in the labels; specifically, the point $(c, p) = (1, 2)$ is twice as ``equitable'' as it is ``inefficient,'' whereas the point $(c, p) = (2, 1)$ is twice as ``inefficient'' as it is ``equitable.'' Similarly, it would be useful if the desired comparative metric $\varphi(c, p)$ emulated the MSS $\beta(c, p)$ in satisfying, e.g., the following preferences: $\varphi(2,-1) = -\varphi(-1, 2)$, $\varphi(1, -2) = -\varphi(-2,1)$, and $\varphi(-1,-2) = -\varphi(-2, -1)$. In other words, it would be desirable to have $\varphi(c', p') = -\varphi(p', c')$ for all $(c', p') \in \R^2$. We can also combine this with the aforementioned desirable property and re-express it as $\varphi(c', p') = -\varphi(p', c') = -\varphi(-c',-p') \implies \varphi(c', p')  = -\varphi(-c',-p')$ for all $(c', p') \in \R^2$.

To summarize, the above discussion suggests that it would be desirable to have the comparative measure $\varphi$ satisfy the following two properties that involve different types of equity--efficiency trade-offs:
$$\varphi(c', p') = -\varphi(p', c') \iff \varphi(c', p') + \varphi(p', c') = 0 \text{ for all } (c', p') \in \R^2 \text{, and}$$
$$\varphi(c', p') = -\varphi(-c', -p') \iff \varphi(c', p') + \varphi(-c', -p') = 0 \text{ for all } (c', p') \in \R^2.$$
If we also require $\varphi$ to satisfy a tautological condition that $\varphi(c,p) = 0$ if and only if $p = c$ (i.e., there is no shortfall or excess in the net fiscal cost or the willingness-to-pay relative to the other), the above properties can be re-expressed as follows: 
$$\varphi(c', p') + \varphi(p', c') = 0 = \varphi(c' + p', p' + c') \text{ for all } (c', p') \in \R^2 \text{, and}$$
$$\varphi(c', p') + \varphi(-c', -p') = 0 = \varphi(c' - c', p' - p') \text{ for all } (c', p') \in \R^2.$$
There are two additional ways to interpret the above equations. The first interpretation involves the notion of ``welfare-balanced'' policy changes, which are a subset of the broader set of ``budget-neutral'' policy changes. The second interpretation involves different types of welfare aggregation.

Suppose a policymaker is considering a policy change from an equiweighted combination of two policies $\{(c_1, p_1), (c_2, p_2)\}$ to a third policy $(c_3, p_3)$. Then, the policy change is considered budget-neutral if $c_1 + c_2 = c_3$. In addition to budget-neutrality, if it is also the case that $p_1 + p_2 = p_3 = c_3$, then we may say that the policy change is ``welfare-balanced,'' meaning that the budget-neutral policy change has zero social surplus. Thus, a policy change from the combination $\{(c', p'), (p', c')\}$ to the policy $(c' + p', p' + c')$ is welfare-balanced. Similarly, a policy change from the combination $\{(c', p'), (-c', -p')\}$ to the policy $(c' - c', p' - p') = (0, 0)$ is also welfare-balanced.

Another related but different interpretation involves two types of welfare aggregation (without invoking welfare-balanced policy changes). As discussed in Section \ref{subsec:limitations_existing}, there are two ways of aggregating welfare of two policies $(c_1, p_1)$ and $(c_2, p_2)$. First, based on the policymaker's importance weights $w_1, w_2 \in \R_{> 0}$ for these policies, a weighted sum of their welfare can be formed as follows: $w_1\,\varphi(c_1, p_1) + w_2\,\varphi(c_2, p_2)$. Second, based on the policymaker's subjective scaling factors $\lambda_1, \lambda_2 \in \R_{> 0}$ for these policies, the welfare of a hypothetical weighted policy can be formed as follows: $\varphi(\lambda_1\,c_1 + \lambda_2\,c_2,\,\lambda_1\,p_1 + \lambda_2\,p_2)$. The above equations say that both aggregation methods should produce the same result if $\lambda_1 = \lambda_2 = w_1 = w_2 = 1$ and if one of the following conditions holds: either $c_2 = p_1$ and $p_2 = c_1$; or $c_2 = -c_1$ and $p_2 = -p_1$. In other words, when there are equal importance weights (or equal scaling factors) and the policies being aggregated are opposites (from both the absolute and comparative perspectives), the choice of the welfare aggregation method does not matter.

The above discussion can be thought of as a search for points of agreement, when they exist, between the relative and absolute approaches to measuring welfare. However, the MVPF (or its augmented version) does not satisfy the above desirable properties that are satisfied by the MSS (since $\beta(c', p') = -\beta(p', c')$ and $\beta(c', p') = -\beta(-c', -p')$ for all $(c', p') \in \R^2$). Thus, there is huge scope for improvement on the MVPF for relative analysis. Nevertheless, the MVPF (minus one) does have the desirable unit-free property (or degree-zero homogeneity, i.e., $m(\lambda\,c, \lambda\,p) - 1 = m(c, p) - 1$ for all $\lambda > 0$) and interpretation as relative shortfall in willingness-to-pay (i.e., $m(c, p) - 1 = (p - c)/c = p/c - 1$) when $c > 0$ and $p \in [-c, c]$, which are two properties not satisfied by the MSS. Unlike the difference-based measure MSS, the MVPF ``is immune to the index-number problem of ranking invariance to the choice of numeraire'' \citep{mayshar1990measures}. This observation naturally leads to the following questions: Is there a measure satisfying all of the four aforementioned desirable properties? If so, is it unique? Section \ref{sec:axiomatic_construction} has my answers to these important questions! 

\clearpage

\section{Axiomatic Construction of the Relative Policy Value}
\label{sec:axiomatic_construction}

The goal of this section is to construct a $(c,p)$-based function $\varphi: \R^2 \to \R$ that satisfies the intuitive and desirable properties discussed in Subsection \ref{subsec:desirable_qualities} for comparative welfare analysis. In Subsection \ref{subsec:rpv_theorem}, I rehash those desirable qualities as ``welfare-symmetry axioms'' and then prove that there is a unique econo-metric that satisfies the axioms. In Subsection \ref{subsec:rpv_interpretations}, I provide various intuitive ways to interpret my new welfare measure that I call the Relative Policy Value (RPV). In Subsection \ref{subsec:general_class}, I axiomatically develop a more general class of comparative welfare indices, although the RPV stands out among them because the other indices do not satisfy all the welfare-symmetry axioms.

\subsection{Welfare-Symmetry Axioms and the RPV Uniqueness Theorem}
\label{subsec:rpv_theorem}

Based on the detailed discussion in Subsection \ref{subsec:desirable_qualities}, I reiterate the desirable properties (for relative welfare measurement) in the form of three ``welfare-symmetry axioms'' as follows, although the first one is a semi-definition rather than a full-fledged axiom.

\begin{axiom}
\label{axiom:1}
Let $(c,p) \in \R^2$. Then, $\varphi(c,p) = 0$ if and only if $c = p$. In addition, if $(c', p') \in \R^2$ such that $c' > 0$ and $p' \in [-c', c']$, then $\varphi(c',p') = \beta(c', p')/c' = (p' - c')/c' = p'/c' -1 = m(c',p') - 1$.  
\end{axiom}

\begin{axiom}
\label{axiom:2}
For all $(c', p') \in \R^2$, $\varphi(c', p') = -\varphi(p', c') \iff \varphi(c', p') + \varphi(p', c') = 0$.
\end{axiom}

\begin{axiom}
\label{axiom:3}
For all $(c', p') \in \R^2$, $\varphi(c', p') = -\varphi(-c', -p') \iff \varphi(c', p') + \varphi(-c', -p') = 0$.
\end{axiom}

The first part of Axiom \ref{axiom:1} is very basic: there is zero welfare if the net fiscal cost breaks even with the willingness-to-pay. The second part of Axiom \ref{axiom:1} relates the new desired welfare measure to both the MSS and MVPF. Thus, Axiom \ref{axiom:1} makes the new measure $\varphi$ interpretable as the shortfall in willingness-to-pay relative to the net fiscal cost when $c > 0$ and $p \in [-c , c]$. As explained in Subsection \ref{subsec:rpv_interpretations}, Axiom \ref{axiom:1} (along with the two other axioms) also implicitly provides $\varphi$ interpretation as the relative shortfall or excess (in equity or efficiency) on other parts of the $(c, p)$-plane as well. In addition, by construction, Axiom \ref{axiom:1} makes the desired welfare measure $\varphi$ inherit the unit-free property of the MVPF, so there is no need to specify zero-degree homogeneity as a separate axiom. Indeed, as stated in Remark \ref{remark:homogeneous_degree_zero}, the welfare-symmetry axioms imply zero-degree homogeneity of $\varphi$. Thus, Axiom \ref{axiom:1} makes $\varphi$ is an ``intuitive measure [that] avoids the impasse between the compensated and equivalent measures of changed surplus'' and is ``a unit free, pure number and is immune to the index-number problem of ranking invariance to the choice of numeraire'' \citep{mayshar1990measures}.

Axiom \ref{axiom:2} is inspired by the MSS-based notion of exchangeability or substitutability between equity ($p$) and efficiency ($-c$). It simply states that $\varphi(c,p) + \varphi(p,c) = 0$, which also holds for the MSS $\beta(c,p) = p - c$. Axiom \ref{axiom:3} states that $\varphi$ is odd symmetric: $\varphi(c,p) + \varphi(-c,-p) = 0 = \varphi(0,0) = \varphi((c,p) + (-c,-p)) \iff \varphi(c,p) = - \varphi(-c,-p)$, which is an asymmetry that is discussed in detail in Subsection \ref{subsec:desirable_qualities}. This property also holds for $\beta(c,p)$. Even though Axiom \ref{axiom:2} only states that $\varphi(c,p) + \varphi(p,c) = 0$, the notion of exchangeability of $p$ and $-c$ is much more apparent when Axioms \ref{axiom:2} and \ref{axiom:3} are used together. Axiom \ref{axiom:3} implies that $\varphi(p,c) = - \varphi(-p,-c)$, and so Axioms \ref{axiom:2} and \ref{axiom:3} together make the notion of exchangeability explicit: $\varphi(c,p) = - \varphi(p,c) = -[ - \varphi(-p,-c)] \iff \varphi(c,p) = \varphi(-p,-c)$, which is another symmetry that is discussed in detail in Subsection \ref{subsec:desirable_qualities}. Figure \ref{figure:symmetrical_points} provides visual intuition and motivation for both Axioms \ref{axiom:2} and \ref{axiom:3}. 

It is easy to see that the standardized versions of MSS and MVPF $\beta(c,p)/c = m(c,p) - 1$ satisfy Axiom \ref{axiom:1} when $c > 0$ but do not satisfy Axioms \ref{axiom:2} and \ref{axiom:3}. Observe that $[m(2,1) - 1] + [m(1,2) - 1] = 1/2 -1 + 2/1 - 1 = 0.5 \neq 0$, violating Axiom \ref{axiom:2}. In addition, $[m(1,-1) -1] + [m(-1,1) -1] = -1-1 + \infty -1 \neq 0$, violating Axiom \ref{axiom:3}. Thus, the MVPF (minus one) does not satisfy these additive symmetries.\footnote{One could argue that additive symmetries are not appropriate for a measure like the MVPF and that multiplicative symmetries need to be used instead. However, there are issues with using multiplicative versions of Axioms \ref{axiom:2} and \ref{axiom:3}. While the MVPF (augmented with the MCF) satisfies $m(c, p)\,m(p, c) = (p/c)\,(c/p) = 1$ and $m(c, p)/m(-c,-p) = (p/c)/[(-p)/(-c)] = 1$ when $(c, p) \in \R^2_{<0} \cup \R^2_{>0}$, it is difficult to more generally implement this notion on the other quadrants when $(c, p) \not\in \R^2_{<0} \cup \R^2_{>0}$ because the MVPF is infinity when $c < 0$ and $p > 0$. Another reason to use additive symmetries rather than multiplicative symmetries is that welfare aggregation for multiple policies (even if they are more than two in number) is much more convenient with additive forms, as shown in Section \ref{sec:welfare_agg}. Multiplicative forms of aggregation have difficulty in combining positive and negative welfare values. Therefore, I work with additive symmetries in Axioms \ref{axiom:2} and \ref{axiom:3}.} In contrast, the MSS $\beta(c,p)$ satisfies Axioms \ref{axiom:2} and \ref{axiom:3} as well as the first part of Axiom \ref{axiom:1} but not the second part of Axiom \ref{axiom:1}, since $\beta(c,p) = p - c \neq (p - c)/c = m(c,p) - 1$ in general. This raises the question of whether there exists a metric satisfying the welfare-symmetry axioms. Indeed, the function $\phi$ in Definition \ref{definition:rpv} that I call the Relative Policy Value (RPV) is such a metric by Lemma \ref{lemma:rpv_axiom}. In addition, Theorem \ref{theorem:rpv} states that the RPV is the unique function that satisfies the welfare-symmetry axioms. The RPV is a real-valued degree-zero homogeneous function such that $\phi(\R^2) = [-2, 2]$. It is an even function about the antidiagonal axis (i.e., $p = -c$) and an odd function about the diagonal axis (i.e., $p = c$). I provide multiple intuitive interpretations and descriptions of the RPV in Subsection \ref{subsec:rpv_interpretations}.

\begin{definition}
\label{definition:rpv}
The Relative Policy Value $\phi$ on $\R^2$ equals zero at the origin but otherwise equals
$$\phi(c,p) = \beta(c, p)\,/\,\lVert (c, p) \rVert_\infty = \frac{p - c}{\mathrm{max}\{\left|\,p \, \right|,\left|\,c\, \right|\}}.$$
\end{definition}

\begin{lemma}
\label{lemma:rpv_axiom}
The Relative Policy Value (RPV) satisfies the welfare-symmetry axioms.
\end{lemma}

\begin{theorem}
\label{theorem:rpv}
The only function obeying the welfare-symmetry axioms is the Relative Policy Value.
\end{theorem}

\begin{remark}
\label{remark:homogeneous_degree_zero}
The Relative Policy Value (RPV) is homogeneous of degree zero.
\end{remark}

\begin{figure}
\begin{center}
\caption{Implications of the Welfare-Symmetry Axioms}
\label{fig:proof_graphical}
	\begin{subfigure}[b]{0.45\textwidth}
         \centering
         \caption{Implication of Axiom \ref{axiom:1} for Contours}
         \includegraphics[scale=.75]{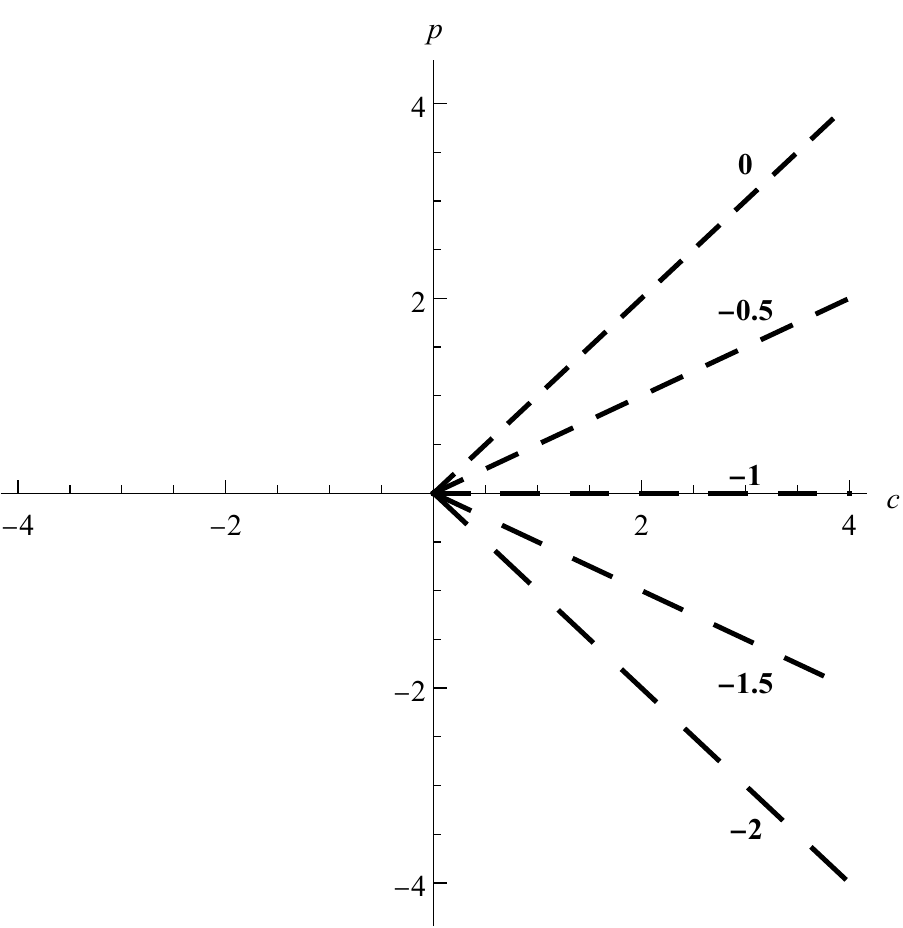}
         \label{fig:axiom1}
     \end{subfigure}
     \begin{subfigure}[b]{0.45\textwidth}
         \centering
         \caption{Implication of Axioms \ref{axiom:1}--\ref{axiom:2} for Contours}
         \includegraphics[scale=.75]{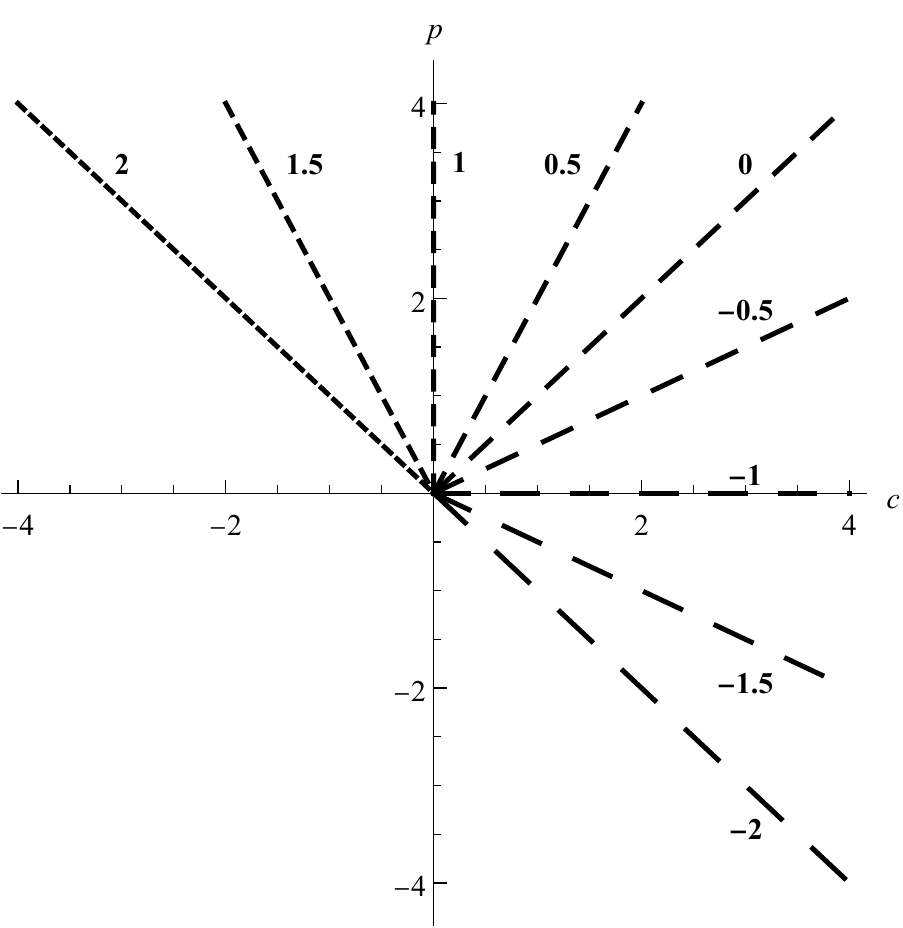}
         \label{fig:axiom2}
     \end{subfigure}
     \begin{subfigure}[b]{0.45\textwidth}
         \centering
         \includegraphics[scale=.75]{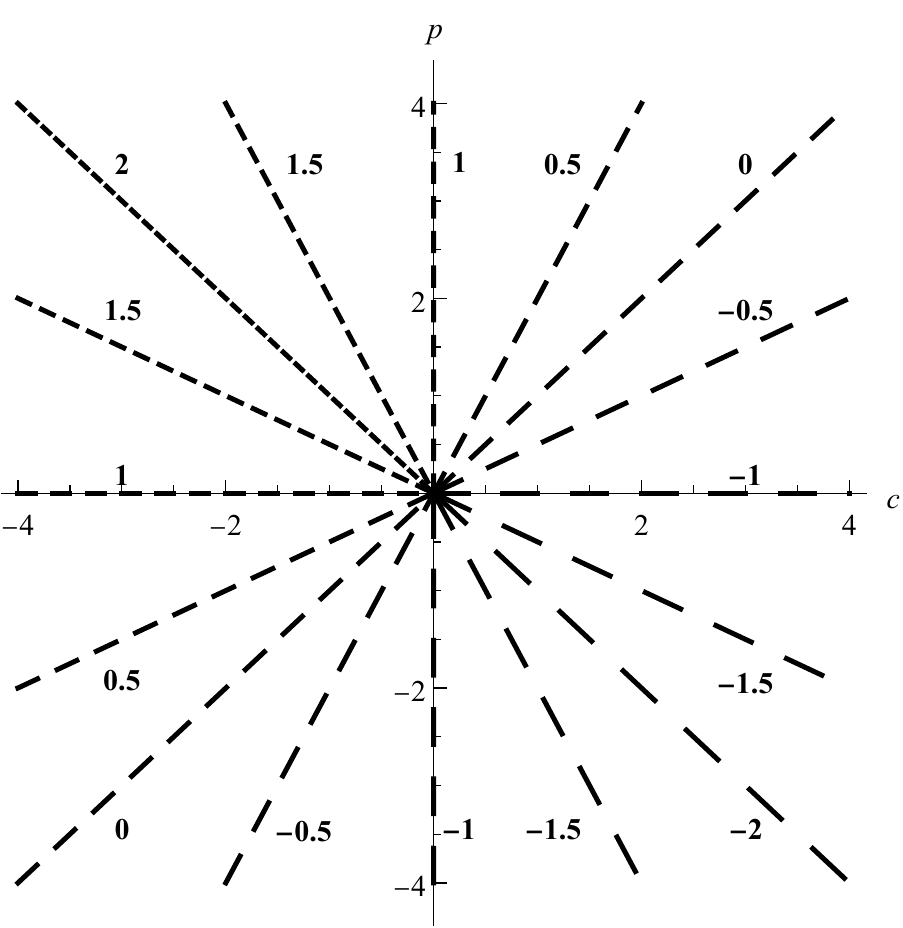}
         \caption{Implication of Axioms \ref{axiom:1}--\ref{axiom:3} for Contours}
         \label{fig:axiom3}
     \end{subfigure}
     \begin{subfigure}[b]{0.45\textwidth}
         \centering
         \includegraphics[scale=.5]{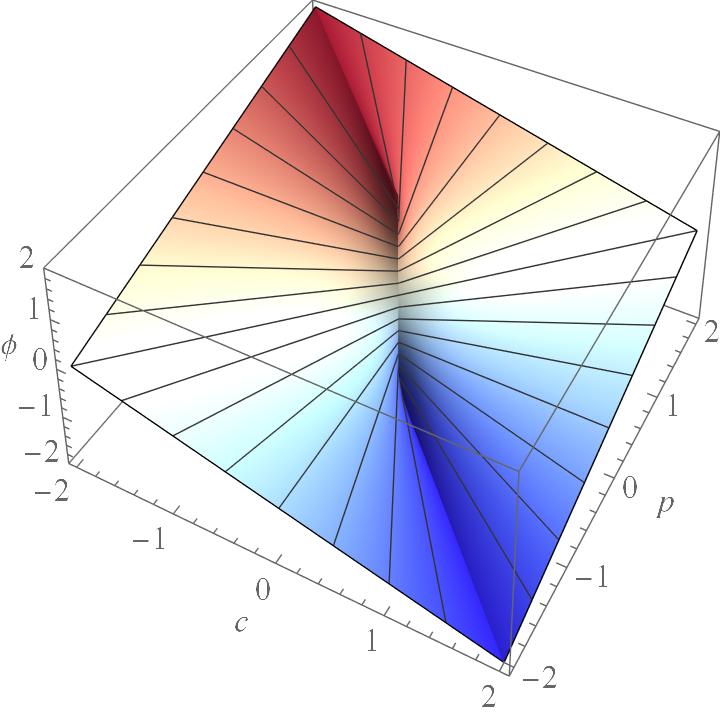}
         \caption{Implication of Axioms \ref{axiom:1}--\ref{axiom:3} for Surface}
         \label{fig:phisp}
     \end{subfigure}
\end{center}
\footnotesize \textit{Note}: Figure \ref{fig:axiom1} shows the implication of Axiom \ref{axiom:1} for the contours of the welfare index $\varphi$. The thick dashed lines represent contours, and their values are indicated near them in boldface. Figure \ref{fig:axiom2} shows the implication of Axioms \ref{axiom:1} and \ref{axiom:2} for the contours of the welfare index $\varphi$. Figure \ref{fig:axiom3} and \ref{fig:phisp} show the implication of Axioms \ref{axiom:1}, \ref{axiom:2}, and \ref{axiom:3} for the contour and surface plots of the welfare index, which coincide with those of the Relative Policy Value (RPV).
\end{figure}

\noindent The proofs of Lemma \ref{lemma:rpv_axiom}, Theorem \ref{theorem:rpv}, and Remark \ref{remark:homogeneous_degree_zero} are given in the \ref{appendix:proofs}. By Remark \ref{remark:homogeneous_degree_zero}, $\phi(\lambda c,\lambda p) =\phi(c,p)$ for all $\lambda > 0$, but this degree-zero homogeneity property is also shared by MVPF when $c > 0$. A consequence of this property is that MVPF and RPV can be used to compare different policies regardless of their scales (i.e., magnitudes of their marginal net fiscal costs). While this feature of RPV and MVPF may not be sensible in many practical contexts, relaxing the simplistic degree-zero homogeneity has a trade-off with the ease of welfare computation and interpretation. Echoing \cite{hendren2020unified}, ``Future work could explore how the MVPF [and RPV] for a given policy change [vary with] a program’s size of spending.'' Another note regarding Remark \ref{remark:homogeneous_degree_zero} (in the context of RPV) is that it can be further generalized when combined with Axiom \ref{axiom:3}. Specifically, $\phi(\lambda c,\lambda p) =\sgn(\lambda)\,\phi(c,p)$ for all $\lambda \in \R$. A disadvantage of Remark~\ref{remark:homogeneous_degree_zero} is that comparisons cannot be made between policies that lie on the same ray. For example, two policies with $(c_1,p_1) = (0, 1)$ and $(c_2,p_2) = (0, 2)$ have the same RPV (equal to one), even though the second policy seems better using the MSS, since $\beta(0,2) = 2 > 1 = \beta(0,1)$. This drawback is shared by the MVPF; in addition, the MVPF does not distinguish between policies with $c < 0$ and those with $c = 0$. Thus, policymakers should ideally use multiple measures (e.g., the RPV together with the MSS, which is homogeneous of degree one) to make their decisions.

It is instructive to use Figure \ref{fig:proof_graphical} to graphically understand Theorem \ref{theorem:rpv} about the uniqueness of RPV. Figure \ref{fig:axiom1} shows an implication of Axiom \ref{axiom:1}. In the region where $c > 0$ and $p \in [-c, c]$, the contours of the desired welfare index $\varphi$ are the same as those of the standardized MVPF, i.e., $m(c, p) - 1$. Next, Axiom \ref{axiom:2} implies that the contours in Figure \ref{fig:axiom1} can be reflected about the diagonal axis ($p = c$) but with a change in the sign of their values, as shown in Figure \ref{fig:axiom2}. Furthermore, Axioms \ref{axiom:2} and \ref{axiom:3} imply that the contours in Figure \ref{fig:axiom2} can be reflected about the antidiagonal axis ($p = -c$), as shown in Figure \ref{fig:axiom3}. Figure \ref{fig:phisp} shows the surface plot of $\varphi$ associated with the contour plot of $\varphi$ shown in Figure \ref{fig:axiom3}. However, the surface plot of $\varphi$, shown in in Figure \ref{fig:phisp}, is identical to the surface plot of $\phi$. Therefore, $\varphi \equiv \phi$, and so the RPV is the unique welfare measure satisfying the welfare-symmetry axioms. As Remark \ref{remark:dsi_uniqueness} states, Axiom \ref{axiom:1} is key to the uniqueness of RPV.

\begin{remark}
\label{remark:dsi_uniqueness}
Without Axiom \ref{axiom:1}, there exist multiple indices that satisfy Axioms \ref{axiom:2} and \ref{axiom:3} as well as degree-zero homogeneity, i.e., $\varphi(\lambda\,c, \lambda\,p) = \varphi(c, p)$ for all $\lambda > 0$ and any $(c, p) \in \R^2$.
\end{remark}

\begin{figure}[ht]
\begin{center}
\caption{Surface Plots of $\zeta$ and $[\phi-\zeta]$}
\label{fig:zeta}
	\begin{subfigure}[b]{0.45\textwidth}
         \centering
         \includegraphics[scale=.5]{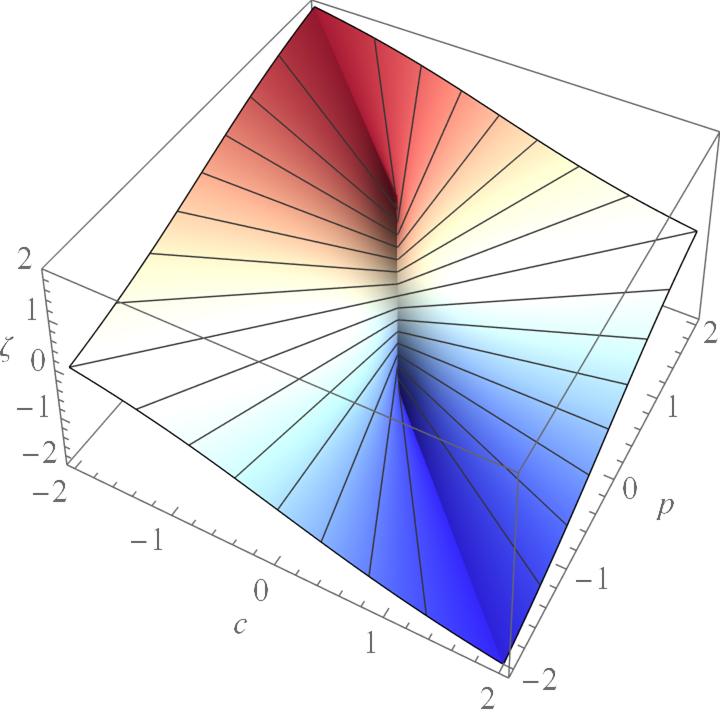}
         \caption{Surface Plot of $\zeta$}
         \label{fig:zeta_3dplot}
     \end{subfigure}
     \begin{subfigure}[b]{0.45\textwidth}
         \centering
         \includegraphics[scale=.5]{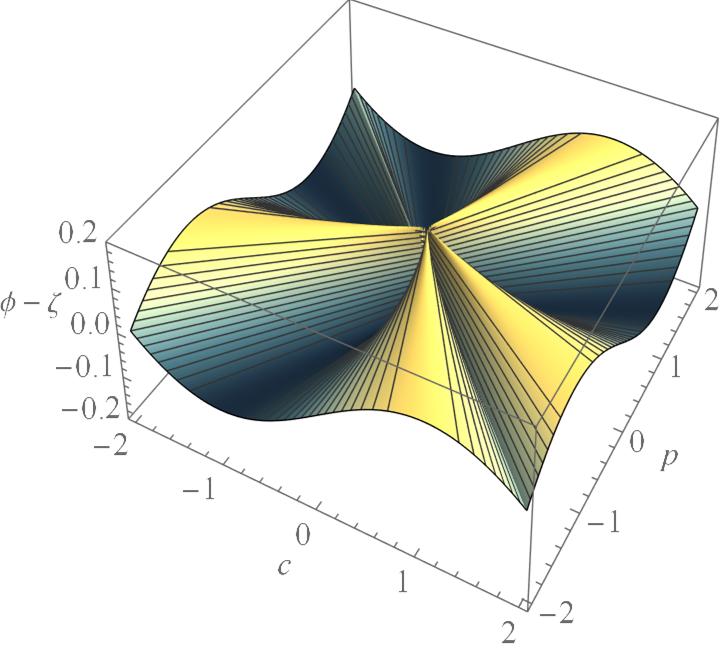}
         \caption{Surface Plot of $[\phi-\zeta]$}
         \label{fig:phi-zeta_3dplot}
     \end{subfigure}
\end{center}
 \footnotesize \textit{Note}: Figure \ref{fig:zeta_3dplot} shows the surface plot of $\zeta$, which looks similar to but is slightly more curved than the surface plot of $\phi$. This is more apparent in Figure \ref{fig:phi-zeta_3dplot}, which shows the surface plot of $[\phi - \zeta]$. 
\end{figure}

The uniqueness of the RPV is pinned down by Axiom \ref{axiom:1}, especially its second sentence. Without this, there exist multiple functions satisfying Axioms \ref{axiom:2} and \ref{axiom:3} as well as the unit-free property. An example demonstrating Remark \ref{remark:dsi_uniqueness} is the function $\zeta: \R^2 \to [-2,2]$ given by\footnote{The function $\zeta$ simply rotates the input vector (by 45 degrees clockwise) and obtains the angle of the rotated vector, before finally scaling the angle appropriately so that the index ranges from $-2$ and 2.}
$$\zeta(c,p) = (4/\pi)\,\sgn(c + p)\,\mathrm{arctan}(p'/c'),$$ where $(c',p') = R(-\pi/4)\cdot (c,p) \text{ and } R(\theta) = \Big[\begin{smallmatrix} \mathrm{cos}(\theta) & -\mathrm{sin}(\theta) \\ \mathrm{sin}(\theta) & \mathrm{cos}(\theta) \end{smallmatrix}\Big]$, which is a rotation matrix. Due to the symmetries inherent in the construction of $\zeta$, this function satisfies Axioms \ref{axiom:2} and \ref{axiom:3} as well as zero-degree homogeneity. However, by Theorem \ref{theorem:rpv}, $\zeta$ does not fulfil all of the welfare-symmetry axioms. In addition, $\zeta$ does not have an intuitive interpretation in terms of relative shortfall or excess in equity or efficiency, which is an interpretation that is unique to the RPV, distinguishing it from the other measures. Nevertheless, $\zeta$ is sometimes very close to the RPV $\phi$. Figure \ref{fig:zeta_3dplot} displays the surface plot of $\zeta$, which is similar to but slightly more curved than that of $\phi$. This is more apparent in Figure \ref{fig:phi-zeta_3dplot}, which shows the surface plot of the function $[\phi - \zeta](c,p) = \phi(c,p) - \zeta(c,p)$.

\begin{figure}
\begin{center}
\caption{Contour Plot of the Relative Policy Value (RPV)}
\vspace{-3mm}
\label{fig:rpv_square}
\includegraphics[scale=1]{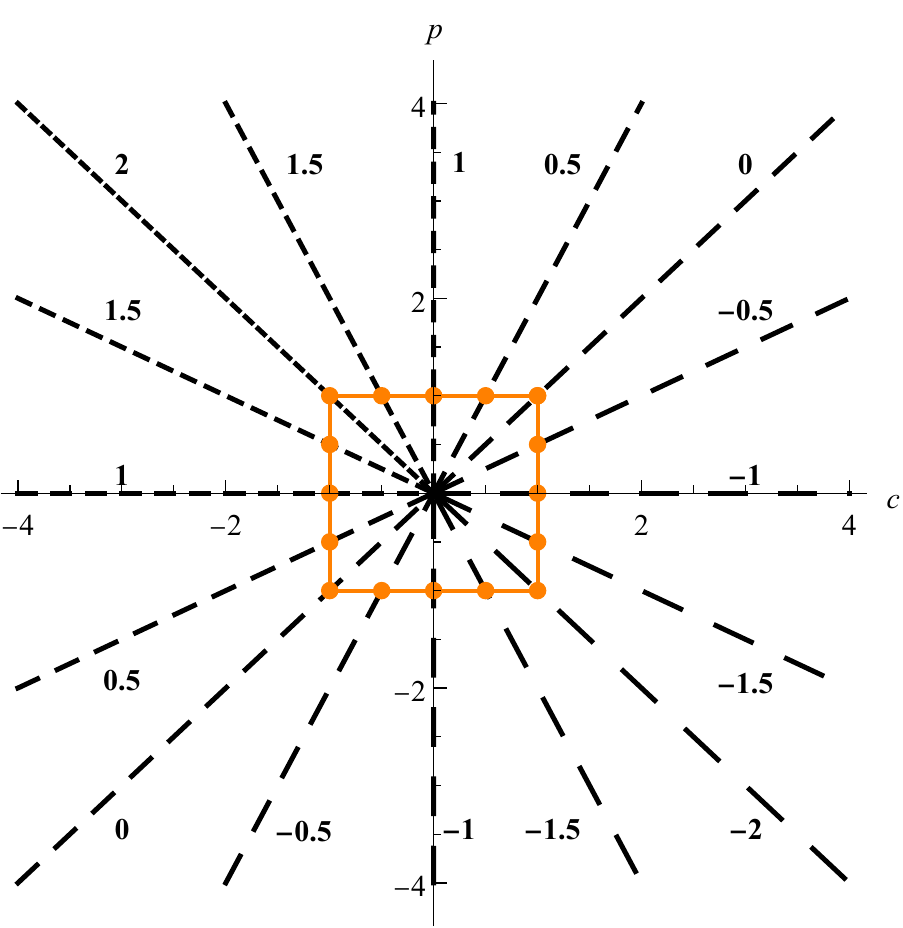}
\end{center}
\footnotesize\vspace{-2mm} \textit{Note}: In the above graph, the thick dashed lines represent contours of the Relative Policy Value (RPV), which normalizes the social surplus $(p - c)$ by the maximum norm $||(c,p)||_\infty = \mathrm{max}\{|\,p\,|,|\,c\,|\}$. The values of the contours are indicated near them in boldface. In addition, the square with vertices $\{-1, 1\} \times \{-1, 1\}$ is superimposed on the RPV contours.
\end{figure}

\begin{figure}
\begin{center}
\caption{Surface Plot of the Relative Policy Value (RPV)}
\label{fig:rpv_surface}
\includegraphics[scale=0.6]{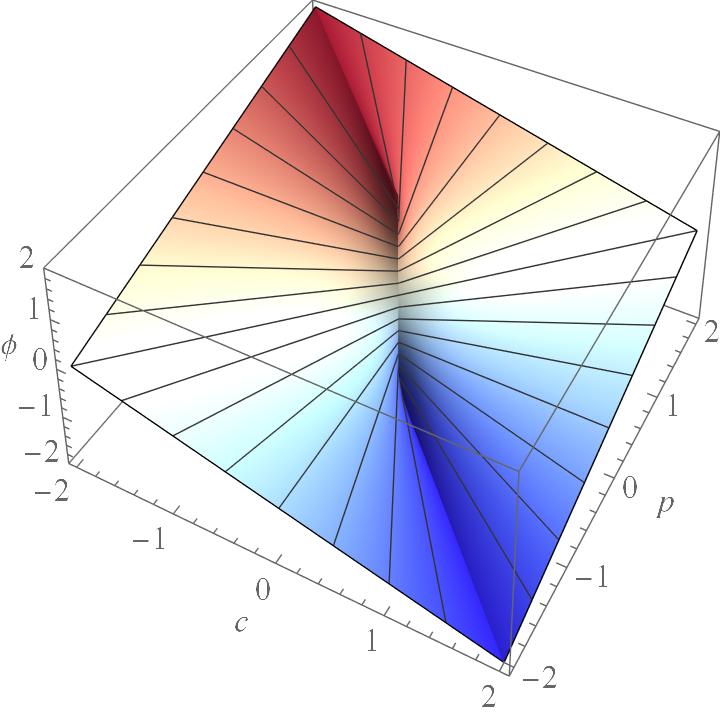}
\end{center}
\footnotesize\vspace{-2mm} \textit{Note}: The above graph shows a three-dimensional picture of the surface plot of the Relative Policy Value (RPV).
\end{figure}

\subsection{Intuitive Interpretations of the Relative Policy Value}
\label{subsec:rpv_interpretations}

There are at least three ways to interpret the Relative Policy Value (RPV). The descriptive interpretation of RPV is based on the notion of substitutability or exchangeability between equity ($p$) and efficiency ($-c$). The RPV compares their sum relative to the most salient or dominant feature, which could be (in)equity or (in)efficiency depending on the values of $|\,p\,|$ and $|\,c\,|$. In other words,
$$\text{Relative Policy Value} = \phi(c, p) = \frac{(p) + (- c)}{\mathrm{max}\{\left|\,p \, \right|,\left|\,c\, \right|\}} = \frac{\mathrm{equity} + \mathrm{efficiency}}{\text{magnitude of the dominant feature}}.$$
Unlike the MVPF, the RPV uses a flexible denominator, exploiting the substitutability between $p$ and $-c$. Since it is obvious that the both sign and magnitude of the social surplus $p - c = (p) + (-c)$ are mostly controlled by the magnitude of the dominant feature, the RPV views the social surplus relative to the magnitude of the dominant feature, resulting in a welfare index with the range $[-2, 2]$.

\clearpage

\begin{table}
\setstretch{1.5}
\caption{Descriptive Interpretation of the Relative Policy Value (RPV)}
\label{table:rpv_interpretation}
\vspace{-5mm}
    \begin{center} \footnotesize
    \begin{tabular}{c|l} 
    \hline\hline
        \textit{RPV} $\phi(c, p) = \tilde{\phi}$ & \textit{Interpretation}: The policy is ``$\dots$'' \\ \hline
        $\tilde{\phi} = 2$ & ``as equitable as it is efficient'' \\ \hline
        $\tilde{\phi} \in (1, 2)$ & either ``more equitable than it is efficient,'' or ``more efficient than it equitable'' \\ \hline
        $\tilde{\phi} = 1$ & either ``equitable with no efficiency/inefficiency,'' or ``efficient with no equity/inequity'' \\ \hline
        $\tilde{\phi} \in (0, 1)$ & either ``more equitable than it is inefficient,'' or ``more efficient than it is inequitable'' \\ \hline
        $\tilde{\phi} = 0$ & either ``as equitable as it is inefficient,'' or ``as efficient as it is inequitable'' \\ \hline
        $\tilde{\phi} \in (-1, 0)$ & either ``more inefficient than it is equitable,'' or ``more inequitable than it is efficient'' \\ \hline
        $\tilde{\phi} = -1$ & either ``inequitable with no efficiency/inefficiency,'' or ``inefficient with no equity/inequity'' \\ \hline
        $\tilde{\phi} \in (-2, -1)$ & either ``more inefficient than it is inequitable,'' or ``more inequitable than it is inefficient'' \\ \hline
        $\tilde{\phi} = -2$ & ``as inequitable as it is inefficient'' \\ \hline \hline
    \end{tabular}
    \end{center}
    \vspace{-1mm}
\setstretch{1}\noindent \footnotesize
\textit{Note}: This table provides interpretations associated with various possible numbers of the Relative Policy Value (RPV).
\end{table}

\begin{figure}[ht]
\begin{center}
\caption{Interpretation of RPV as Relative Shortfall or Excess in Efficiency or Equity}
\label{fig:rpv_interpretation_shortfall}
\begin{tikzpicture}[scale=0.9]
\begin{axis}[xlabel={$c$}, ylabel={$p$}, ymin=-2,ymax=2,xmax=2,xmin=-2, axis lines = middle]
\draw (-2,-2) to (2,2);
\draw (-2,2) to (2,-2);
\node[rotate=90] at (0.15,0.85) {\scriptsize -- more equitable $\to$};
\node[rotate=90] at (0.15,-1) {\scriptsize $\longleftarrow$ less equitable ---};
    \addplot[
        scatter/classes={a={white}, b={red}},
        scatter, mark=, only marks, 
        scatter src=explicit symbolic,
        nodes near coords*={\Label},
        visualization depends on={value \thisrow{label} \as \Label} 
    ] table[meta=class, row sep=\\]
    {
        x y class label \\
        1 0 a {\scriptsize --- less efficient $\longrightarrow$} \\
         -1 0 a {\scriptsize $\longleftarrow$ more efficient ---} \\
    };
    \addplot[
        scatter/classes={a={black}, b={red}},
        scatter, mark=, only marks, 
        scatter src=explicit symbolic,
        nodes near coords*={\Label},
        visualization depends on={value \thisrow{label} \as \Label} 
    ] table[meta=class, row sep=\\]
    {
        x y class label \\
        1.5 0.5 a {\footnotesize $\phi = \frac{p - c}{c}$} \\
        1 1.5 a {\footnotesize $\phi = \frac{p - c}{p} $} \\
        -1 -2 a {\footnotesize $\phi = \frac{p - c}{-p}$} \\
        -1.5 -1 a {\footnotesize $\phi = \frac{p - c}{-c}$} \\
        -1 1.5 a {\footnotesize $\phi = \frac{p-c}{p}$} \\
        -1.5 0.5 a {\footnotesize $\phi = \frac{p-c}{-c}$} \\
        1 -2 a {\footnotesize $\phi = \frac{p-c}{-p}$} \\
        1.5 -1 a {\footnotesize $\phi = \frac{p - c}{c}$} \\
    }; 
\end{axis}
\end{tikzpicture}
\hspace{5mm}
\begin{tikzpicture}[scale=0.9]
\begin{axis}[xlabel={$c$}, ylabel={$p$}, ymin=-2,ymax=2,xmax=2,xmin=-2, axis lines = middle]
\draw (-2,-2) to (2,2);
\draw (-2,2) to (2,-2);
\node[rotate=90] at (0.15,0.85) {\scriptsize -- more equitable $\to$};
\node[rotate=90] at (0.15,-1) {\scriptsize $\longleftarrow$ less equitable ---};
    \addplot[
        scatter/classes={a={white}, b={red}},
        scatter, mark=, only marks, 
        scatter src=explicit symbolic,
        nodes near coords*={\Label},
        visualization depends on={value \thisrow{label} \as \Label} 
    ] table[meta=class, row sep=\\]
    {
        x y class label \\
        1 0 a {\scriptsize --- less efficient $\longrightarrow$} \\
         -1 0 a {\scriptsize $\longleftarrow$ more efficient ---} \\
    };
    \addplot[
        scatter/classes={a={black}, b={red}},
        scatter, mark=, only marks, 
        scatter src=explicit symbolic,
        nodes near coords*={\Label},
        visualization depends on={value \thisrow{label} \as \Label} 
    ] table[meta=class, row sep=\\]
    {
        x y class label \\
        1.5 0.5 a {\footnotesize $\phi = \frac{p}{c} -1$} \\
        1 1.5 a {\footnotesize $\phi = 1 - \frac{c}{p} $} \\
        -1 -2 a {\footnotesize $\phi = \frac{c}{p} - 1$} \\
        -1.5 -1 a {\footnotesize $\phi = 1 -\frac{p}{c}$} \\
        -1 1.5 a {\footnotesize $\phi = 1 - \frac{c}{p}$} \\
        -1.5 0.5 a {\footnotesize $\phi = 1 - \frac{p}{c}$} \\
        1 -2 a {\footnotesize $\phi = \frac{c}{p}-1$} \\
        1.5 -1 a {\footnotesize $\phi = \frac{p}{c} - 1$} \\
    }; 
\end{axis}
\end{tikzpicture}
\end{center}
\footnotesize \textit{Note}: The above figure illustrates the interpretation of the RPV as relative shortfall or excess in efficiency or equity.
\end{figure}
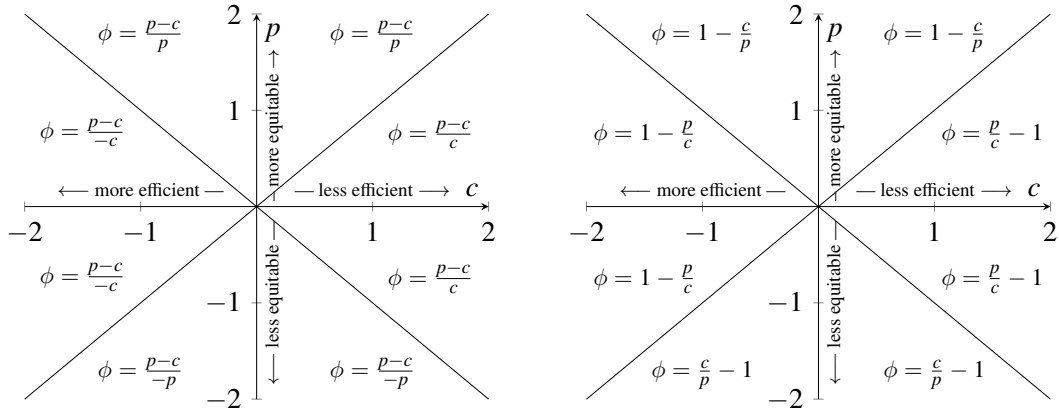

Hence, the RPV can be described as a normalized social surplus, i.e., $\beta(c,p) = p - c$ divided the maximum norm $\mathrm{max}\{\left|\,p \, \right|,\left|\,c\, \right|\}$. Figure \ref{fig:rpv_square} shows the associated contour plot of the RPV, and Figure \ref{fig:rpv_surface} shows RPV's surface plot. As shown more formally in Subsection \ref{subsec:general_class}, the reciprocal of the maximum norm maps any point $(c, p)$ on the real plane (except the origin) to the point $(c, p)/||\,(c,p)\,||_\infty$, which always lies on the square with vertices $\{-1, 1\}\times\{-1,1\}$, which is superimposed on RPV's contours in Figure \ref{fig:rpv_square}. Then, the welfare of a point on the square can be measured simply using $p/||\,(c,p)\,||_\infty - c/||\,(c,p)\,||_\infty = \phi(c, p)$, which ranges from $-2$ to $2$. As shown in Table \ref{table:rpv_interpretation}, the various possible numbers or sub-ranges of the Relative Policy Value (RPV) have intuitive interpretations on a comparative basis. Thus, it is policy-relevant to know whether the RPV of a policy falls between $-2$ and $-1$, or between $-1$ and $0$, or between $0$ and $1$, or between $1$ and $2$. Thus, the RPV can be interpreted as relative shortfall or excess in willingness-to-pay $p$ (equity) or the net fiscal revenue $-c$ (efficiency). This interpretation is transparent in Figure \ref{fig:rpv_interpretation_shortfall}, which provides sub-quadrant-specific simplifications of the RPV formula. Table \ref{table:rpv_excess_shortfall} provides further clarifications regarding this interpretation of the RPV as relative shortfall or excess in $p$ or $-c$. Therefore, the RPV can be thought of as indicating the degree of Pareto superiority (on a relative basis), which is highest for policies with RPV $=$ 2 and lowest for policies with RPV $ = -2$.

\begin{table}
\setstretch{2}
\caption{Interpretation of the RPV as Relative Shortfall or Excess in Efficiency or Equity}
\label{table:rpv_excess_shortfall}
\vspace{-5mm}
    \begin{center} 
    \begin{tabular}{c|c|c} 
    \hline\hline
        \textit{Region} & \textit{Simplified RPV Formula} & \textit{Interpretation} \\ \hline
        $c > 0$ and $p \in [-c, c]$ & $\phi(c, p) = \frac{p - c}{c} = \frac{p}{c} - 1$ & Relative shortfall in equity \\ \hline
        $p > 0$ and $c \in [-p, p]$ & $\phi(c, p) = \frac{p - c}{p} = 1 + \frac{(-c)}{p}$ & Relative excess in efficiency \\ \hline
        $c < 0$ and $p \in [c, -c]$ & $\phi(c, p) = \frac{p - c}{-c} = 1 + \frac{p}{(-c)}$ & Relative excess in equity \\ \hline
        $p < 0$ and $c \in [p, -p]$ & $\phi(c, p) = \frac{p - c}{-p} = \frac{(-c)}{(-p)} - 1$ & Relative shortfall in efficiency \\ \hline
    \end{tabular}
    \end{center}
    \vspace{-1mm}
\setstretch{1}\noindent \footnotesize
\textit{Note}: The above table interprets the Relative Policy Value (RPV) as relative shortfall or excess in efficiency or equity.
\end{table}

The second interpretation of the RPV is based on its relationship with the MVPF. In the region where $c > 0$ and $p \geq -c$ (i.e., the union of the sub-quadrants I-B, I-A, and IV-B), the RPV can be expressed in terms of the MVPF:
$$\phi(c, p) = 1\{p \leq c\}\,[m(c, p) - 1] + 1\{p > c\}\,[1 - 1/m(c,p)] \text{ when } c > 0 \text{ and } p \geq -c,$$
as shown in Figure \ref{fig:rpv_interpretation_shortfall}. Thus, there is a one-to-one relationship between the RPV and the MVPF in this region. However, in the rest of the $(c, p)$-plane, the RPV and MVPF differ in economically significant ways. The RPV also has an interpretation in terms of ``fiscal externalities'' \citep{finkelstein2020welfare} when $c > 0$ and $p > c$. In this case, note that $$m(c,p) = \frac{p}{c} = \frac{p}{c - p + p} = \frac{p/p}{(c - p)/p + p/p} = \frac{1}{1 - (p - c)/p} = \frac{1}{1 - (1 - c/p)} = \frac{1}{1 - \phi(c, p)}.$$
Thus, $-\phi(c,p) = c/p - 1$ can be interpreted as the fiscal externality associated with one unit of willingness-to-pay for a policy (when equity exceeds inefficiency). Thus, \citeauthor{finkelstein2020welfare}'s (\citeyear{finkelstein2020welfare}) notion of a ``fiscal externality'' implicitly involves the ratio of $c$ to $p$ (as opposed to the ratio of $p$ to $c$) when $p > c$ and $c > 0$. However, this notion of ``fiscal externality'' can be extended to more general externalities in efficiency and equity when $c$ and $p$ are not restricted to quadrant I-B. Thus, $\phi(c, p)$ can be interpreted generally as an externality (in equity or efficiency). Based on the above discussion, it is easy to see how this general notion of externality relates to the relative shortfall (or excess) interpretation in Table \ref{table:rpv_excess_shortfall}.

Finally, the Relative Policy Value (RPV) can be interpreted as a fundamental quantity that can be weighted to construct various types of comparative and absolute welfare measures, as shown in Table \ref{table:rpv_weights}. Weighting the RPV by the maximum norm gives back the social surplus $\beta(c, p)$, suggesting that the RPV and the maximum norm can serve as sufficient statistics for welfare analysis. Multiplying the RPV by the magnitude of the social surplus produces an unbounded curved hybrid measure, which is based on a specific combination of comparative and absolute notions. A bounded hybrid welfare measure can be obtained by multiplying the RPV by the hyperbolic tangent of the magnitude of the social surplus. Several comparative welfare measures can be interpreted as weighted versions of the RPV. The RPV is trivially the special comparative measure resulting from weighting the RPV simply by unity. However, when the RPV is weighted by the ratio of the maximum norm (i.e., the Chebyshev distance from the oirign) to the Euclidean norm (i.e., the Euclidean distance from the origin), a slightly different comparative measure arises. Similarly, weighting the RPV by the ratio of the maximum norm to a more general norm, such as the taxicab norm (i.e., Manhattan norm), produces yet another kind of comparative welfare metric. Subsection \ref{subsec:general_class} provides an axiomatic basis for this general class of relative welfare indices, although only one of them (i.e., the RPV) satisfies the welfare-symmetry axioms.

\begin{table}
\setstretch{2}
\caption{Interpretation of the RPV as Fundamental Component of Cost--Benefit Analysis}
\label{table:rpv_weights}
\vspace{-5mm}
    \begin{center} 
    \begin{tabular}{ccccc|c|c} 
    \hline\hline
       \textit{RPV} & $\times$ & \textit{Weight}  & $=$ & \textit{Measure} & \textit{Range} & \textit{Type of Measure} \\ \hline
       $\phi(c,p)$ & $\times$ & 1 & $=$ & $\frac{p - c}{\mathrm{max}\{|\,p\,|,|\,c\,|\}}$ & $[-2, 2]$ & Comparative \\ \hline
       $\phi(c,p)$ & $\times$ & $\frac{\sqrt{2}\,\mathrm{max}\{|\,p\,|,|\,c\,|\}}{\sqrt{p^2 + c^2}}$ & $=$ & $\frac{\sqrt{2}(p - c)}{\sqrt{p^2 + c^2}}$ & $[-2, 2]$ & Comparative \\ \hline
       $\phi(c,p)$ & $\times$ & $\frac{2\,\mathrm{max}\{|\,p\,|,|\,c\,|\}}{|\,p\,| + |\,c\,|}$ & $=$ & $\frac{2\,(p - c)}{|\,p\,| + |\,c\,|}$ & $[-2, 2]$ & Comparative \\ \hline
       $\phi(c,p)$ & $\times$ & $\mathrm{tanh}(|\,p - c\,|)$ & $=$ & $\frac{(p - c)\,\mathrm{tanh}(|\,p - c\,|)}{\mathrm{max}\{|\,p\,|,|\,c\,|\}}$ & $(-2, 2)$ & Hybrid \\ \hline
       $\phi(c,p)$ & $\times$ & $|\,p - c\,|$ & $=$ & $\frac{\mathrm{sgn}(p - c)\,(p - c)^2}{\mathrm{max}\{|\,p\,|,|\,c\,|\}}$ & $(-\infty, \infty)$ & Hybrid \\ \hline
       $\phi(c,p)$ & $\times$ & $\mathrm{max}\{|\,p\,|,|\,c\,|\}$ & $=$ & $p - c$ & $(-\infty, \infty)$ & Absolute \\ \hline
    \end{tabular}
    \end{center}
    \vspace{-1mm}
\setstretch{1}\noindent \footnotesize
\textit{Note}: The above table interprets the Relative Policy Value (RPV) as a fundamental component of cost--benefit analysis.
\end{table}

\clearpage

\subsection{A Related General Class of Relative Welfare Indices}
\label{subsec:general_class}

\begin{figure}[ht]
\caption{Relative Policy Value Equals the Social Surplus on a Special Square}
\label{fig:rpv_equals_ss}
\begin{center}
\begin{tikzpicture}
\begin{axis}
[    xlabel={$c$},
    xlabel style={sloped},
    ylabel={$p$},
    ylabel style={sloped},
    zlabel={$\phi(c,p)$}
]
\addplot3[color=blue] table[x = col1, y = col2,  z = col3, row sep=\\] 
{
col1 col2 col3 \\
1 -1 -2 \\
1 -0.5 -1.5 \\
1 0 -1 \\
1 0.5 -0.5 \\
1 1 0 \\
0.5 1 0.5 \\
0 1 1 \\
-0.5 1 1.5 \\
-1 1 2 \\
-1 0.5 1.5 \\
-1 0 1 \\
-1 -0.5 0.5 \\
-1 -1 0 \\
-0.5 -1 -0.5 \\
0 -1 -1 \\
0.5 -1 -1.5 \\
1 -1 -2 \\
};
\end{axis}
\end{tikzpicture}
\end{center}
\footnotesize \textit{Note}: The above figure plots the RPV $\phi(c, p)$ on the square $\mathcal{S}$ with vertices $\{-1, 1\} \times \{-1, 1\}$, on which $\phi(c, p) = p - c$.
\end{figure}
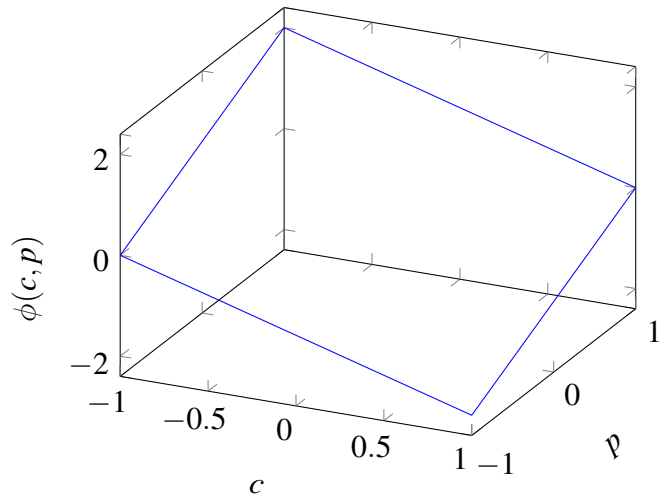

Before axiomatically developing a more general class of comparative welfare measures, it is useful to further understand the intuition behind the Relative Policy Value (RPV), which divides the social surplus by the maximum norm. Consider the square $\mathcal{S}$ with vertices $\{-1,1\} \times \{-1, 1\}$, i.e., $\mathcal{S} = (\{-1,1\} \times [-1, 1]) \cup ([-1, 1] \times \{-1, 1\})$. Figure \ref{fig:rpv_equals_ss} shows the three-dimensional graph of the RPV $\phi(c,p)$ on the square $\mathcal{S}$. The RPV $\phi(c,p) = (p - c)/1 = \beta(c, p)$ equals the MSS on the square $\mathcal{S}$. The square $\mathcal{S}$ is special in a sense because the preference orderings based on an absolute welfare measure like $\beta(c, p)$ would be compatible with preference orderings based on a comparative welfare measure on the square $\mathcal{S}$. Even more importantly, on square $\mathcal{S}$, relative and absolute notions of shortfall or excess in equity or efficiency coincide, making $\mathcal{S}$ a special square that is very useful for cost--benefit analysis.

\begin{figure}[ht]
\caption{Contour Plot of the Relative Policy Value}
\vspace{-3mm}
\label{fig:rpv_contour_ss}
\begin{center}
\includegraphics[scale=0.75]{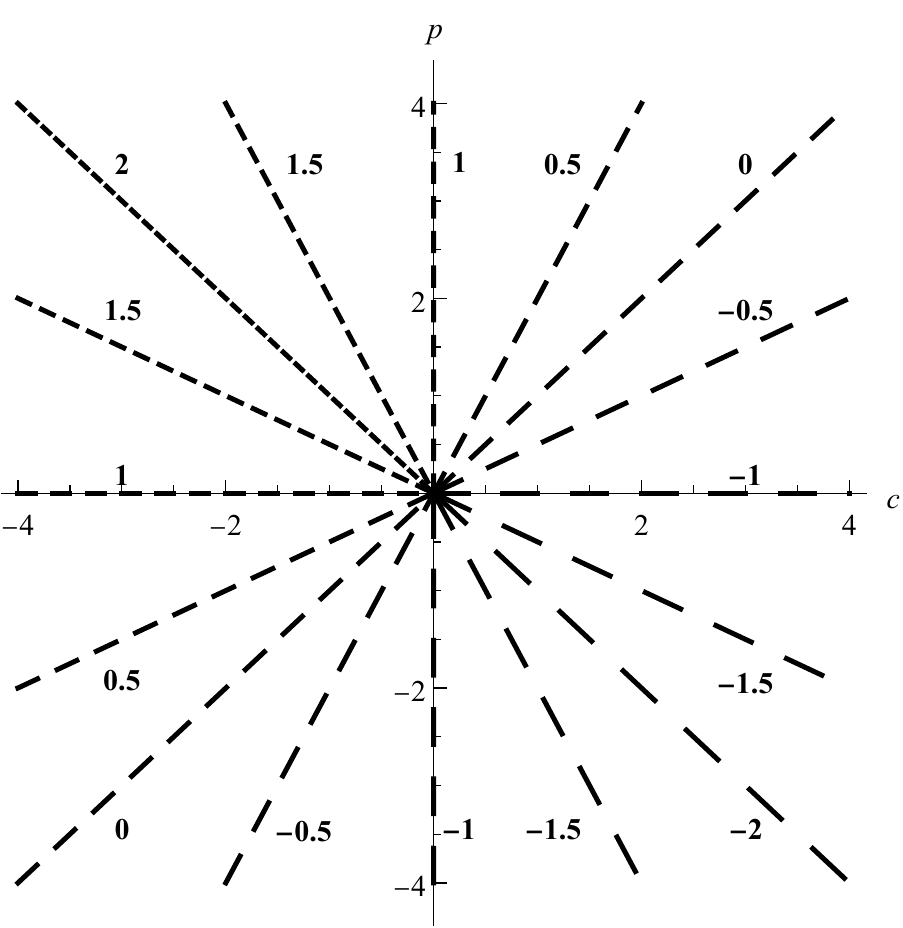}
\end{center}
\footnotesize \textit{Note}: The above figure displays the contour plot of the Relative Policy Value, showing RPV's degree-zero homogeneity.
\end{figure}

Note that $\mathcal{S} = \{(c, p): \mathrm{max}\{|\,p\,|,|\,c\,|\} = 1\}$, and so the RPV equals the MSS when the maximum norm equals one. Then, imposing the unit-free property, i.e., degree-zero homogeneity, on the welfare measure leads to the RPV, which standardizes the MSS $(p - c)$ using the maximum norm $\mathrm{max}\{|\,p\,|,|\,c\,|\}$. The resulting contours are shown in Figure \ref{fig:rpv_contour_ss}. The reciprocal $1/\mathrm{max}\{|\,p\,|,|\,c\,|\}$ serves as a multiplicative factor that makes the RPV scale-free. In addition, a useful feature of the RPV is that $\phi(c, p) \times ||(c, p)||_\infty = p - c$. In other words, the RPV and the maximum norm are sufficient statistics for both comparative and absolute welfare analyses!

The above intuition can be formalized, resulting a mathematically equivalent but a mechanical alternative version of the welfare-symmetry axioms. The alternative equivalent axiomatic framework, containing what I call the ``surplus-normalization axioms,'' consists of three intuitive axioms: the desired comparative welfare measure $\varphi(c, p)$ equals zero if and only if $p$ and $c$ break even, i.e., $p = c$; the function $\varphi(c, p)$ equals the social surplus $\beta(c, p) = p - c$ when $(c, p) \in \mathcal{S}$; and the measure $\varphi(c, p)$ is homogeneous of degree zero, i.e., $\varphi(\lambda\,c, \lambda\,p) = \varphi(c, p)$ for all $\lambda > 0$. These surplus-normalization axioms for the desired welfare measure $\varphi: \R^2 \to \R$ are stated more formally below. The RPV is also the unique measure satisfying these axioms.

\begin{axiom}
\label{axiom:4}
Let $(c,p) \in \R^2$. Then, $\varphi(c,p) = 0$ if and only if $c = p$.
\end{axiom}

\begin{axiom}
\label{axiom:5}
For all $(c, p) \in \mathcal{S}$, where $||(c,p)||_\infty = \mathrm{max}\{|\,p\,|, |\,c\,|\} = 1$, $\varphi(c,p)= p - c$.
\end{axiom}

\begin{axiom}
\label{axiom:6}
For all $\lambda > 0$ and any $(c,p) \in \R^2$, $\varphi(\lambda\,c,\lambda\,p) = \varphi(c,p)$.
\end{axiom}

\begin{theorem}
\label{theorem:rpv_Linfinity}
The Relative Policy Value $\phi(c,p) = (p-c)/||(c,p)||_\infty = (p-c)/\mathrm{max}\{|\,p\,|, |\,c\,|\}$ is the only function satisfying both the welfare-symmetry axioms and surplus-normalization axioms.
\end{theorem}

The proof of Theorem \ref{theorem:rpv_Linfinity} is given in \ref{appendix:proofs}. Therefore, the set of Axioms \ref{axiom:1}, \ref{axiom:2}, and \ref{axiom:3} is mathematically equivalent to the set of Axioms \ref{axiom:4}, \ref{axiom:5}, and \ref{axiom:6}. The above axiomatization using the maximum norm (i.e., the $L^\infty$-norm) can be modified by replacing the $L^\infty$-norm with another $L^q$-norm for $1 \leq q < \infty$, resulting in a class of welfare indices, although they do not satisfy the welfare-symmetry axioms or the surplus-normalization axioms (by Theorems \ref{theorem:rpv} and \ref{theorem:rpv_Linfinity}). For each $\infty > q \geq 1$, the desired ``$L^q$-normalized welfare index'' $\varphi_q(c,p): \R^2 \to \R$ satisfies three ``$L^q$-normalization axioms,'' which are similar to surplus-normalization axioms except that the $L^q$-norm is used instead of the maximum norm. More formally, the ``$L^q$-normalization axioms'' are below.

\begin{axiom}
\label{axiom:7}
Let $(c,p) \in \R^2$. Then, $\varphi_q(c,p) = 0$ if and only if $c = p$.
\end{axiom}

\begin{axiom}
\label{axiom:8}
For all points $(c, p)$ such that $||(c,p)||_q = (|\,c\,|^q + |\,p\,|^q)^{1/q} =  1$, $\varphi_q(c,p)= 2^{1/q}\,(p - c)$. 
\end{axiom}

\begin{axiom}
\label{axiom:9}
For all $\lambda > 0$ and any $(c,p) \in \R^2$, $\varphi_q(\lambda\,c,\lambda\,p) = \varphi_q(c,p)$.
\end{axiom}

\begin{theorem}
\label{theorem:rwi_Lq}
For each $q \in [1, \infty)$, there is a unique comparative welfare index $\varphi_q: \R^2 \to \R$ that satisfies the associated $L^q$-normalization axioms: $\varphi_q \equiv \phi_q$, where $\phi_q$ is the $L^q$-normalized welfare index that equals zero at the origin $(0,0)$ but otherwise equals $\phi_q(c,p) = 2^{1/q}\,(p-c)/||(c,p)||_q$.
\end{theorem}

The proof of Theorem \ref{theorem:rwi_Lq} is given in \ref{appendix:proofs}. Note that $L^1$-normalized and $L^2$-normalized welfare indices, i.e., $\varphi_1(c,p) = 2\,(p-c)/(|\,p\,|+|\,c\,|)$ and $\varphi_2(c,p) = \sqrt{2}\,(p-c)/\sqrt{p^2 + c^2}$, respectively, are listed in Table \ref{table:rpv_weights} as examples of weighted forms of the RPV. The $L^1$-normalized welfare index $\varphi_1$ is an interesting metric because $\varphi_1(c,p) = 2$ for all $c < 0$ and $p > 0$, similar to how the MVPF is infinity everywhere on the second quadrant. In addition, $\varphi_1(c,p) = -2$ for all $c > 0$ and $p < 0$. Thus, the preferences underlying the relative welfare index $\varphi_1$ can be used to suggest a possible way to ``fix'' the MVPF. As discussed in Subsection \ref{subsec:desirable_qualities}, $(c,p)$ and $(-c,-p)$ are opposites in a sense for all $(c, p) \in \R^2$. Thus, from a multiplicative perspective, $\tilde{m}(c,p)\,\tilde{m}(-c,-p) = 1$ would hold for a ``fixed'' MVPF $\tilde{m}(c, p)$, motivating the following axioms.

\begin{axiom}
\label{axiom:10}
Let $(c,p) \in \R^2$. Then, $\tilde{m}(c,p) = 1$ if and only if $c = p$.
\end{axiom}

\begin{axiom}
\label{axiom:11}
For all $(c, p) \in \R \times \R_{\geq 0}$, where $\tilde{m}(c,p) = m(c,p)$, where $m(c,p)$ is the MVPF.
\end{axiom}

\begin{axiom}
\label{axiom:12}
For all $(c, p) \in \R^2$, $\tilde{m}(c,p)\,\tilde{m}(-c,-p) = 1$, and let ``$0 \times \infty$'' $=$ ``$\infty \times 0$'' $= 1$.
\end{axiom}

\begin{theorem}
\label{theorem:fixed_mvpf}
The ``fixed'' MVPF $\tilde{m}: \R^2 \to [0, \infty]$ satisfying the above Axioms \ref{axiom:10}, \ref{axiom:11}, \ref{axiom:12} is given by $\tilde{m}(c,p) = p/c$ if $(c,p) \in \R^2_{>0}$, $\tilde{m}(c,p) = c/p$ if $(c,p) \in \R^2_{<0}$, $\tilde{m}(c,p) = \infty$ if $(c,p) \in \R^2_{\leq 0} \times \R^2_{\geq 0}$ (excluding the origin), $\tilde{m}(c,p) = 0$ if $(c,p) \in \R^2_{\geq 0} \times \R^2_{\leq 0}$ (excluding the origin), and $\tilde{m}(0,0) = 1$.
\end{theorem}

The proof of Theorem \ref{theorem:fixed_mvpf} is given in \ref{appendix:proofs}. Axioms \ref{axiom:10}, \ref{axiom:11}, \ref{axiom:12} essentially define a multiplicative group (in the terminology of group theory) in which the identity element is the break-even line, and any other group element $\tilde{M}$ is a set of points with a common value for the ``fixed'' MVPF, which is a value between 0 and $\infty$. It is isomorphic to the multiplicative group of $\R_{\geq 0} \cup \{\infty\}$, letting or assuming that ``$\infty \times a$'' $=$ ``$a \times \infty$'' is equal to 1 if $a = 0$ but equal to $\infty$ if $a > 0$. However, the ``fixed'' MVPF is redundant because there is a bijection between the ``fixed'' MVPF $\tilde{m}(c,p)$ and the $L^1$-normalized welfare index $\varphi_1(c,p)$, which is bounded and does not have non-numerical values. Furthermore, the $L^q$-normalized welfare indices are also largely of theoretical interest, because they do not satisfy the desirable welfare-symmetry axioms, which are practically useful for welfare aggregation, as discussed in Subsection \ref{subsec:desirable_qualities}. Therefore, the rest of the paper only focuses on my main measure, the Relative Policy Value (RPV).

\clearpage

\section{Welfare Aggregation Using the Relative Policy Value}
\label{sec:welfare_agg}

Having axiomatically constructed the Relative Policy Value (RPV), I now discuss how it overcomes many of the issues in MVPF-based welfare aggregation (see Subsection \ref{subsec:limitations_existing}). To reiterate, suppose $\{(c_l, p_l)\}_{l \in \mathcal{L}}$ are the values of net fiscal cost and willingness-to-pay for elements $l$ in general policy collection $\mathcal{L}$ that is of interest to the policymaker. I assume that empirical researchers are able to operationalize $\mathcal{L}$ in a reasonable manner, but I allow $\mathcal{L}$ to be a general collection of public policies. For example, if $l' \neq l''$ are two elements of $\mathcal{L}$, then $(c_{l'},p_{l'})$ and $(c_{l''}, p_{l''})$ may represent the net fiscal cost and willingness-to-pay either for separate policies or for a single policy by different population subgroups. Although it is possible to let $\mathcal{L}$ be a continuum (or a hybrid union of discrete elements and continua), the notation can get unnecessarily cumbersome. Thus, for expositional ease, I work with a discrete set $\mathcal{L}$.

As discussed in Subsection \ref{subsec:limitations_existing}, there are two different ways to aggregate welfare across the policy collection $\mathcal{L}$ using a welfare measure $\gamma(c,p)$. \cite{hendren2020unified} suggest calculating welfare $\gamma(C(\lambda_\mathcal{L}), P(\lambda_\mathcal{L}))$ of a ``category average'' $(C(\lambda_\mathcal{L}), P(\lambda_\mathcal{L})) = \sum_{l \in \mathcal{L}} \lambda_l\times(c_l, p_l)$ where $\lambda_\mathcal{L} = \{\lambda_l\}_{l \in \mathcal{L}}$ are subjective non-negative scaling factors. However, welfare of scaling-factor-weighted (i.e., $\lambda_\mathcal{L}$-weighted) policy average is difficult to interpret unless, e.g., $\{(c_l, p_l)\}_{l \in \mathcal{L}}$ are measured per-capita for a federal policy implemented across states and $\lambda_\mathcal{L}$ contains the population weights. However, if alternatively $w_\mathcal{L} = \{w_l\}_{l \in \mathcal{L}}$ represent subjective importance weights a policymaker attaches to welfare from different policies (or population subgroups), then it is straightforward to interpret the importance-weighted welfare sum $\sum_{l \in \mathcal{L}} w_l\,\gamma(c_l, p_l)$ as the policymaker's total utility. Note that I allow the scaling factors $\lambda_\mathcal{L}$ and the importance weights $w_\mathcal{L}$ to be very general, and so $\lambda_{l'}$ and $w_{l'}$ are allowed to depend on $\{(c_l, p_l)\}_{l \in \mathcal{L}}$, based on the policymaker's preferences, for all $l' \in \mathcal{L}$. Thus, $\lambda_\mathcal{L}$ and $w_\mathcal{L}$ can be more generally thought of as well-defined functions from $\prod_{l \in \mathcal{L}} \R^2$ to $\prod_{l \in \mathcal{L}} \R_{\geq 0}$ that map $\{(c_l, p_l)\}_{l \in \mathcal{L}}$ to $\{\lambda_l\}_{l \in \mathcal{L}}$ and $\{w_l\}_{l \in \mathcal{L}}$, respectively.

Using the above general concept of the subjective scaling factors, the following definition defines the the ``Joint Policy Value'' (JPV) to formalize the first notion of welfare aggregation discussed above. It is an adaptation and generalization of \citeauthor{hendren2020unified}'s (\citeyear{hendren2020unified}) concept of welfare of a ``category average.''

\begin{definition}
\label{definition:jpv}
The Joint Policy Value, based on a well-defined map $\lambda_\mathcal{L}: \prod_{l \in \mathcal{L}} \R^2 \to \prod_{l \in \mathcal{L}} \R_{\geq 0}$ that fully specifies the policymaker's chosen subjective scaling factors, is a real-valued function $\Upsilon: \prod_{l \in \mathcal{L}} \R^2 \to \R$ given by
$$\textstyle \Upsilon(\{(c_l, p_l)\}_{l \in \mathcal{L}}; \lambda_\mathcal{L}) = \phi\big(\sum_{l \in \mathcal{L}} \lambda_l\times(c_l, p_l)\big),$$
where $\{\lambda_l\}_{l \in \mathcal{L}} \equiv \lambda_\mathcal{L} \equiv \lambda_\mathcal{L}(\{(c_l, p_l)\}_{l \in \mathcal{L}})$.
\end{definition}

Similarly, using the aforementioned general concept of the subjective or objective importance weights, the following definition defines the the ``Total Policy Value'' (TPV) to formalize the second notion of welfare aggregation, which has a natural interpretation as the policymaker's (importance-weighted) total utility/welfare of policies that were actually implemented, unlike a hypothetical policy that is used to define the Joint Policy Value (JPV). Following the below definition of the Total Policy Value (TPV), I provide some examples of importance weighting functions and the associated forms of the TPV. Depending on the specified importance weights, the TPV can be interpreted from an absolute or a comparative perspective (or a hybrid perspective), just as the weighted RPV can give rise to various absolute or comparative measures, as shown in Table \ref{table:rpv_weights}.

\begin{definition}
\label{definition:tpv}
The Total Policy Value, based on a well-defined map $w_\mathcal{L}: \prod_{l \in \mathcal{L}} \R^2 \to \prod_{l \in \mathcal{L}} \R_{\geq 0}$ that fully specifies the policymaker's chosen subjective importance weights, is a real-valued function $\Psi: \prod_{l \in \mathcal{L}} \R^2 \to \R$ given by
$$\textstyle \Psi(\{(c_l, p_l)\}_{l \in \mathcal{L}}; w_\mathcal{L}) = \sum_{l \in \mathcal{L}} w_l\times\phi(c_l, p_l),$$
where $\{w_l\}_{l \in \mathcal{L}} \equiv w_\mathcal{L} \equiv w_\mathcal{L}(\{(c_l, p_l)\}_{l \in \mathcal{L}})$.
\end{definition}

\begin{example}
If $w_\mathcal{L}(\{(c_l, p_l)\}_{l \in \mathcal{L}}) = \{\tilde{w}_l\,\mathrm{max}\{|\,p_l\,|, |\,c_l\,|\}\}_{l \in \mathcal{L}}$, where $\{\tilde{w}_l\}_{l\in\mathcal{L}} \in \prod_{l \in \mathcal{L}}\R_{\geq 0}$ are some constants, then the Total Policy Value (TPV) is simply the weighted sum of social surpluses (i.e., the weighted sum of MSS values). In other words, $\Psi(\{(c_l, p_l)\}_{l \in \mathcal{L}}; w_\mathcal{L}) = \sum_{l \in \mathcal{L}} \tilde{w}_l\,(p_l - c_l)$, which is perhaps the most conventional welfare aggregation method in the public finance literature.
\end{example}

\begin{example}
If $w_\mathcal{L}(\{(c_l, p_l)\}_{l \in \mathcal{L}}) = \{\tilde{w}_l\}_{l \in \mathcal{L}}$, where $\{\tilde{w}_l\}_{l\in\mathcal{L}}$ is a point on the standard $(|\mathcal{L}|-1)$-simplex, then the Total Policy Value (TPV) is simply the weighted average of the RPVs $\{\phi(c_l, p_l)\}_{l \in \mathcal{L}}$. In other words, $\Psi(\{(c_l, p_l)\}_{l \in \mathcal{L}}; w_\mathcal{L}) = \sum_{l \in \mathcal{L}} \tilde{w}_l\,\phi(c_l, p_l)$, which is bounded within $[-2, 2]$.
\end{example}

\begin{example}
If $w_\mathcal{L}(\{(c_l, p_l)\}_{l \in \mathcal{L}}) = \{2^{1/q}\,\tilde{w}_l\,||(c_l,p_l)||_\infty/||(c_l,p_l)||_q\}_{l \in \mathcal{L}}$, where $q \geq 1$ and $\{\tilde{w}_l\}_{l\in\mathcal{L}}$ is a point on the standard $(|\mathcal{L}|-1)$-simplex, then the Total Policy Value (TPV) is simply the weighted average of the $L^q$-normalized welfare indices $\{\varphi_q(c_l, p_l)\}_{l \in \mathcal{L}}$. In other words, $\Psi(\{(c_l, p_l)\}_{l \in \mathcal{L}}; w_\mathcal{L}) = \sum_{l \in \mathcal{L}} \tilde{w}_l\,\varphi_q(c_l, p_l)$, which is bounded within $[-2, 2]$.
\end{example}

\begin{example}
If $w_\mathcal{L}(\{(c_l, p_l)\}_{l \in \mathcal{L}}) = \{|\,p_l - c_l\,|/[\sum_{k \in \mathcal{L}} |\,p_k - c_k\,|]\}\}_{l \in \mathcal{L}}$, then the Total Policy Value (TPV) is simply the weighted average of the RPVs such that the sum of weights is normalized to one, and the importance weight for each policy is proportional to the magnitude of its social surplus (MSS). In other words, $\Psi(\{(c_l, p_l)\}_{l \in \mathcal{L}}; w_\mathcal{L}) = [\sum_{l \in \mathcal{L}}|\,p_l - c_l\,|\,\phi(c_l,p_l)]/[\sum_{l \in \mathcal{L}}|\,p_l - c_l\,|]$, which is bounded within $[-2,2]$, assuming that $\sum_{l \in \mathcal{L}}|\,p_l - c_l\,| \neq 0$.
\end{example}

\begin{example}
If $w_\mathcal{L}(\{(c_l, p_l)\}_{l \in \mathcal{L}}) = \{|\,c_l\,|/[\sum_{k \in \mathcal{L}} |\,c_k\,|]\}\}_{l \in \mathcal{L}}$, then the Total Policy Value (TPV) is simply the cost-magnitude-weighted average of the RPVs. In other words, $\Psi(\{(c_l, p_l)\}_{l \in \mathcal{L}}; w_\mathcal{L}) = [\sum_{l \in \mathcal{L}}|\,c_l\,|\,\phi(c_l,p_l)]/[\sum_{l \in \mathcal{L}}|\,c_l\,|]$, which is bounded within $[-2,2]$, assuming that $\sum_{l \in \mathcal{L}}|\,c_l\,| \neq 0$.
\end{example}

\clearpage

\begin{figure}
\begin{center}
\caption{Aggregate Welfare from Hope Tax Credit for Joint Filers (JF) and Single Filers (SF)}
\label{fig:htc_aggregation}
	\begin{subfigure}[b]{0.45\textwidth}
         \centering
         \caption{\centering Total Policy Value with Weighted RPVs $\hphantom{--} w\,\phi(c_\mathrm{SF},p_\mathrm{SF}) + (1-w)\,\phi(c_\mathrm{JF},p_\mathrm{JF})$}
         \includegraphics[scale=.77]{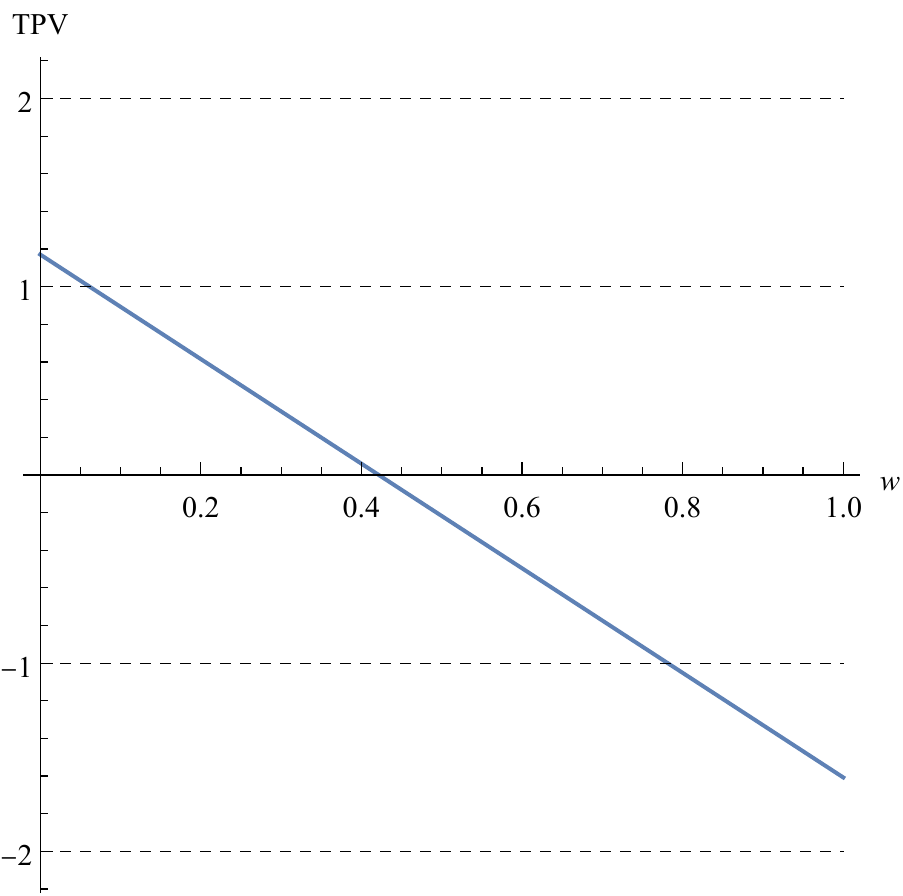}
         \label{fig:htc_tpv}
     \end{subfigure}
     \begin{subfigure}[b]{0.45\textwidth}
         \centering
         \caption{\centering Joint Policy Value of Weighted Policies $\hphantom{--}\phi(w\,(c_\mathrm{SF},p_\mathrm{SF}) + (1-w)\,(c_\mathrm{JF},p_\mathrm{JF}))$}
         \includegraphics[scale=.75]{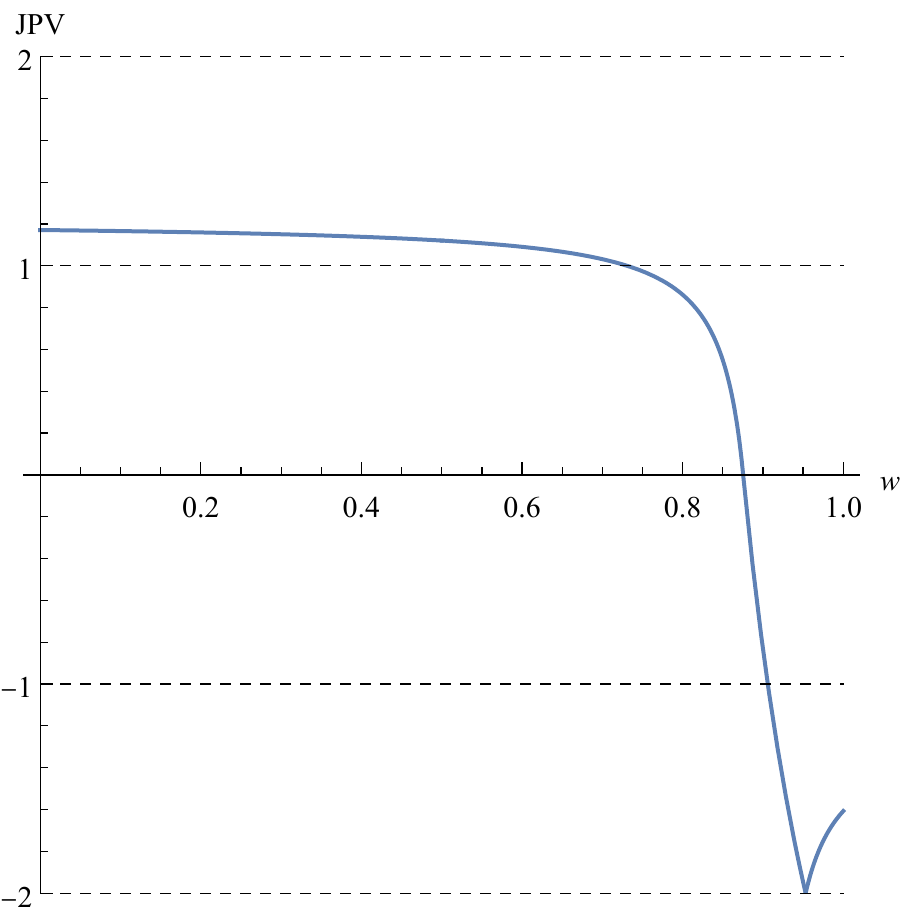}
         \label{fig:htc_jpv}
     \end{subfigure}
     \begin{subfigure}[b]{0.45\textwidth} \vspace{10mm}
         \centering
         \caption{\centering Maximum Norm of Weighted Policies $\hphantom{--}||\,w\,(c_\mathrm{SF},p_\mathrm{SF}) + (1-w)\,(c_\mathrm{JF},p_\mathrm{JF})\,||_\infty$}
         \includegraphics[scale=.75]{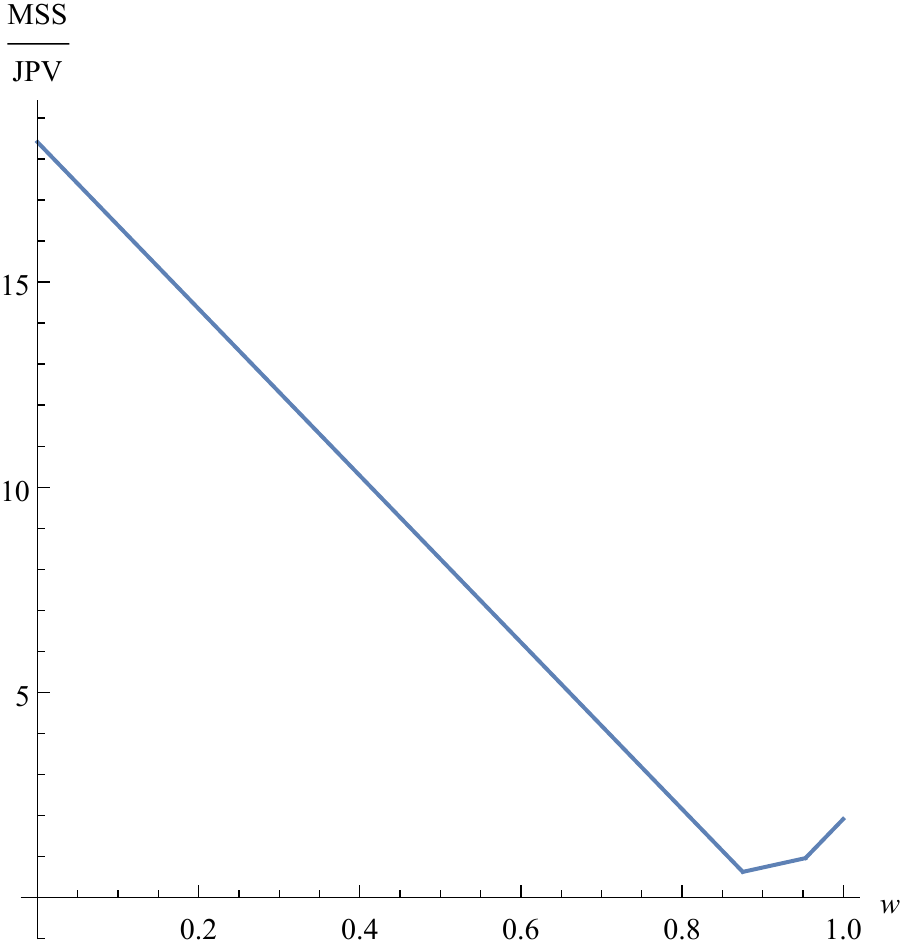}
         \label{fig:htc_max}
     \end{subfigure}
     \begin{subfigure}[b]{0.45\textwidth}
         \centering
         \caption{\centering Weighted Sum of Social Surpluses $\hphantom{---} w\,(p_\mathrm{SF} - c_\mathrm{SF}) + (1-w)\,(p_\mathrm{JF}- c_\mathrm{JF})$}
         \includegraphics[scale=.79]{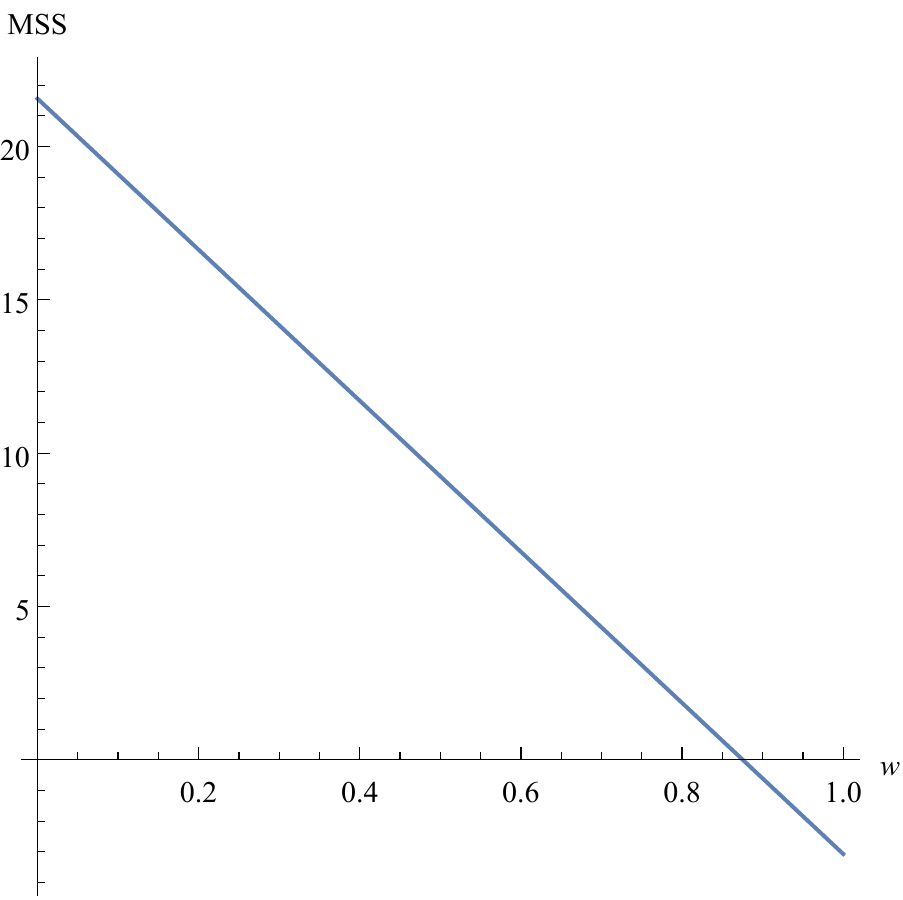}
         \label{fig:htc_mss}
     \end{subfigure}
\end{center}
\footnotesize \textit{Note}: The above figures illustrate different welfare aggregation methods using \citeauthor{hendren2020unified}'s (\citeyear{hendren2020unified}) estimates of net fiscal cost and willingness-to-pay for the Hope Tax Credit (HTC), which was implemented in 2007, for Joint Filers (JF) and Single Filers (SF) at Phase Start: $(c_\mathrm{JF}, p_\mathrm{JF}) = (-3.15, 18.41)$ for the Joint Filers (JF), and $(c_\mathrm{SF}, p_\mathrm{SF}) = (1.16, -1.91)$ for the Single Filers (SF). Statistical uncertainty in these estimates is ignored above because the main purpose of the above exercise is to illustrate how the various aggregation methods differ. Figure \ref{fig:htc_tpv} shows the Total Policy Value (TPV) in the form of weighted sum of the Relative Policy Values (RPVs) for SF and JF using the importance weights $w \in [0, 1]$ and $(1-w)$, respectively. Figure \ref{fig:htc_jpv} shows the Joint Policy Value (JPV) when the scaling factors are set equal to the importance weights $w$ and $1-w$. Figure \ref{fig:htc_max} shows the maximum norm of the weighted policies. Figure \ref{fig:htc_mss} shows the weighted sum of the Marginal Social Surplus (MSS) for JF and SF.
\end{figure}

\clearpage

Figure \ref{fig:htc_aggregation} illustrates different welfare aggregation methods using \citeauthor{hendren2020unified}'s (\citeyear{hendren2020unified}) estimates of net fiscal cost and willingness-to-pay related to the Hope Tax Credit (HTC) policy, which was implemented in 2007, for Joint Filers (JF) and Single Filers (SF) at Phase Start: $(c_\mathrm{JF}, p_\mathrm{JF}) = (-3.15, 18.41)$ for the Joint Filers (JF), and $(c_\mathrm{SF}, p_\mathrm{SF}) = (1.16, -1.91)$ for the Single Filers (SF). Statistical uncertainty in these estimates is ignored for now because the main purpose of Figure \ref{fig:htc_aggregation} is to illustrate how the various aggregation methods differ. Figure \ref{fig:htc_tpv} shows the Total Policy Value (TPV) in the form of weighted sum of the Relative Policy Values (RPVs) for SF and JF using the importance weights $w \in [0, 1]$ and $(1-w)$, respectively. The TPV changes linearly from 1.17 (when $w = 0$) to $-1.61$ (when $w = 1$). Figure \ref{fig:htc_mss} shows the weighted sum of the Marginal Social Surplus (MSS) for JF and SF. Note that this weighted sum of the MSS values is a form of TPV using importance weights equal to $w\, ||\,(c_\mathrm{SF}, p_\mathrm{SF})\,||_\infty$ for SF and  $(1-w)\, ||\,(c_\mathrm{JF}, p_\mathrm{JF})\,||_\infty$ for JF. This weighted sum of MSS values also changes linearly from 21.56 (when $w = 0$) to $-3.07$ (when $w=1$). However, as shown in \ref{fig:htc_jpv}, the associated Joint Policy Value (JPV) is a non-linear function of the scaling factors, which are set equal to the importance weights $w$ and $1-w$. The non-smooth change in the JPV near the right end of Figure \ref{fig:htc_jpv} may seem odd at first glance, but it can be explained by the non-linear change in the maximum norm (associated with the JPV) near the right end of Figure \ref{fig:htc_max}, which shows the maximum norm of the weighted HTC policies. On the other hand, it is easier to interpret the linear functions in Figures \ref{fig:htc_tpv} and \ref{fig:htc_mss}, which aggregate welfare using comparative and absolute notions, although both are just different applications of the general Total Policy Value concept.

Figure \ref{fig:htc_aggregation} shows how a change in the welfare aggregation method can drastically change policy conclusions, even in the simple case of the HTC policy (at the phase end) with just two heterogeneous population subgroups. Consider the case with equal scaling factors or importance weights, i.e., $w = 1 - w = 0.5$. The associated TPV is below zero, but the associated JPV is above one. If the policymaker weights the two population subgroups (the Joint Filers and the Single Filers) equally, then the aggregate welfare of these policies from a comparative perspective is negative because of the harmful effect of the policy on the Single Filers, with the policy implication that the HTC policy may need to be modified in order to reduce the harmful effects on Single Filers so that the overall welfare from the HTC policy across the population can be improved. However, using the JPV would suggest that the hypothetical combined policy is Pareto superior, implying that the policymaker need not rethink the design of the HTC policy (at the phase end). Of course, this makes sense from an absolute welfare perspective, but it ignores the equity and efficiency considerations for the separate subgroups. Thus, if the policymaker cares about the equity and efficiency of each of the two subgroups, then it may make more sense for the policymaker to make decisions based on the TPV rather than the JPV. Following this discussion, I reexamine \citeauthor{hendren2020unified}'s (\citeyear{hendren2020unified}) ``College Adult'' policy category, which I initially inspected in Subsection \ref{subsec:limitations_existing}.

\begin{table}
\setstretch{1.5}
\caption{Two Different Types of Welfare Aggregation Across ``College Adult'' Policies}
\label{table:college_adult_example}
\vspace{-5mm}
    \begin{center}
    \begin{tabular}{l|c|c|c|c|c|c|c}
    \hline\hline
\textit{Program} & $\cL_1$ & $\cL_2$ & $\hat{c}_j$ & $\hat{p}_j$ &  \textit{MSS+$1$} &  \textit{MVPF} & \textit{RPV} \\ \hline
AOTC (SI) & \checkmark &\checkmark & 0.53   & 5.36 & 5.83  & 10.05 & 0.90    \\ \hline
HOPE Cred. & \checkmark &\checkmark  & 0.42   & 5.27 & 5.85  & 12.58 & 0.92   \\ \hline
HOPE/LLC  & \checkmark & $\times$     & 4.86  & $-$42.82 & $-$46.68  & $-$8.81 & $-$1.11 \\ \hline
HOPE/LLC $\times$ 0.1  & $\times$ &\checkmark & 0.49  & $-$4.28 & $-$3.77  & $-$8.81 & $-$1.11 \\ \hline
Adult Pell  & \checkmark &\checkmark    & 1.57  & 3.42 & 2.85   & 2.18  & 0.54   \\ \hline
Tuition Deduc.~(JE)  & \checkmark &\checkmark & 1.29  & 1.00 & 0.71     & 0.77   & $-$0.23  \\ \hline
Tuition Deduc.~(JS)  & \checkmark &\checkmark  & 1.38  & $-$0.03 & $-$0.41  & $-$0.02  & $-$1.02 \\ \hline
Tuition Deduc.~(SE)  & \checkmark &\checkmark  & $-$1.13 & 1.00 & 3.13     & $\infty$ & 1.89  \\ \hline
Tuition Deduc.~(SS)  & \checkmark &\checkmark & $-$5.10  & 5.38 & 11.47   & $\infty$ & 1.95   \\\hline\hline 
\multicolumn{3}{l|}{\textit{Welfare of $[\frac{1}{8}\sum_{j \in \mathcal{L}_1} (\hat{c}_j, \hat{p}_j)]$ }} & 0.48 & $-$2.68 & $-$2.16 & $-$5.59 & $-$1.18 \\\hline
\multicolumn{3}{l|}{\textit{Welfare of $[\frac{1}{8}\sum_{j \in \mathcal{L}_2} (\hat{c}_j, \hat{p}_j)]$}} & $-$0.07 &  2.14 & 3.21 & $\infty$ & 1.03 \\\hline\hline
\multicolumn{5}{l|}{\textit{Equal-weighted average welfare of policies in $\mathcal{L}_1$}} & $-$2.16 & ? & 0.48 \\ \hline
\multicolumn{5}{l|}{\textit{Equal-weighted average welfare of policies in $\mathcal{L}_2$}} & 3.21 & ? & 0.48 \\ \hline\hline
    \end{tabular}
    \end{center}
    \vspace{-1mm}
\setstretch{1}\noindent \footnotesize
\textit{Note}: The checkmark(s) next to a policy indicate(s) which of the collections $\cL_1$ or $\cL_2$ contain(s) it. For each policy $j$ in $\cL_1$ or $\cL_2$ above, $\hat{c}_j$ and $\hat{p}_j$ in the above table refer to the estimates of the net fiscal cost $c_j$ and the willingness-to-pay $p_j$, respectively, for each policy $j$. The ``College Adult'' policy category, as defined by \cite{hendren2020unified}, consists of the policies in the policy collection $\mathcal{L}_1$. (Although the authors consider additional related policies, there is much more ambiguity about their net fiscal costs and willingness-to-pay values, resulting in their exclusion from $\mathcal{L}_1$.) AOTC (SI) refers to Ameican Opportunity Tax Credit, Simulated Instrument. HOPE Cred.~refers to Hope Tax Credit. HOPE/LLC refers to Hope and Lifetime Learners Tax Credits. (HOPE/LLC $\times$ 0.1 represents a hypothetical policy that has a net fiscal cost and willingness-to-pay equal to  10\% of those of HOPE/LLC.) Adult Pell refers to the Introduction of Pell Grants to Adults. Tuition deduc refers to Tax Deduction for Postsecondary Tuition. JE refers to Joint Filers at Phase End. JS refers to Joint Filers at Phase Start. SE refers to Single Filers at Phase End. SS refers to Single Filers at Phase Start. The third last column reports the benefit-to-cost ratio BCR ($=$ MSS$+$1), which is simply equal to one plus the marginal social surplus because of the way \cite{hendren2020unified} operationalize $c_j$ and $p_j$. The last two columns report the Marginal Value of Public Funds (MVPF) and the Relative Policy Value (RPV), respectively, and the associated aggregate welfare measures. Using the RPV, the welfare of the weighted policies $[\frac{1}{8}\sum_{j \in \mathcal{L}_1} (\hat{c}_j, \hat{p}_j)]$ and $[\frac{1}{8}\sum_{j \in \mathcal{L}_2} (\hat{c}_j, \hat{p}_j)]$ refer to the Joint Policy Value (JPV) using equal scaling factors for those two hypothetical weighted policies. In addition, the equal-weighted average welfare values of policies in $\mathcal{L}_1$ and $\mathcal{L}_2$ refer to the Total Policy Values (TPVs) of the sets $\mathcal{L}_1$ and $\mathcal{L}_2$, respectively, using equal importance weights. Since the average MVPF is ambigious for both the policy collections $\mathcal{L}_1$ and $\mathcal{L}_2$, this ambiguity is represented by a question mark (?).
\end{table}

Table \ref{table:college_adult_example}, which incorporates the information in Figure \ref{fig:welfare_agg_example}, illustrates the problems with MVPF-based welfare aggregation for the ``College Adult'' policy category, which consists of eight policies, as defined by \cite{hendren2020unified} in their second table. If we replace the HOPE/LLC policy with another policy that has the same program cost but only a tenth of its net fiscal cost and willingness-to-pay, the BCR ($=$ MSS$+$1) increases but the MVPF and RPV of this hypothetical modified policy do not change (because of their degree-zero homogeneity). Replacing HOPE/LLC with the modified policy dramatically changes the average BCR of the ``College Adult'' category from $-2.16$ to $3.21$. The average MVPF of the ``College Adult'' category is undefined or ambiguous regardless of which version of HOPE/LLC is included. In contrast, the average RPV remains unchanged, as should be the case because the rescaling of the HOPE/LLC policy does not affect its RPV, which is scale-free. In the cases where the average MVPF is defined, it would also be invariant to scaling of policies, but the average MVPF can be ambiguous and thus cannot be computed in many cases, and so the average RPV is a reliable alternative. 

The point estimate (0.48) of the average RPV, or the Total Policy Value (TPV) using equal importance weights, suggests that the ``College Adult'' policies generate positive aggregate welfare from a comparative perspective (when the inequitable and inefficient policies, such as ``HOPE/LLC'' and ``Tuition Deduc.~(JS),'' are weighted against equitable and efficient policies, such as ``Tuition Deduc.~(SE)'' and ``Tuition Deduc.~(SS),'' as well as other intermediate policies).\footnote{Of course, there is statistical uncertainty in Table \ref{table:college_adult_example}, but its main purpose is not to draw strong conclusions but to show a real example that demonstrates the issues one encounters in practice when using the BCR (or the MSS) and the MVPF for welfare aggregation.}
This is in contrast to the negative Joint Policy Value (JPV) using equal scaling factors for a hypothetical ``category average.'' The estimated JPV for the ``College Adult'' policy category is $-1.18$, which is associated with an MVPF of $-5.59$. Even though this JPV is negative, it is unclear what the hypothetical policy $[\frac{1}{8}\sum_{j \in \mathcal{L}_1} (\hat{c}_j, \hat{p}_j)]$ represents, given that it is an average across different policies implemented in different years (and not even an average across population subgroups). Thus, in cases like this, it may be more reasonable to use the Total Policy Value (TPV) to perform welfare aggregation across policies, because it is the weighted sum of the RPVs of policies that were actually implemented.

In summary, the MVPF-based welfare aggregation methods, such as the MVPF of the ``category average'' \citep{hendren2020unified}, have several conceptual and mathematical issues. However, there is a growing need for reliable, general, and flexible methods to aggregate welfare not just across different policies but also across population subgroups. This is especially true in contexts where the heterogeneity in welfare effects is empirically important and very policy-relevant \citep[see, e.g., ][]{allcott2019welfare}. The discussion in this section shows that the Joint Policy Value (JPV) and the Total Policy Value (TPV), which are defined using RPVs, are general tools that can be used to aggregate welfare from either comparative, absolute, or hybrid perspectives.

\clearpage

\section{Statistical Inference for Cost--Benefit Analysis}
\label{sec:stat_inference}

The previous sections formulate the Relative Policy Value (RPV) $\phi$ and discuss how it can used to evaluate individual policies or to aggregate welfare across policies. However, in practice, policymakers rarely know the true net fiscal cost $c$ and willingness-to-pay $p$ for any given policy to be able to even compute $\phi(c,p)$. Nevertheless, it is often possible to use data from experimental or observational studies to at least approximately learn about $c$ and $p$ and thus about the welfare associated with the policy. Since $c$ and $p$ are usually abstract objects and heavily dependent on the socioeconomic context, the analyst may need to make numerous assumptions and judgment calls to operationalize $c$ and $p$ in data analysis. Fortunately, the work of \cite{hendren2020unified} provides excellent guidance for tackling such operationalization challenges. However, even after this step, there exist significant statistical hurdles, because the RPV and MVPF are not ``well-behaved'' functions, and so conventional methods are not applicable for conducting statistical inference, as Subsection \ref{subsec:stat_irregularities} explains. This motivates the need for making uniformly valid inferences regarding public policies. Section \ref{subsec:uniform_inference} provides statistical tools for this purpose. Before proceeding to these discussions, Section \ref{subsec:parameters} clarifies the policy-relevant parameters of interest, expresses many of those parameters in terms of the RPV, and discusses how to statistically and economically interpret the confidence sets for the RPV and related parameters.

\subsection{Parameters of Interest and Statistical Versus Economic Significance}
\label{subsec:parameters}

Building on the economic concepts developed in the previous sections, I now discuss the parameters of interest for statistical inference. I first provide a general form of parameters that can be used to quantify relative, absolute, or aggregate welfare for cost--benefit analysis. For evaluating individual policies from a comparative welfare perspective, I argue that the estimates and confidence intervals for the Relative Policy Value (RPV) serve as sufficient statistics for a wide class of policy-relevant parameters. In addition, I clarify the notions of statistical and economic significance in the empirical analysis of RPV, which is carried out in Section \ref{section:reanalysis}.

Suppose $\mathcal{L}$ is a collection of policies of interest. For expositional ease, I assume that $\mathcal{L}$ is discrete, but appropriate modifications can be made to extend the theory to the setting where $\mathcal{L}$ is a continuum (or a hybrid union of continua and discrete sets). A general cost--benefit analysis of $\mathcal{L}$ can be based on the parameters $X_\mathcal{L}\equiv \{(c_l, p_l)\}_{l \in \mathcal{L}}$, where $c_l$ is the net fiscal cost and $p_l$ is the willingness-to-pay for policy $l \in \mathcal{L}$. Then, a general cost--benefit analysis based on $X_\mathcal{L}$ would involve making inferences on a general parameter set such as $\{\Theta_k(X_\mathcal{L})\}_{k \in \mathcal{K}}$, where $\mathcal{K}$ is a general index/subscript set, and $\Theta_k$ is a well-defined function from $\prod_{l \in \mathcal{L}}\,\R^2$ to $\R$ for each $k \in \mathcal{K}$. For example, $\Theta_k(X_\mathcal{L})$ for each $k \in \mathcal{K}$ can be the Relative Policy Value (RPV), the Joint Policy Value (JPV), or more generally the Total Policy Value (TPV).

For comparative analysis of a single policy represented by $(c,p)$, the parameter of interest can be any of the economically coherent measures discussed in Section \ref{sec:axiomatic_construction}, such as the RPV $\phi(c,p)$, the $L^q$-normalized welfare index $\varphi_q(c,p)$, or the ``fixed'' MVPF $\tilde{m}(c,p)$. However, I argue that the latter two can be expressed in terms of the RPV, and so it is not necessary for empirical researchers to separately report inferences on the other comparative welfare measures for individual policies. This notion of sufficiency of the RPV is stated more formally below.

\begin{theorem}
\label{theorem:rpv_mvpf_conversion}
Let $(c, p) \in \R^2$ and $\phi(c,p) = \tilde{\phi}$. Then, $\tilde{m}(c,p) = 0$ if $\tilde{\phi} \leq -1$, $\tilde{m}(c,p) = \infty$ if $\tilde{\phi} \geq 1$, and $\tilde{m}(c,p) = 1\{-1 < \tilde{\phi} < 0\}\,(\tilde{\phi} + 1) + 1\{0 \leq \tilde{\phi} < 1\}\,(1 - \tilde{\phi})^{-1}$ if $-1 < \tilde{\phi} < 1$. 
\end{theorem}

\begin{theorem}
\label{theorem:rpv_Lp_conversion}
Let $(c, p) \in \R^2$ and $\phi(c,p) = \tilde{\phi}$. Then, for any $q \geq 1$, $\varphi_q(c,p)$ can be expressed in terms of $\tilde{\phi}$. Specifically, $\varphi_q(c,p) = \varphi_q(1\{\tilde{\phi} < 0\}\,(1,\,\tilde{\phi} +1) + 1\{\tilde{\phi} \geq 0\}\,(1-\tilde{\phi},\,1))$.
\end{theorem}

The proofs of Theorems \ref{theorem:rpv_mvpf_conversion} and \ref{theorem:rpv_Lp_conversion} are given in \ref{appendix:proofs}. Note that $\tilde{m}$ and $\varphi_q$ (for $q = 1$) are not necessarily sufficient for the RPV. For example, if we know that $\tilde{m}(c,p) = 0$ or $\varphi_1(c,p) = -2$, then we can conclude that $\phi(c,p) \leq -1$ but we cannot pin down the exact value of $\phi(c,p)$ unless we know the value of $(c,p)$. Similarly, if we know that $\tilde{m}(c,p) = \infty$ or $\varphi_1(c,p) = 2$, then we can conclude that $\phi(c,p) \geq 1$ but we cannot pin down the exact value of $\phi(c,p)$ unless we know the value of $(c,p)$. However, it is possible to express the RPV $\phi(c,p)$ in terms of $\varphi_q(c,p)$ if $q > 1$. Nevertheless, it is enough to report inferences on the RPV, which is the only metric that satisfies the welfare-symmetry axioms, for comparative welfare analysis of individual policies if they are treated separately.

\begin{figure}
\begin{center}
\caption{Examples of Confidence Intervals for the Relative Policy Value (RPV)}
\vspace{-1mm}
\label{fig:ci_explanation_graph}
\includegraphics[scale=1]{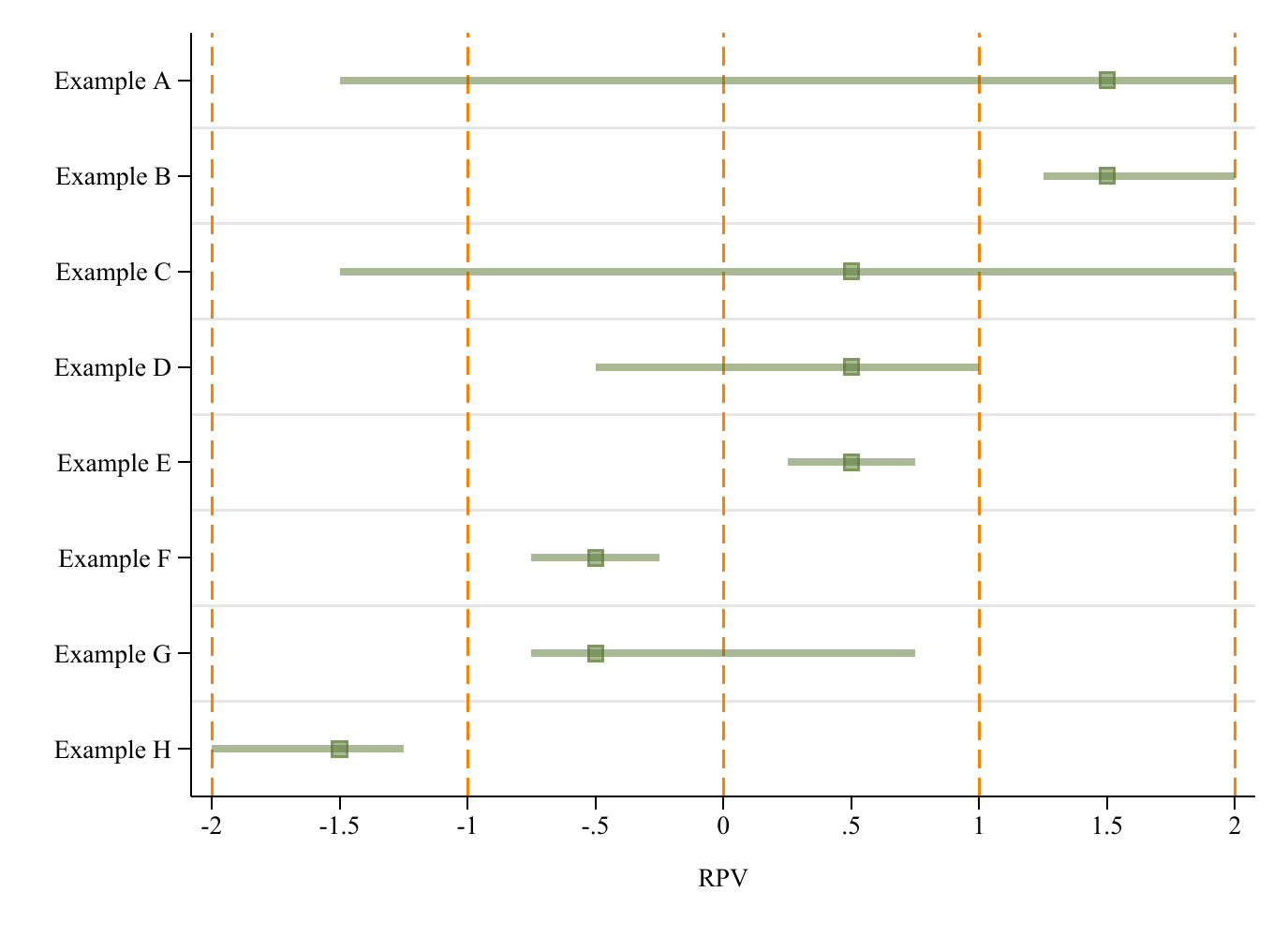} \vspace{-4mm}
\begin{tabular}{cc}
   \hphantom{---}  & \hphantom{---}  \\
   \hphantom{---}   & \hphantom{---} \\
\end{tabular}
\begin{tabular}{|c|c|c|c|c|} \hline
    Range of RPV & $[-2, -1]$ & $(-1, 0)$ & $[0, 1)$ & $[1, 2]$  \\ \hline
   Interpretation & Pareto  & \quad Below \quad & At or above & Pareto \\ 
   of RPV's Range & inferior & break-even & break-even & superior \\ \hline
\end{tabular}
\end{center}
\vspace{2mm}
\footnotesize \textit{Note}: The above graph shows several artificial examples of possible confidence intervals for the Relative Policy Value (RPV) to clarify the notions of statistical versus economic significance in this context.
\end{figure}

I now clarify the notions of economic and statistical significance when interpreting the empirical results of hypothesis tests on the RPV. Because of the obvious relationship between hypothesis tests and confidence intervals (using the same significance level), it is enough to discuss how to interpret the confidence intervals for the RPV. For purely illustrative purposes, Figure \ref{fig:ci_explanation_graph} shows several artificial examples of confidence intervals for the RPV. Since the RPV (like the MVPF) is not a regular function, its estimate may not be at the center of a valid confidence interval.\footnote{In fact, it is generally quite difficult or even impossible to get unbiased estimates for irregular functions (like the RPV or the MVPF), as \cite{hirano2012impossibility} prove.} For example, in Examples A and B of Figure \ref{fig:ci_explanation_graph}, the estimates (equal to 1.5) indicate Pareto superiority, but only the policy associated with Example B can be said to be Pareto superior after accounting for statistical uncertainty (at the chosen significance level). On the other hand, the confidence interval for the RPV of policy in Example A ranges from $-1.5$ to $2$, and so it is not possible to reject Pareto superiority or Pareto inferiority in this case.\footnote{As I show later in Section \ref{section:reanalysis}, there are some important policies for which the confidence interval for the MVPF is a singleton $\{\infty\}$, but the RPV-based confidence intervals are like the one in Example A.} Even though the point estimate of the RPV for Example C is lower than that for Example A, their confidence intervals are the same after accounting for statistical uncertainty.

The policy associated with Example D is a more interesting case. Its confidence interval is $[-0.50,0.99]$, which includes 0, and so the RPV of the policy is not statistically significant based on a hypothesis test of the traditional null hypothesis (of whether the RPV is different from 0). However, it is economically significant in a sense because we can conclude (at the chosen significance level) that it is neither Pareto inferior nor Pareto superior. We can also reject the hypothesis that the relative shortfall in either equity or efficiency is lower than fifty percent. Because each possible value of the RPV (ranging from $-2$ to $2$) has an economic meaning, as shown in the legend of Figure \ref{fig:ci_explanation_graph}, the confidence intervals for the RPV can be economically meaningful, even if the hypothesis tests of the traditional null hypothesis do not result in a rejection. It would be more economically meaningful to conduct hypothesis tests of whether the RPV is the Pareto inferior range $[-2, -1]$, Pareto superior range $[1, 2]$, or in other ranges $(-1, 0)$ or $[0,1)$ that are below and above the break-even point, respectively. In the same way, conclusions about economic versus statistical significance of Examples E, F, G, and H can be made using the legend in Figure \ref{fig:ci_explanation_graph}.

\clearpage

\subsection{Statistical Irregularities in Comparative Welfare Measures}
\label{subsec:stat_irregularities}

In the rest of this section, I discuss which statistical procedures fail and which succeed in producing valid confidence sets. Uniformly valid inference concerns constructing confidence sets such that their asymptotic coverage is controlled uniformly over a reasonably large class of data generating processes. See \cite{romano2012uniform} for background on this topic. It is relatively easy to construct such sets for the marginal social surplus (MSS) $\beta(c, p) = p - c$ using some standard procedures. One can also use more recent methods such as the bootstrap-based calibrated projection procedure \citep{kaido2019confidence} to construct confidence intervals for $\beta(c, p)$ that are not overly conservative but have correct uniform coverage. However, such procedures are not applicable for RPV or MVPF because they are not well-behaved on several regions of the real plane, as stated in Remark \ref{remark:not_lipschitz_or_convex}, which is proved in \ref{appendix:proofs}.

\begin{remark}
\label{remark:not_lipschitz_or_convex}
The Marginal Value of Public Funds $m(c,p)$ and the Relative Policy Value $\phi(c,p)$ are functions that satisfy neither Lipschitz continuity nor convexity nor full differentiability.
\end{remark}

There are several reasons why the RPV and MVPF are not Lipschitz continuous. For example, there are regions where RPV and MVPF have unbounded derivatives. Convexity also fails for both functions because it is possible to take two points on different contour rays, say, $(1,2)$ and $(2,10)$, such that the average of the two points falls on a contour ray with a higher RPV (0.75) or MVPF (4) than the average of the RPVs (0.65) or MVPFs (3.5) of the two points. The RPV and MVPF are also not fully differentiable. For example, the RPV is not differentiable on regions such as the anti-diagonal axis. In addition to not being well-defined on the third quadrant ($\R^2_{<0}$), the MVPF is also not differentiable on the vertical axis.

However, one advantage of RPV over MVPF is that the RPV, unlike the MVPF, is fully differentiable on the vertical axis (except at the origin). In other words, unlike MVPF $m(c,p)$ that goes off to infinity as $c$ goes from 1 to $-1$ (and stays at infinity between $c = 0$ and $c = -1$) when $p$ is fixed at 1, the RPV increases linearly from 0 to 2. Although the non-differentiability of the RPV on the anti-diagonal axis ($p = -c$) is still an issue, it is perhaps less severe than that of the MVPF from the perspective of a policymaker who cares more about knowing whether the policy is Pareto inferior (i.e., with an RPV below $-1$) or Pareto superior (i.e., with an RPV above 1) rather than the exact value of RPV when its magnitude is above one. Of course, this argument does not diminish the need to deal with the lack of full differentiability in the RPV function.

The limiting distribution of the estimator (say, the sample mean vector) of $(c,p)$ is typically Gaussian in many empirical scenarios. However, since the RPV and MVPF are not fully differentiable (by Remark \ref{remark:not_lipschitz_or_convex}), Theorem 3.1 of \cite{fang2019inference} implies that the usual bootstrap procedure, such as the percentile bootstrap confidence interval or a bias-corrected variant (providing second-order corrections) of it, generally fails to be consistent for $m(c,p)$ and $\phi(c,p)$ in this case. The impossibility results of \cite{hirano2012impossibility} can also be used to make this argument. This is particularly true for $m(c,p)$ because it is either undefined or not real-valued when $c \leq 0$, whereas $\phi(c, p)$ is at least well-defined on the entire real plane. Thus, the modified bias-corrected bootstrap\footnote{The bias-corrected bootstrap implicitly assumes consistency of the bootstrap in order to provide second-order correctness in a certain set of problems under some assumptions.}
confidence interval \citep{efron1987better} that \cite{hendren2020unified} use\footnote{Specifically, \cite{hendren2020unified} use the following procedure to construct a $(1 - \alpha)$-confidence interval for $m(c, p)$. Let $\{(c^*_b, p^*_b)\}_{b \in \mathcal{B}}$ be the set of resampled estimates of $(c,p)$ and, in addition, let $m^*_b = m(c^*_b,p^*_b)$ for $b \in \mathcal{B}$, and let $(\hat{c}, \hat{p})$ be the original estimate of $(c,p)$ and $\hat{m} = m(\hat{c}, \hat{p})$. Let $\hat{f} = |\tilde{\mathcal{B}}|^{-1}  \sum_{b \in \tilde{\mathcal{B}}} 1\{m^*_b < \hat{m}\}$, where $\tilde{\mathcal{B}} = \{b: (c^*_b, p^*_b) \not\in \R^2_{< 0}\}$, and $\hat{g} = \Phi^{-1}(\hat{f})$; also let $\lambda(d) = \Phi(2 \hat{g} + \Phi^{-1}(d))$. Then, the confidence interval used by \cite{hendren2020unified} is $[Q^*(\lambda(\gamma/2)), Q^*(\lambda(1 - \gamma/2))]$, where $Q^*(d)$ is the $d$-quantile of the bootstrap-based simulated distribution $\hat{J}(q) = |\tilde{\mathcal{B}}|^{-1}  \sum_{b \in \tilde{\mathcal{B}}} 1\{m^*_b < q\}$, and $(1 - \gamma) = \mathrm{min}\{1, (1 - \alpha) + (1- |\tilde{\mathcal{B}}|/|\mathcal{B}|)\}$. However, this does not resolve bootstrap failure. Even in the simple case where $\tilde{\mathcal{B}} = \mathcal{B}$ almost surely, the non-differentiability of the MVPF $m(c,p)$ on the vertical axis results in failure of the usual bootstrap procedures and its variants (such as the bias-corrected bootstrap method). This is a consequence of Theorem 3.1 of \cite{fang2019inference}.}
is also generally not theoretically valid, despite the simulation results reported in their Online Appendix H, which use a modified coverage criterion because of the specific way $m(c,p)$ is defined (or undefined). One could consider alternative methods such as the recently developed variants of the delta method \citep{fang2019inference}, e.g., the numerical delta method \citep{hong2018numerical}, to conduct inference. However, due to Remark \ref{remark:not_lipschitz_or_convex}, uniform inference is generally not possible using those methods because ``the Lipschitz and convexity properties of [the function] are key to establishing uniform size control'' \citep{hong2018numerical}.

Even though the usual bootstrap procedures (as well as the recent variants discussed above) do not generally yield uniform confidence intervals for MVPF, an alternative method provides uniform coverage but under a particular assumption. Specifically, if we know a priori that $(c, p) \not\in \R^2_{\leq 0}$, then it is possible to use the generalized
moment selection (GMS) procedure of \cite{andrews2010inference} to get uniformly valid confidence interval for $m(c, p)$. Let $(\tilde{C}, \tilde{P}) \sim \tilde{F}$ be a random vector such that $\mathbb{E}_{\tilde{F}}[(\tilde{C}, \tilde{P})] = (c,p)$. Define $g_1(\tilde{C}, \tilde{P}, m) = -\tilde{C}$ and $g_2(\tilde{C}, \tilde{P}, m) = \tilde{P} - m\,\tilde{C}$. Then, let $\theta(\tilde{F}) = \{m \in \R: \mathbb{E}_{\tilde{F}}[g_1(\tilde{C}, \tilde{P}, m)] < 0 \text{ and } \mathbb{E}_{\tilde{F}}[g_2(\tilde{C}, \tilde{P}, m)] = 0 \}$ so that either $\theta(\tilde{F}) \in \R$ or $\theta(\tilde{F})  = \varnothing$. Since $(c, p) \not\in \R^2_{\leq 0}$ in this setting, $\theta(\tilde{F})  = \varnothing$ implies $\mathbb{E}_{\tilde{F}}[\tilde{C}] = c \leq 0$, i.e., $m(c, p) = \infty$. On the other hand, if $\theta(\tilde{F}) \in \R$, then $\mathbb{E}_{\tilde{F}}[-\tilde{C}] < 0$ and $\mathbb{E}_{\tilde{F}}[\tilde{P} - m \,\tilde{C}] = 0$, i.e., $c > 0$ and $p - mc = 0$, i.e., $m = p/c$, implying that $\theta(\tilde{F}) = m(c,p)$. Therefore, a GMS-based confidence set (such that it is assigned the object $\infty$ if empty), constructed by inverting Anderson--Rubin-type tests, provides uniform coverage \citep[see][]{andrews2010inference}, assuming that $(c, p) \not\in \R^2_{\leq 0}$. However, this is a testable assumption and would be untenable if the data do not support it.

\begin{figure}
\begin{center}
\caption{Uniform Confidence Sets for Four ``College Child'' Policies}
\label{fig:uniform_cs_college_child}
    \begin{subfigure}[b]{0.45\textwidth}
         \centering
         \caption{``FIU GPA'' Policy}
         \vspace{0.75mm}
         \includegraphics[scale=.225]{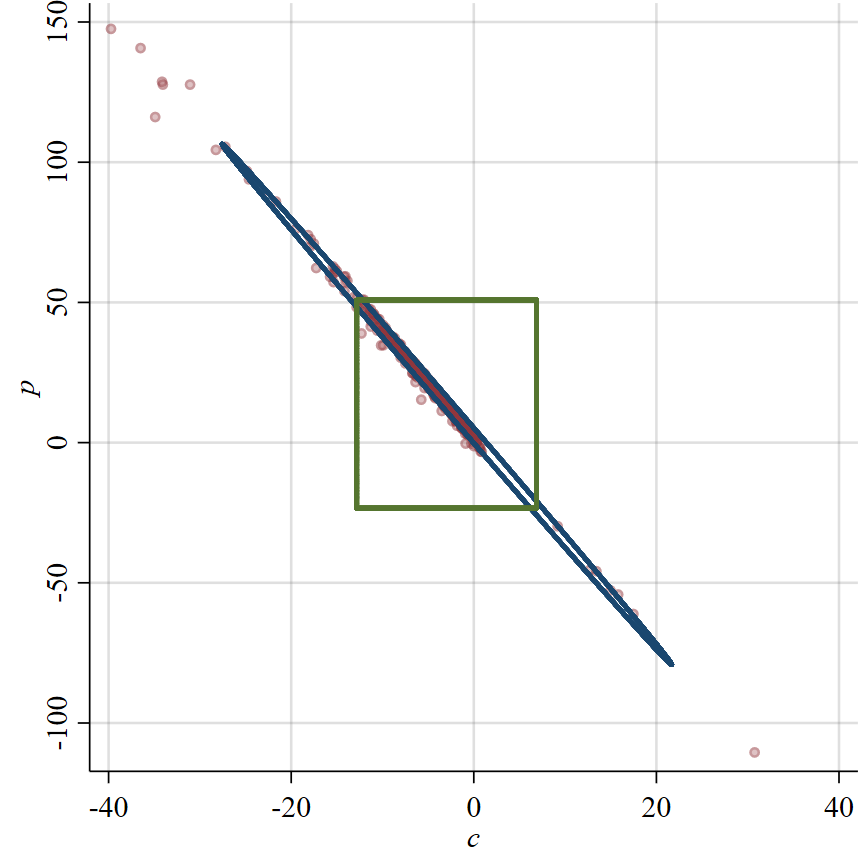}
         \label{fig:fiu_unif_cs}
     \end{subfigure}
     	\begin{subfigure}[b]{0.45\textwidth}
         \centering
         \caption{``CC Texas'' Policy}
         \includegraphics[scale=.225]{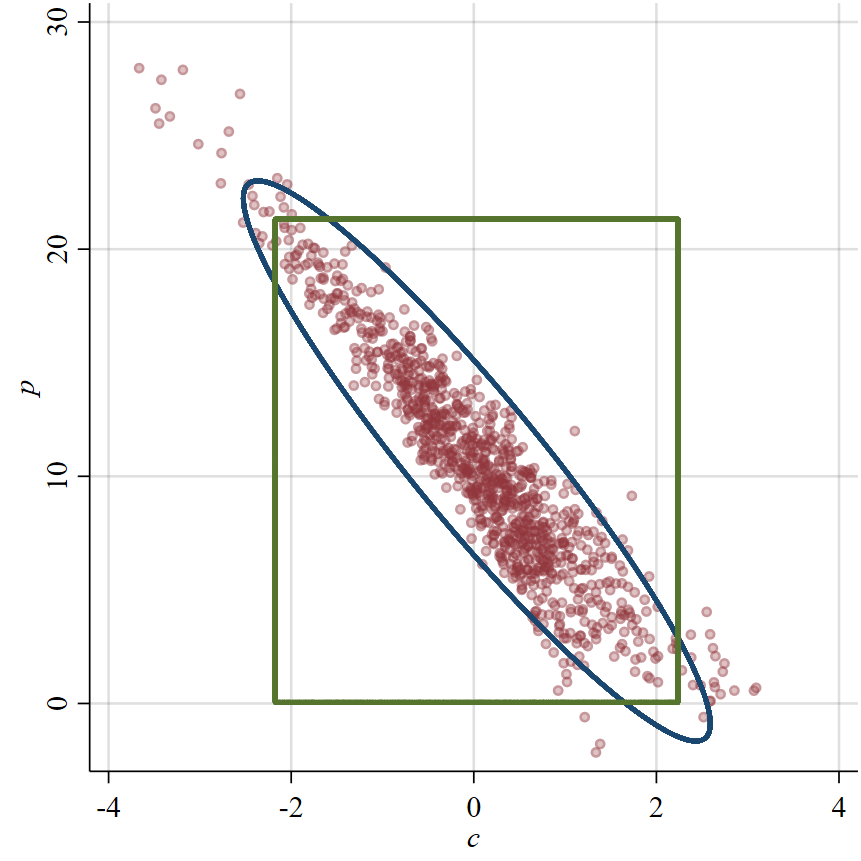}
         \label{fig:cc_texas_unif_cs}
     \end{subfigure}
     \begin{subfigure}[b]{0.45\textwidth}
         \centering
         \vspace{3mm}
         \includegraphics[scale=.225]{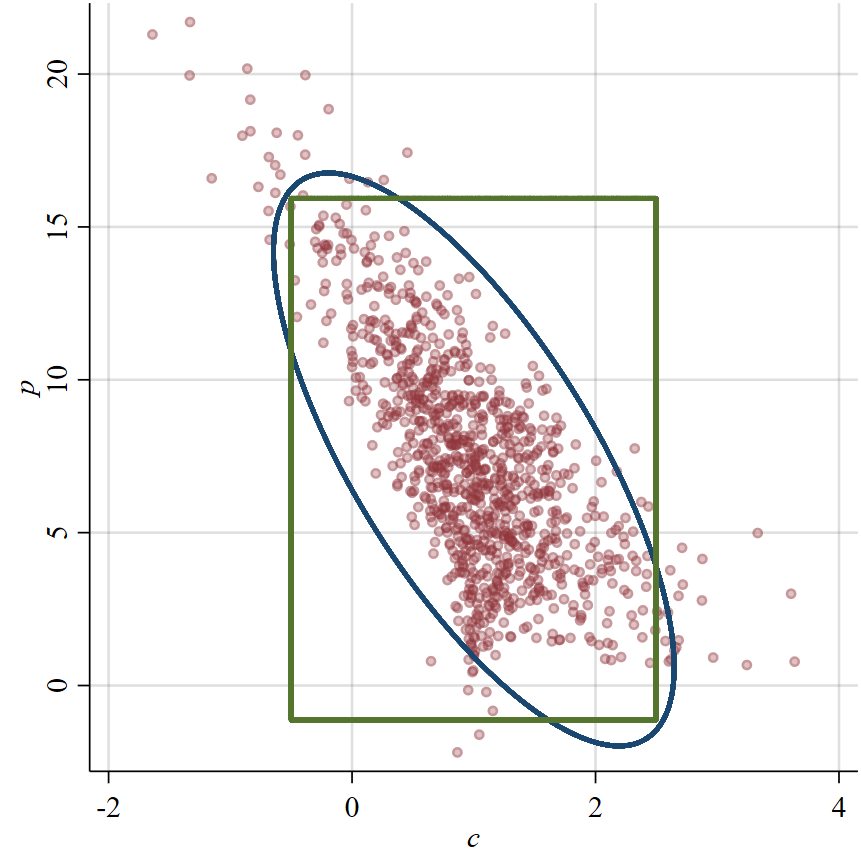}
         \caption{``Florida Grant'' Policy}
         \label{fig:fsag_unif_cs}
     \end{subfigure}
     \begin{subfigure}[b]{0.45\textwidth}
         \centering
         \vspace{3mm}
         \includegraphics[scale=.225]{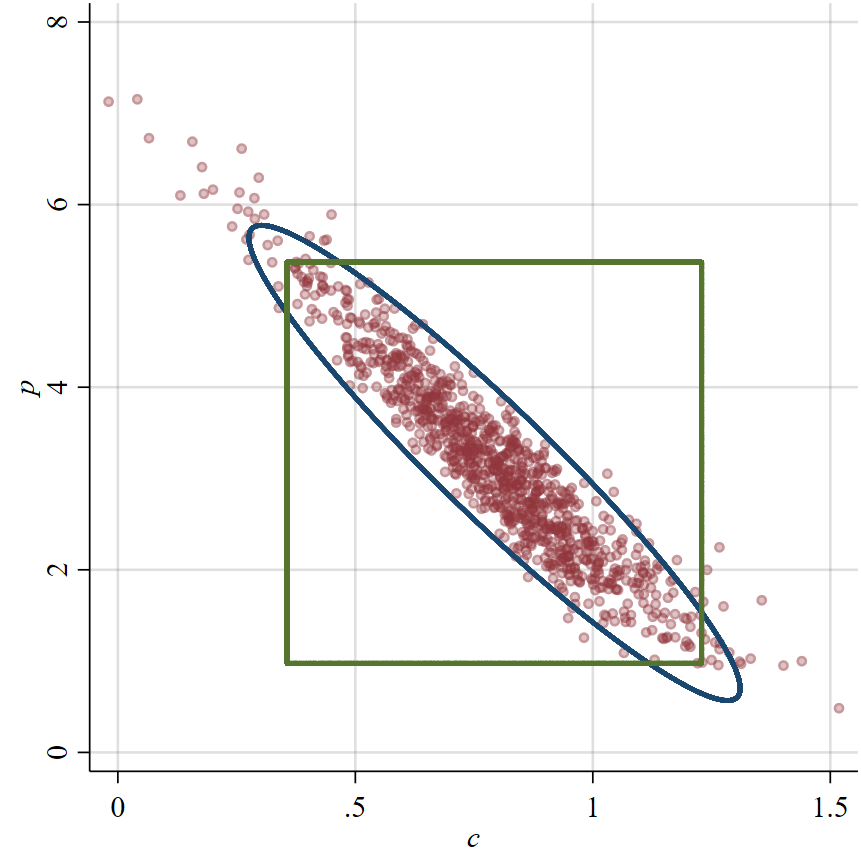}
         \caption{``College Spend'' Policy}
         \label{fig:nat_spend_unif_cs}
     \end{subfigure}
\end{center}
\footnotesize \textit{Note}: The above figures show the boundaries of the rectangular and ellipsoidal 95\% uniform confidence sets (constructed using Lemma \ref{lemma:cp_uniform_cs} in Section \ref{subsec:uniform_inference}) for the vector $(c,p)$ containing the net fiscal cost and willingness-to-pay of four policies within the ``College Child'' policy category considered by \cite{hendren2020unified}. The dots in the above figures represent \citeauthor{hendren2020unified}'s (\citeyear{hendren2020unified}) resampled estimates of $(c,p)$. A few of the resampled estimates fall far outside the confidence regions and are thus not displayed in the above figures. ``FIU GPA,'' ``CC Texas,'' ``Florida Grant,'' and ``College Spend'' refer to ``Florida International University Admissions at GPA Threshold,'' ``Community College Tuition Changes in Texas,'' ``Florida Student Access Grant,'' and ``Spending at Colleges from State Appropriations,'' respectively \citep{hendren2020unified}.
\end{figure}

Figure \ref{fig:fiu_unif_cs} shows rectangular and elliptical 95\% uniform confidence regions (constructed using Lemma \ref{lemma:cp_uniform_cs} in Subsection \ref{subsec:uniform_inference}) for the vector $(c,p)$ containing the net fiscal cost and willingness-to-pay of the ``FIU GPA (Florida International University Admissions at GPA Threshold)'' policy, which \cite{hendren2020unified} use as a prominent example to illustrate the computation of MVPF. Both the rectangular and ellipsoidal confidence sets in this case span all four quadrants.\footnote{While it is visually clear that the rectangular confidence set in Figure \ref{fig:fiu_unif_cs} spans all four quadrants, this property is not visually obvious for the thin ellipsoidal confidence region. However, the ellipse shown in the figure includes points such as $(-0.0131057, -0.1443748)$ and $(-0.0480095, -0.0133246)$, which are in the third quadrant.} Therefore, at the 95\% confidence level, one cannot reject the possibility that $(c, p) \in \R^2_{\leq 0}$. This is not unique to the FIU GPA policy; there are several other policies for which negative net fiscal costs together with negative willingness-to-pay values cannot be rejected. In these cases, even the procedure of \cite{andrews2010inference} is not applicable, and so an alternative is required.

One simple way to understand the inconsistency of the conventional bootstrap procedures in the case of the RPV or MVPF is to inspect the distribution of $\phi(\overline{x}^*)$ or $m(\overline{x}^*)$, where $\overline{x}^*$ is a bootstrap resample of the estimate of $x = (c,p)$. The extreme regions of the distribution of $\phi(\overline{x}^*)$ or $m(\overline{x}^*)$ need not always correspond to the extreme regions of the bivariate distribution of $\overline{x}^*$ (that fall outside of the confidence set for $x$), especially since $\phi$ and $m$ are not well-behaved. Thus, by using the distribution of $\phi(\overline{x}^*)$ or $m(\overline{x}^*)$ to construct a confidence interval for $\phi(x)$ or $m(x)$, the analyst may potentially exclude the RPV or MVPF values of some reasonable points within the confidence set for $x = (c,p)$. This, of course, leads to the failure of the conventional bootstrap methods in producing valid confidence intervals for the RPV or MVPF. This discussion highlights several analytical and statistical challenges in using MVPF to conduct credible inference regarding welfare generated by public policies. Although the RPV is well-defined unlike the MVPF, the RPV is also a statistically challenging function. As Remark \ref{remark:not_lipschitz_or_convex} states, the RPV is neither Lipschitz continuous nor convex nor fully differentiable. Thus, procedures such as the numerical delta method \citep{hong2018numerical} cannot be used to conduct uniform inference for RPV. Nevertheless, one can use results in Section \ref{subsec:uniform_inference} to make uniformly valid inferences on the very general parameter set $\{\Theta_k(X_\mathcal{L})\}_{k \in \mathcal{K}} \equiv \{\Theta_k(\{(c_l,p_l)\}_{l \in \mathcal{L}})\}_{k \in \mathcal{K}}$ discussed in Subsection \ref{subsec:parameters}.

\subsection{Uniformly Valid Inference Methods for Cost--Benefit Analysis}
\label{subsec:uniform_inference}

It is useful to first construct a uniform $(1 - \alpha)$-confidence set for $X_\mathcal{L} \equiv \{(c_l, p_l)\}_{l \in \cL}$ in order to produce uniform $(1 - \alpha)$-confidence sets for $\{\Theta_k(X_\mathcal{L})\}_{k \in \mathcal{K}}$ at a chosen significance level $\alpha \in (0,1)$ (and thus the chosen confidence level $1 - \alpha$). Let $\{(x_{l,1}, \dots, x_{l,n_l})\}_{l \in \cL} \equiv \{x_l^{(n_l)}\}_{l \in \cL}$, where $n_l$ denotes the sample size, be a collection of independent and identically distributed random variables such that $x_{l,i} = (c_{l,i}, p_{l,i}) \sim F_l \in \mathbb{F}_l$ on $\R^2$ for all $i \in \{1, \dots, n_l\} $ for all $l \in \cL$. For expositional ease, assume that random variables $\{x_{l,i}\}_{i = 1}^{n_l}$ have the mean $(c_l, p_l) \equiv (\mu_c(F_l), \mu_p(F_l)) = \mathbb{E}[(c_{l,i}, p_{l,i})]$ for all $l \in \mathcal{L}$, although it is possible to make appropriate modifications to extend the theory to the case where $(c_l, p_l)$ may depend on other moments of a broader set of random observations. Assume that each of $\{F_l\}_{l \in \cL}$ satisfies the standardized uniform integrability condition \citep[as defined by][]{romano2012uniform}. In addition, assume that $\Sigma(F_l)$ is a positive definite covariance matrix, with the associated correlation matrix $\Omega(F_l)$, for $F_l$ for all $l \in \cL$. In addition, define $n = \mathrm{min}_{l \in \cL} \,n_l$. Using this setup, it is possible to adapt and apply the general results of \cite{romano2012uniform} to obtain a uniform confidence set for $X_\mathcal{L} \equiv \{(c_l, p_l)\}_{l \in \cL}$, based on which a uniform confidence set for $\{\Theta_k(X_\mathcal{L})\}_{k 
\in \mathcal{K}}$ can be constructed. To this end, first define the root as
$$R_n \equiv R_n(\{x_l^{(n_l)}\}_{l \in \cL}, F) = \mathrm{sup}_{l \in \cL}\,\, R^l_{n_l}(x_l^{(n_l)}, F_l),$$
where $F \equiv \{F_l\}_{l \in \mathcal{L}} \in \mathbb{F}$ is the set of the marginal distributions and
$$R^l_{n_l}(x^{(n_l)}_l,F_l) = \mathrm{max}\{\sqrt{n_l}|\,\overline{c}_{l,n_l} - \mu_c(F_l)|/s_{c,l,n_l},\,\sqrt{n_l}|\,\overline{p}_{l,n_l} - \mu_p(F_l)|/s_{p,l,n_l}\},$$
with $(\overline{c}_{l,n_l},\overline{p}_{l,n_l})$ and $(s_{c,l,n_l}, s_{p,l,n_l})$ denoting the sample means and sample standard deviations of the sample $\{(c_{l,i},p_{l,i})\}_{i = 1}^{n_l}$, respectively, for all $l \in \mathcal{L}$. In case $(c_l, p_l)$ are estimated using an alternative method (rather than computing sample means), in many practical settings we can replace $(\overline{c}_{l,n_l},\overline{p}_{l,n_l})$ and $(s_{c,l,n_l}, s_{p,l,n_l})/\sqrt{n_l}$ with the estimates $(\hat{c}_{l,n_l},\hat{p}_{l,n_l})$ of $(c_l, p_l)$ and their standard errors, respectively, assuming there are no other theoretical issues in that particular empirical context. Define a confidence set $\hat{S}_n(d)$ based on the sample using a critical value $d$ as follows:
$$\textstyle \hat{S}_n(d) = \{\{(\mu_c(\tilde{F}_l),\mu_p(\tilde{F}_l))\}_{l \in \cL} \in \prod_{l \in \cL} \R^2: R_n(\{x^{(n_l)}_l\}_{l \in \cL}, \tilde{F}) \leq d\}.$$
Note that the confidence set $\hat{S}_n(d)$ is a hyperrectangle:
$$\textstyle \hat{S}_n(d) = \prod_{l \in \cL} \bigtimes_{v \in \{c,p\}} [\overline{v}_{l,n_l} - d\,s_{v,l,n_l}/\sqrt{n_l}, \,\,\overline{v}_{l,n_l} + d\,s_{v,l,n_l}/\sqrt{n_l}].$$
We could also construct an alternative confidence set $\hat{E}_n(t)$, which reduces to an ellipse when $\mathcal{L}$ is just a singleton, using a critical value $t$:
$$\textstyle \hat{E}_n(t) = \{\{(\mu_c(\tilde{F}_l),\mu_p(\tilde{F}_l))\}_{l \in \cL} \in \prod_{l \in \cL} \R^2: R^\circ_n(\{x^{(n_l)}_l\}_{l \in \cL}, \tilde{F}) \leq t\},$$
where
$$R^\circ_n \equiv R^\circ_n(\{x_l^{(n_l)}\}_{l \in \cL}, F) = \mathrm{sup}_{l \in \cL}\,\, R^{\circ, l}_{n_l}(x_l^{(n_l)}, F_l)$$
and
$$R^{\circ, l}_{n_l}(x^{(n_l)}_l,F_l) = (z^c_{l,n_l}(F_l), z^p_{l,n_l}(F_l))'\,\Omega^{-1}(F_l)\,(z^c_{l,n_l}(F_l), z^p_{l,n_l}(F_l)),$$
with $z_{l,n_l}(F_l) = (z^c_{l,n_l}(F_l), z^p_{l,n_l}(F_l)) = \sqrt{n_l}\,([\overline{c}_{l,n_l} - \mu_c(F_l)]/s_{c,l,n_l}, \,[\overline{p}_{l,n_l} - \mu_p(F_l)]/s_{p,l,n_l})$. Using these roots, the general results of \cite{romano2012uniform} apply for any $\alpha \in (0, 1)$.

\begin{lemma}
\label{lemma:cp_uniform_cs}
Suppose $F_l$ satisfies the standardized uniform integrability condition for all $l \in \cL$. Let $H_n(\cdot, F)$ be the distribution of the root $R_n$. Let $\hat{d}^\alpha_n = H^{-1}_n(1 - \alpha, \hat{F}_n)$, where $\hat{F}_n = \{\hat{F}_{l,n_l}\}_{l \in \mathcal{L}}$ and $\hat{F}_{l,n_l}$ is the empirical distribution function based on the sample $x_l^{(n_l)}$ for all $l \in \cL$. Then, the hyperrectangle $\hat{S}_n(\hat{d}^\alpha_n)$ is a uniformly valid asymptotic $(1 - \alpha)$-confidence set for $\{(c_l, p_l)\}_{l \in \cL}$, i.e., $\mathrm{lim}_{n \to \infty} \,\mathrm{inf}_{F \in \mathbb{F}} \,\mathbb{P}_F\{\{(c_l, p_l)\}_{l \in \cL} \in \hat{S}_n(\hat{d}^\alpha_n)\} \geq 1 - \alpha$ for any $\alpha \in (0, 1)$. If $H^\circ_n(\cdot, F)$ is the distribution of the root $R^\circ_n$, and $\hat{t}^\alpha_n = H^{\circ-1}_n(1 - \alpha, \hat{F}_n)$, then $\hat{E}_n(\hat{t}^\alpha_n)$ is also a uniform $(1 - \alpha)$-confidence set for $\{(c_l, p_l)\}_{l \in \cL}$, i.e., $\mathrm{lim\,inf}_{n \to \infty} \,\mathrm{inf}_{F \in \mathbb{F}} \,\mathbb{P}_F\{\{(c_l, p_l)\}_{l \in \cL} \in \hat{E}_n(\hat{t}^\alpha_n)\} \geq 1 - \alpha.$
\end{lemma}

The proof of Lemma \ref{lemma:cp_uniform_cs} is given in \ref{appendix:proofs}. The above lemma and the other econometric results in this section are essentially applied in nature and should not be considered technically novel because they rely on established theoretical results, especially those of \cite{romano2012uniform}. However, one of the major contributions of this paper is to connect ideas and to adapt and apply recent methods to solve an economically important policy-relevant problem. Therefore, the proposed statistical procedures are novel from an applied econometric perspective in the context of cost--benefit analysis. Figure \ref{fig:uniform_cs_college_child} shows examples of (single, non-simultaneous) confidence sets constructed separately for four policies using Lemma \ref{lemma:cp_uniform_cs}, which could also be used to construct simultaneous confidence sets. Based on a confidence set $B$ for $X_\mathcal{L} \equiv \{(c_l, p_l)\}_{l \in \cL}$, it is possible to construct a confidence set $\hat{C}_n(B)$  for $\{\Theta_k\}_{k \in \mathcal{K}} \equiv \{\Theta_k(X_\mathcal{L})\}_{k \in \mathcal{K}}$ using the projection method as follows:

$$\hat{C}_n(B) = \prod_{k \in \mathcal{K}} \,\Big[\,\underset{\tilde{X}_\mathcal{L} \,\in\, B}{\mathrm{inf}}\, \Theta_k(\tilde{X}_\mathcal{L}), \,\,\,\underset{\tilde{X}_\mathcal{L} \,\in\, B}{\mathrm{sup}}\, \Theta_k(\tilde{X}_\mathcal{L})\,\Big].$$

\vspace{2mm}

\begin{theorem}
\label{theorem:param_uniform_cs}
Suppose $F_l$ satisfies the standardized uniform integrability condition for all $l \in \cL$, and suppose $\hat{B}^\alpha_n$ is chosen to be either $\hat{S}_n(\hat{d}^\alpha_n)$ or $\hat{E}_n(\hat{t}^\alpha_n)$, as defined in Lemma \ref{lemma:cp_uniform_cs}. Then, $\hat{C}_n(\hat{B}^\alpha_n)$ is a uniform $(1 - \alpha)$-confidence set, i.e., $\mathrm{lim\,inf}_{n \to \infty} \,\mathrm{inf}_{F \in \mathbb{F}} \,\mathbb{P}_F\{\{\Theta_k(X_\mathcal{L})\}_{k \in \mathcal{K}} \in \hat{C}_n(\hat{B}^\alpha_n)\} \geq 1 - \alpha.$
\end{theorem}

\begin{remark}
\label{remark:minimalist_projection}
Since the asymptotically conservative set $\hat{C}_n(\hat{B}^\alpha_n)$ described in Theorem \ref{theorem:param_uniform_cs} may be too large and could potentially over-cover $\{\Theta_k(X_\mathcal{L})\}_{k \in \mathcal{K}}$ (in addition to being computationally difficult sometimes), it may be worthwhile to report a ``minimalist'' projection-based confidence set $\hat{C}^\odot_{n,\mathcal{B}}(\hat{B}^\alpha_n) = \mathrm{conv}\{\{\Theta_k(X^{*,b}_\mathcal{L})\}_{k \in \mathcal{K}}: b \in \mathcal{B}, X^{*,b}_\mathcal{L} \in \hat{B}^\alpha_n\}$, where $\{X^{*,b}_\mathcal{L}\}_{b \in \mathcal{B}}$ is a set of $|\mathcal{B}|$ resampled estimates of $X_\mathcal{L}$.\footnote{Alternatively, instead of using $\hat{C}^\odot_{n,\mathcal{B}}(\hat{B}^\alpha_n)$, one may also possibly project $\mathrm{conv}\{\hat{B}^\alpha_n \cap \{X^{*,b}_\mathcal{L}\}_{b \in \mathcal{B}} \}$ using $\{\Theta_k(\cdot)\}_{k \in \mathcal{K}}$.} In the case where there is just a single parameter of interest, $\hat{C}^\odot_{n,\mathcal{B}}(\hat{B}^\alpha_n)$ reduces to an interval on the real line. Although $\hat{C}^\odot_{n,\mathcal{B}}(\hat{B}^\alpha_n)$ may not be uniformly valid by itself, it can be considered a ``sub''-uniformly-valid confidence interval, because $\hat{C}^\odot_{n,\mathcal{B}}(\hat{B}^\alpha_n) \subset \hat{C}_n(\hat{B}^\alpha_n)$ by construction. Therefore, any values in $\hat{C}^\odot_{n,\mathcal{B}}(\hat{B}^\alpha_n)$ must also be present in $\hat{C}_n(\hat{B}^\alpha_n)$, implying that the values in the minimalist set $\hat{C}^\odot_{n,\mathcal{B}}(\hat{B}^\alpha_n)$ are statistically plausible values of the parameter $\{\Theta_k(X_\mathcal{L})\}_{k \in \mathcal{K}}$, and so they cannot be rejected during hypothesis testing at the significance level $\alpha$.
\end{remark}

\begin{theorem}
\label{theorem:param_rpv}
Suppose $\mathcal{L}$ and $\mathcal{K}$ are singletons and $\phi(c,p)$ is the parameter of interest for a single policy with $X_\mathcal{L} \equiv (c,p)$, and suppose $\hat{B}^\alpha_n$ is chosen to be either $\hat{S}_n(\hat{d}^\alpha_n)$ or $\hat{E}_n(\hat{t}^\alpha_n)$ defined in Lemma~\ref{lemma:cp_uniform_cs}. Then, $\hat{C}_n(\hat{B}^\alpha_n) = \phi(\partial \hat{B}^\alpha_n)$, where $\partial \hat{B}^\alpha_n$ represents the boundary of the region $\hat{B}^\alpha_n$, is a uniform $(1 - \alpha)$-confidence set for $\phi(c,p)$, i.e., $\mathrm{lim\,inf}_{n \to \infty} \,\mathrm{inf}_{F \in \mathbb{F}} \,\mathbb{P}_F\{\phi(c,p) \in \phi(\partial \hat{B}^\alpha_n)\} \geq 1 - \alpha.$
\end{theorem}

In the above case where the Relative Policy Value (RPV) $\phi(c,p)$ of a single policy is a parameter of interest, the uniform confidence interval $\phi(\partial \hat{B}^\alpha_n)$ can be computed almost exactly using numerical or analytic optimization techniques. Alternatively, one may use simulation methods to approximate $\phi(\partial \hat{B}^\alpha_n)$ up to desired precision as stated in the following theorems, which use the notation and assumptions in Lemma \ref{lemma:cp_uniform_cs} and Theorem \ref{theorem:param_rpv}. 

\begin{theorem}
\label{theorem:rpv_rectangle_ci}
Suppose $\hat{B}^\alpha_n = \hat{S}_n(\hat{d}^\alpha_n)$ such that its center is $\overline{x}_n = (\overline{c}_n, \overline{p}_n)$, its length equals $2\hat{r}^c_n$, and its width equals $2\hat{r}^p_n$. Let $\{(b^1_k,b^2_k,u_k)\}_{k = 1}^K$ be $K$ random vectors such that $(b^1_k,b^2_k,u_k)$ are mutually independent, $b^1_k, b^2_k \sim \mathrm{Bernoulli}(0.5)$, and $u_k \sim \mathrm{Uniform}[-1,1]$ for $k \in \{ 1, \dots, K\}$. Let $\upsilon_k \equiv \overline{x}_n + b^1_k((-1)^{b^2_k}\hat{r}^c_n,u_k\hat{r}^p_n) + (1 - b^1_k)(u_k\hat{r}^c_n,(-1)^{b^2_k}\hat{r}^p_n)$ for all $k \in \{ 1, \dots, K\}$. Then, $\hat{C}^{\dagger}_{n,K}(\hat{B}^\alpha_n) = [\mathrm{min}\{\phi(\upsilon_k)\}_{k = 1}^K,\,\mathrm{max}\{\phi(\upsilon_k)\}_{k = 1}^K]$ is a rectangle-based approximate asymptotically valid uniform confidence interval such that $\hat{C}^{\dagger}_{n,K}(\hat{B}^\alpha_n)\to \phi(\partial \hat{B}^\alpha_n)$ almost surely as $K \to \infty$.
\end{theorem}

\begin{theorem}
\label{theorem:rpv_ellipse_ci}
Suppose $\hat{B}^\alpha_n = \hat{E}_n(\hat{t}^\alpha_n) = \{\tilde{x} \in \R^2: (\overline{x}_n - \tilde{x})'[\Sigma^{-1}(\hat{F}_n)](\overline{x}_n - \tilde{x}) \leq \hat{t}^\alpha_n\}$. Let $u_1, \dots, u_K$ be $K$ independent $\mathrm{Uniform}[0, 2\pi]$ random variables,  and let $v_k = \sqrt{\hat{t}^\alpha_n}(\mathrm{cos}(u_k), \mathrm{sin}(u_k))$ for all $k \in \{1,\dots, K\}$. Let $\tilde{x}_k = \overline{x}_n + [\Sigma^{1/2}(\hat{F}_n)] v_k$ for all random draws $k \in \{1,\dots, K\}$. Then, $\hat{C}^{\dagger}_{n,K}(\hat{B}^\alpha_n) = [\mathrm{min}\{\phi(\tilde{x}_k)\}_{k = 1}^K,\,\mathrm{max}\{\phi(\tilde{x}_k)\}_{k = 1}^K]$ is an ellipse-based approximate asymptotically valid uniform confidence interval such that $\hat{C}^{\dagger}_{n,K}(\hat{B}^\alpha_n) \to \phi(\partial \hat{B}^\alpha_n)$ almost surely as $K \to \infty$.
\end{theorem}

\begin{table}[ht]
\setstretch{1.5}
\caption{Confidence Intervals for MVPFs and RPVs of Four ``College Child'' Policies}
\label{table:college_child_example}
\vspace{-5mm}
    \begin{center} \footnotesize
    \begin{tabular}{c|c|c|c|c|c}
    \hline\hline
\textit{Description} & \textit{Statistic} & \textit{\,\,\,\,\,\,FIU GPA\,\,\,\,\,\,} & \textit{\,\,\,\,\,\,CC Texas\,\,\,\,\,\,} & \textit{Florida Grant} & \textit{College Spend} \\\hline
MVPF & $m(\hat{c}, \hat{p})$ & $ \infty $ & $ 349.51 $ & $ 7.42 $ & $ 4.00 $ \\ \hline
Bias-Corrected & $\mathrm{min}\{\text{Efron CI}\}$ & $ \infty $ & $ 1.61 $ & $ 1.09 $ & $ 1.25 $ \\
Bootstrap CI & $\mathrm{max}\{\text{Efron CI}\}$ & $ \infty $ & $ \infty $ & $ \infty $ & $ 20.44 $ \\ \hline\hline
RPV & $\phi(\hat{c},\hat{p})$ & $ 1.22 $ & $ 1.00 $ & $ 0.87 $ & $ 0.75 $ \\ \hline
Rectange-Based & $\mathrm{min}\{\hat{C}^{\dagger}_{n,K}(\hat{S}_n(\hat{q}^\alpha_n))\}$ & $ -2.00 $ & $ -0.98 $ & $ -2.00 $ & $ -0.21 $ \\
Uniform CI & $\mathrm{max}\{\hat{C}^{\dagger}_{n,K}(\hat{S}_n(\hat{q}^\alpha_n))\}$ & $ 2.00 $ & $ 2.00 $ & $ 2.00 $ & $ 0.93 $ \\ \hline
Ellipse-Based & $\mathrm{min}\{\hat{C}^{\dagger}_{n,K}(\hat{E}_n(\hat{t}^\alpha_n))\}$ 
& $ -2.00 $ & $ -1.68 $ & $ -1.93 $ & $ -0.56 $ \\
Uniform CI & $\mathrm{max}\{\hat{C}^{\dagger}_{n,K}(\hat{E}_n(\hat{t}^\alpha_n))\}$ & $ 1.90 $ & $ 1.12 $ & $ 1.05 $ & $ 0.95 $ \\ \hline
Rectangle-Based & $\mathrm{min}\{\hat{C}^\odot_{n,\mathcal{B}}(\hat{S}_n(\hat{q}^\alpha_n))\}$ & $ -1.37 $ & $ -0.54 $ & $ -1.72 $ & $ -0.20 $ \\
Minimalist CI & $\mathrm{max}\{\hat{C}^\odot_{n,\mathcal{B}}(\hat{S}_n(\hat{q}^\alpha_n))\}$ & $ 1.38 $ & $ 1.11 $ & $ 1.04 $ & $ 0.93 $ \\ \hline
Ellipse-Based & $\mathrm{min}\{\hat{C}^\odot_{n,\mathcal{B}}(\hat{E}_n(\hat{t}^\alpha_n))\}$ & $ -1.37 $ & $ -1.24 $ & $ -0.70 $ & $ -0.24 $ \\
Minimalist CI & $\mathrm{max}\{\hat{C}^\odot_{n,\mathcal{B}}(\hat{E}_n(\hat{t}^\alpha_n))\}$ & $ 1.27 $ & $ 1.12 $ & $ 1.04 $ & $ 0.95 $ \\
\hline\hline
    \end{tabular}
    \end{center}
    \vspace{-1mm}
\setstretch{1}\noindent \footnotesize
\textit{Note}: The above table shows the MVPFs, RPVs, and their associated confidence intervals for four policies within the ``College Child'' category considered by \cite{hendren2020unified}. ``FIU GPA,'' ``CC Texas,'' ``Florida Grant,'' and ``College Spend'' refer to ``Florida International University Admissions at GPA Threshold,'' ``Community College Tuition Changes in Texas,'' ``Florida Student Access Grant,'' and ``Spending at Colleges from State Appropriations,'' respectively \citep{hendren2020unified}. The estimate of $(c,p)$ for each policy is denoted by $(\hat{c},\hat{p})$. Efron CI refers to the modified bias-corrected bootstrap 95\% confidence interval \citep{efron1987better} that \cite{hendren2020unified} use to report statistical uncertainty in the MVPF estimates. The confidence intervals (CIs) for the RPVs reported above (using $|\mathcal{B}| = 10^3$ and $K = 10^5$) are described in Theorems \ref{theorem:rpv_rectangle_ci} and \ref{theorem:rpv_ellipse_ci} and Remark \ref{remark:minimalist_projection}.
\end{table}

The proofs of Theorems \ref{theorem:param_uniform_cs}, \ref{theorem:param_rpv}, \ref{theorem:rpv_rectangle_ci}, and \ref{theorem:rpv_ellipse_ci} are given in \ref{appendix:proofs}. Note that I do not discuss the other comparative welfare measures, such as the ``fixed'' MVPF or the $L^q$-normalized welfare indices, above because the RPV is sufficient in the sense formalized in Theorems \ref{theorem:rpv_mvpf_conversion} and \ref{theorem:rpv_Lp_conversion}. Table \ref{table:college_child_example} illustrates how statistical inference using the above results can sometimes vastly differ from conclusions based on standard procedures.
A reanalysis of the ``FIU GPA'' policy, which is one of the main examples of \cite{hendren2020unified}, and three other policies (in the ``College Child'' category) illustrates how and why bootstrap failure occurs for MVPF and also why the methods presented in this paper are useful. Table \ref{table:college_child_example} shows the welfare estimates and the associated confidence intervals for the four examples. The associated rectangular and elliptical uniform confidence sets for the net fiscal costs and willingness-to-pay values of these policies are shown in Figure \ref{fig:uniform_cs_college_child}. The lower end points of the bias-corrected bootstrap 95\% confidence intervals (or Efron CIs) for MVPFs of the four policies constructed by \cite{hendren2020unified} are all above 1. In particular, the Efron CI for the MVPF of the FIU GPA policy is degenerate at $\{\infty\}$. These 95\% Efron CIs are formed by removing ``extreme'' 5\% of the resampled MVPF estimates. However, as can be seen in Figure \ref{fig:uniform_cs_college_child}, some of these ``extreme'' MVPF estimates are associated with ``reasonable'' values (i.e., those within the uniform confidence set) for the net fiscal costs and willingness-to-pay values of the policies, contributing to the invalidity of these Efron CIs.

In contrast, the minimalist projection-based confidence interval $\hat{C}^\odot_{n,\mathcal{B}}(\hat{B}^\alpha_n)$, which is not too conservative but also possibly not uniformly valid (as described in Remark \ref{remark:minimalist_projection}), selects 95\% of the resampled welfare estimates based on the ``reasonable'' values of the net fiscal costs and the willingness-to-pay values (WTPs) of the policies. Similarly, the (more conservative) uniform confidence interval $\hat{C}_n(\hat{B}^\alpha_n) = \phi(\partial \hat{B}^\alpha_n)$ described in Theorem \ref{theorem:param_rpv} also uses ``reasonable'' values for the net fiscal costs and WTPs of the policies to form confidence intervals for their RPVs. Regardless of which projection-based method is used to construct confidence intervals for the RPVs of the four policies, the results in Table \ref{table:college_child_example} show that we cannot reject either positivity or negativity of the RPVs of the four policies at the 95\% confidence level. In other words, the available data on these four policies do not allow us to make strong conclusions about the welfare generated by these four policies. In fact, the uniform 95\% confidence interval for the RPV of the FIU GPA policy is $[-2, 2]$, i.e., the entire range of the RPV function, despite the single-point 95\% confidence interval $\{\infty\}$ that \cite{hendren2020unified} report for the policy's MVPF. These concrete empirical examples demonstrate the challenges associated with statistical inference for the MVPF using conventional methods. Thus, it is important to use uniformly valid confidence sets and well-defined welfare measures like the RPV to make inferences about the welfare of policies.

\subsection{Practical Performance of the Proposed Inference Methods}
\label{subsec:simulations}

\begin{figure}
\begin{center}
\caption{Simulation Evidence on Average Coverage Probability versus Distance from Origin}
\label{fig:coverage_amax}
	\begin{subfigure}[b]{1\textwidth}
         \centering
         \caption{$n = 100$}
         \includegraphics[scale=0.8]{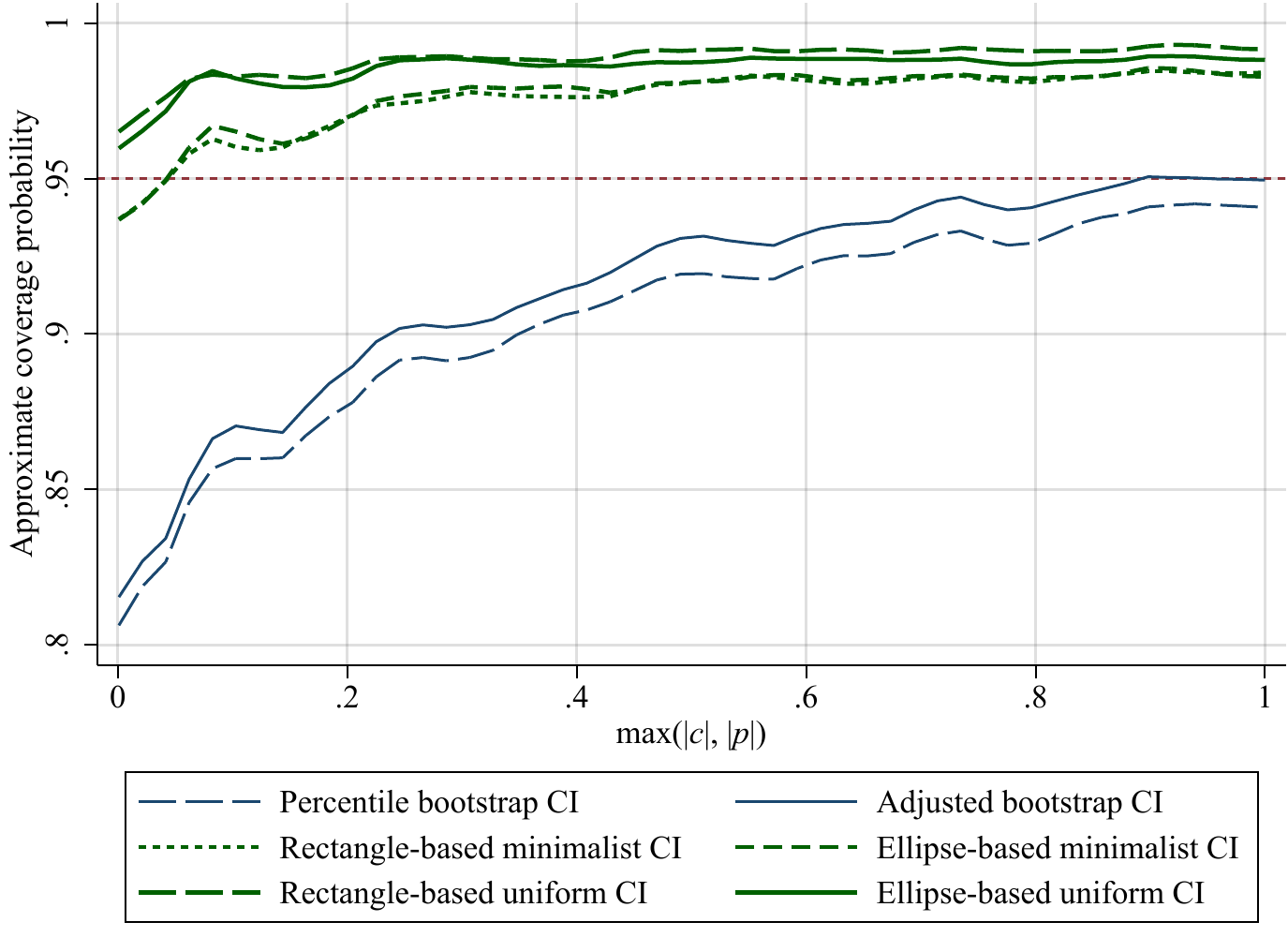}
         \label{fig:coverage_amax_100}
     \end{subfigure}
    \begin{subfigure}[b]{1\textwidth}
         \centering
         \includegraphics[scale=0.8]{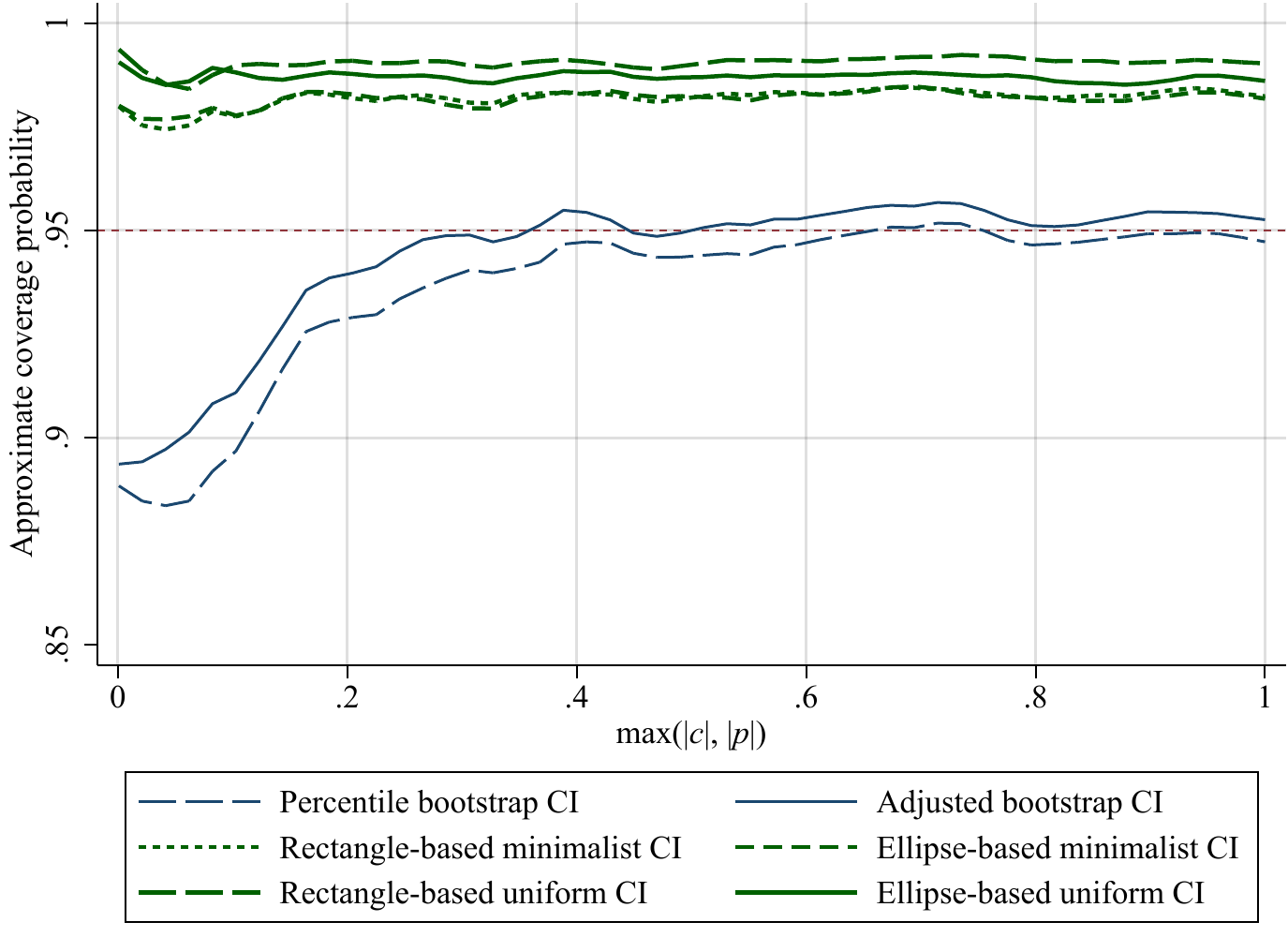}
         \caption{$n = 1000$}
         \label{fig:coverage_amax_1000}
     \end{subfigure}
\end{center}
\footnotesize \textit{Note}: The above figures plot the approximate average coverage probability (of various confidence intervals for the RPV) for two sample sizes ($n = 100$ and $n = 1000$) as a function of $\mathrm{max}\{|\,c\,|,\,|\,p\,|\}$, the Chebyshev distance from the origin. These graphs are produced by carrying out the following procedure a quarter million times: (i) randomly draw $(c,p)$ from $\mathrm{Unif}([-1,1]^2)$; (ii) use $\mathcal{N}((c,p),\bigl( \begin{smallmatrix} \hphantom{-}20 & -10\\ -10 & \hphantom{-}20\end{smallmatrix}\bigr))$ to obtain a sample, using which $(c,p)$ can be estimated and a thousand resampled estimates can be generated; and (iii) use the estimate and resampled estimates to form six confidence intervals (CIs) for $\phi(c,p)$ using a 95\% confidence level. The six CIs are the usual percentile bootstrap CI, the adjusted bootstrap CI (which is the union of the percentile bootstrap CI and the bias-corrected bootstrap CI), the minimalist CIs $\hat{C}^\odot_{n,\mathcal{B}}(\hat{E}_n(\hat{t}^\alpha_n))$ and $\hat{C}^\odot_{n,\mathcal{B}}(\hat{S}_n(\hat{q}^\alpha_n))$, and the uniform CIs $\hat{C}^{\dagger}_{n,K}(\hat{E}_n(\hat{t}^\alpha_n))$ and $\hat{C}^{\dagger}_{n,K}(\hat{S}_n(\hat{q}^\alpha_n))$.
\end{figure}

\begin{figure}
\begin{center}
\caption{Simulation Evidence on Average Coverage Probability versus Magnitude of the RPV}
\label{fig:coverage_adsi}
	\begin{subfigure}[b]{1\textwidth}
         \centering
         \caption{$n = 100$}
         \includegraphics[scale=0.8]{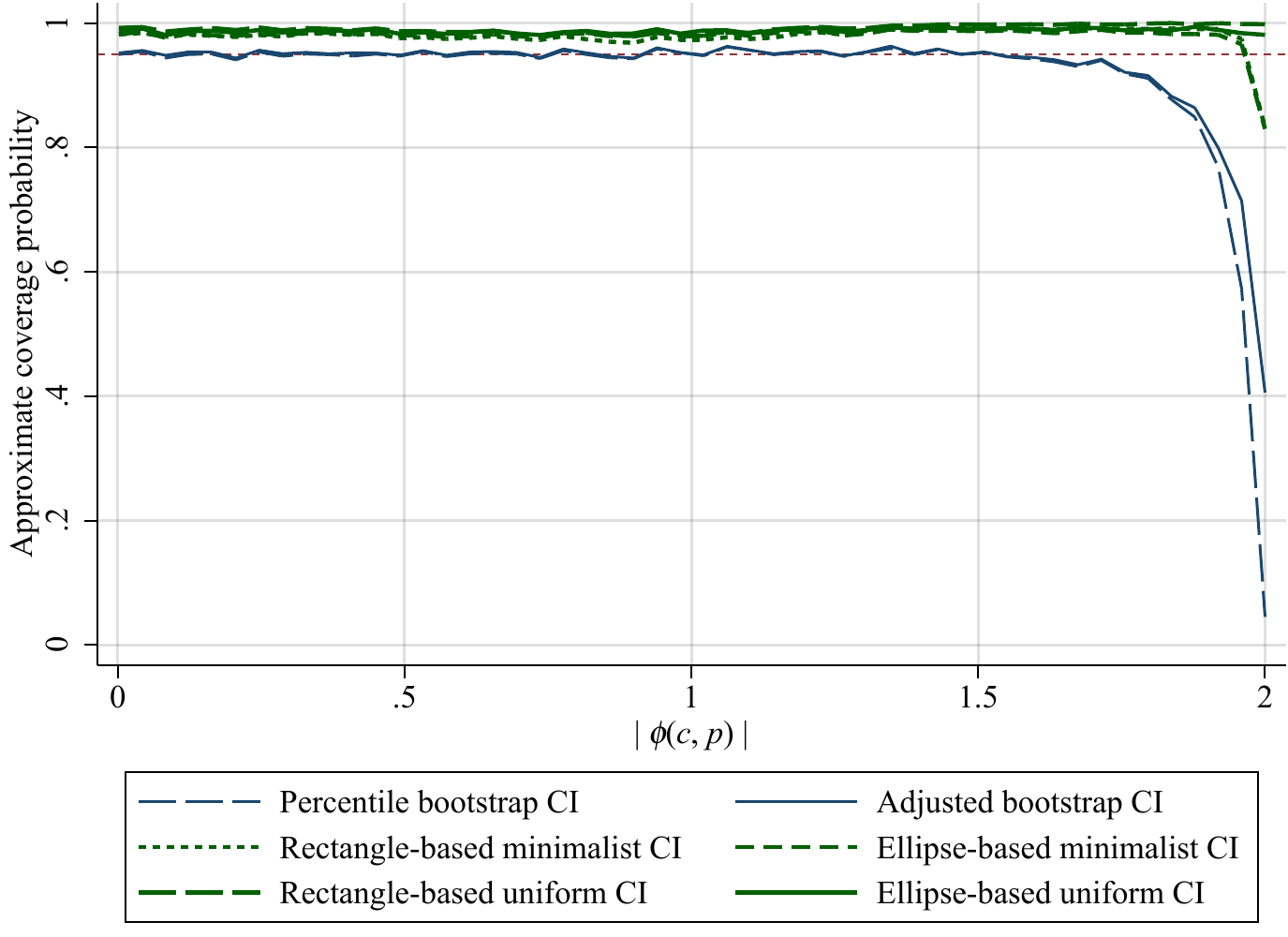}
         \label{fig:coverage_adsi_100}
     \end{subfigure}
    \begin{subfigure}[b]{1\textwidth}
         \centering
         \includegraphics[scale=0.8]{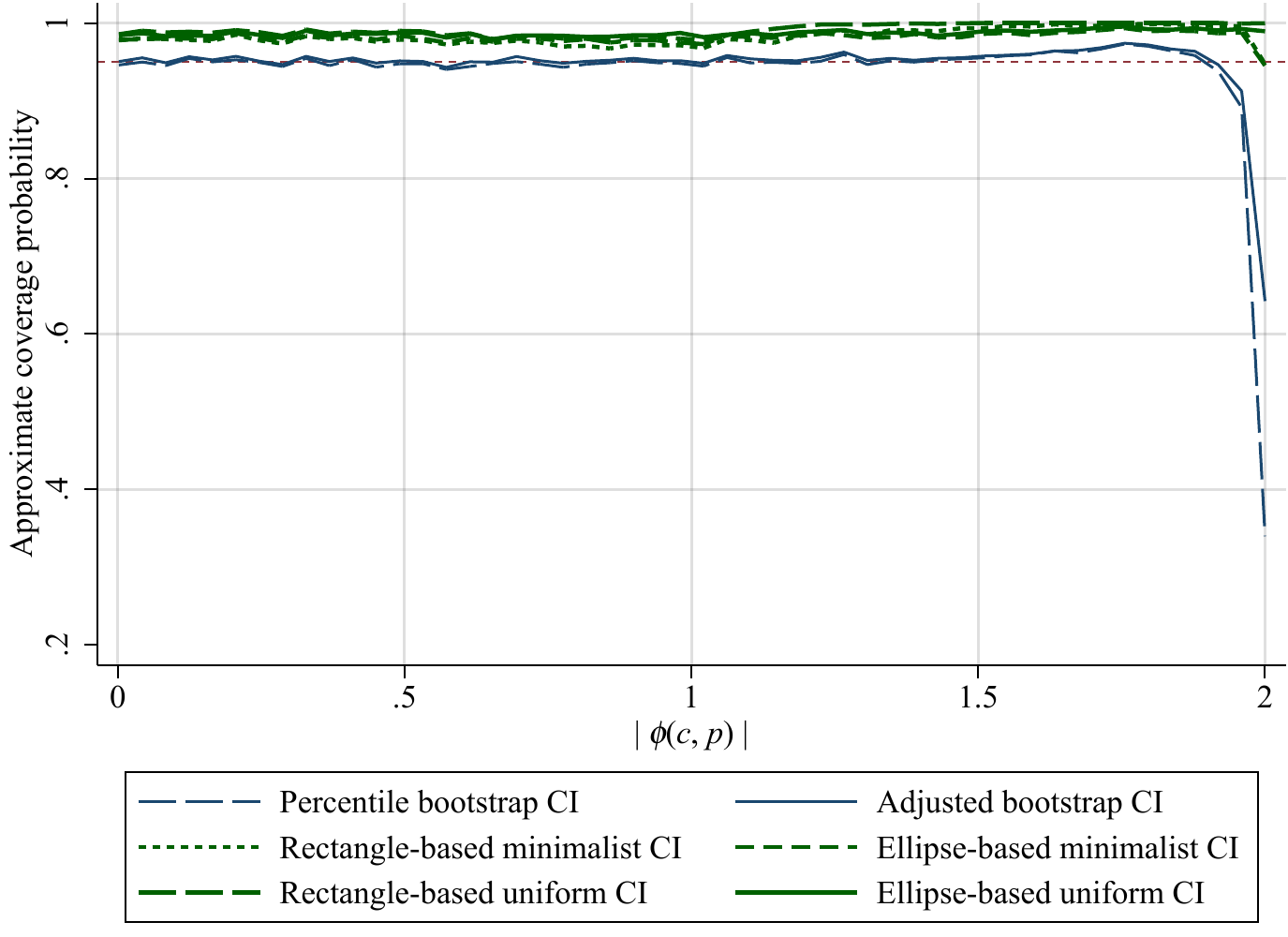}
         \caption{$n = 1000$}
         \label{fig:coverage_adsi_1000}
     \end{subfigure}
\end{center}
\footnotesize \textit{Note}: The above figures plot the approximate average coverage probability (of various confidence intervals for the RPV) for two sample sizes ($n = 100$ and $n = 1000$) as a function of $|\,\phi(c,p)\,|$, the magnitude of the true RPV parameter value. These graphs are produced by carrying out the following procedure a quarter million times: (i) randomly draw $(c,p)$ from $\mathrm{Unif}([-1,1]^2)$; (ii) use $\mathcal{N}((c,p),\bigl( \begin{smallmatrix} \hphantom{-}20 & -10\\ -10 & \hphantom{-}20\end{smallmatrix}\bigr))$ to obtain a sample, using which $(c,p)$ can be estimated and a thousand resampled estimates can be generated; and (iii) use the estimate and resampled estimates to form six confidence intervals (CIs) for $\phi(c,p)$ using a 95\% confidence level. The six CIs are the usual percentile bootstrap CI, the adjusted bootstrap CI (which is the union of the percentile bootstrap CI and the bias-corrected bootstrap CI), the minimalist CIs $\hat{C}^\odot_{n,\mathcal{B}}(\hat{E}_n(\hat{t}^\alpha_n))$ and $\hat{C}^\odot_{n,\mathcal{B}}(\hat{S}_n(\hat{q}^\alpha_n))$, and the uniform CIs $\hat{C}^{\dagger}_{n,K}(\hat{E}_n(\hat{t}^\alpha_n))$ and $\hat{C}^{\dagger}_{n,K}(\hat{S}_n(\hat{q}^\alpha_n))$.
\end{figure}

\begin{figure}
\begin{center}
\caption{Simulation Evidence on Average CI Width versus Distance from Origin}
\label{fig:width_amax}
	\begin{subfigure}[b]{1\textwidth}
         \centering
         \caption{$n = 100$}
         \includegraphics[scale=0.8]{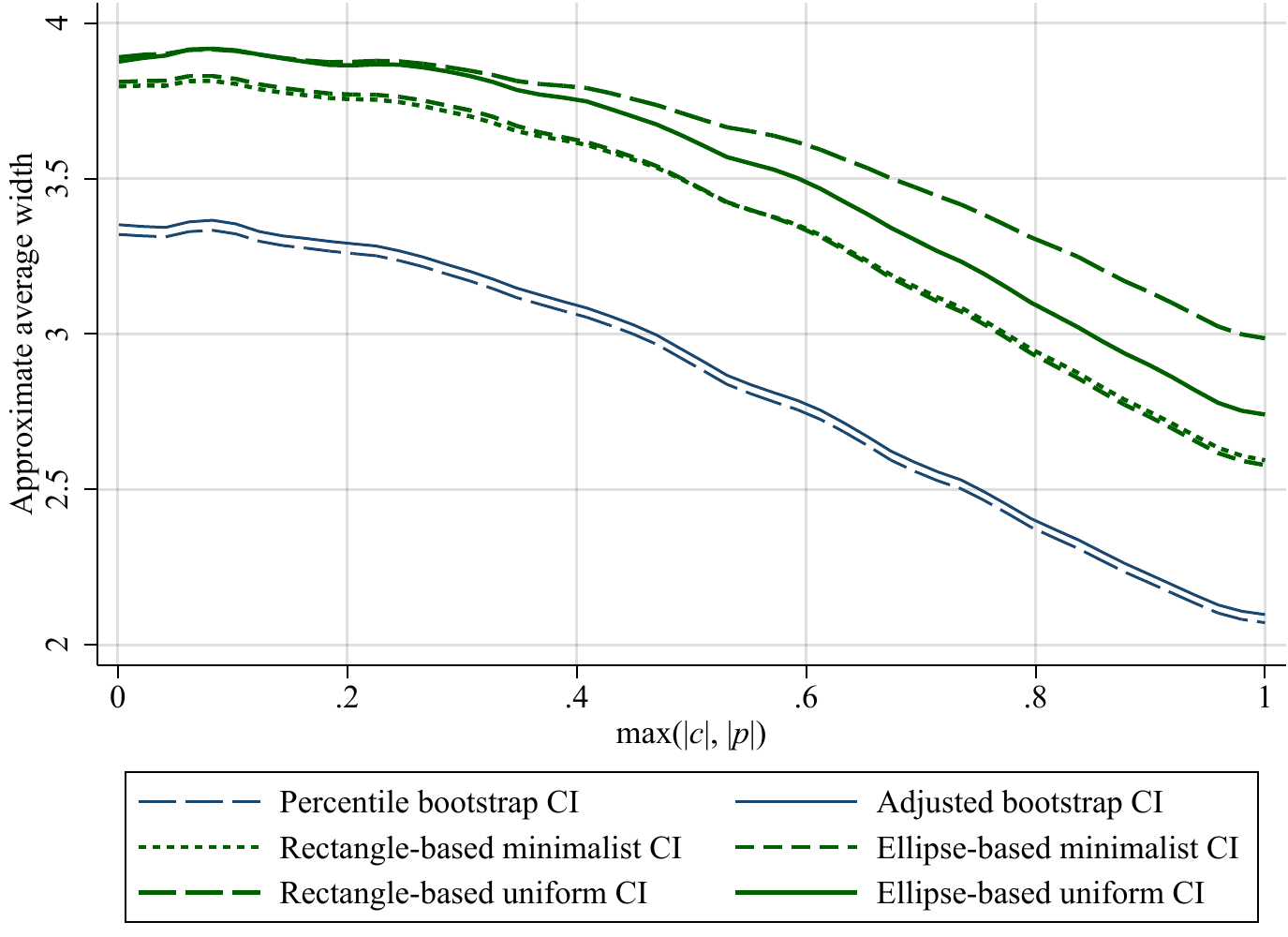}
         \label{fig:width_amax_100}
     \end{subfigure}
    \begin{subfigure}[b]{1\textwidth}
         \centering
         \includegraphics[scale=0.8]{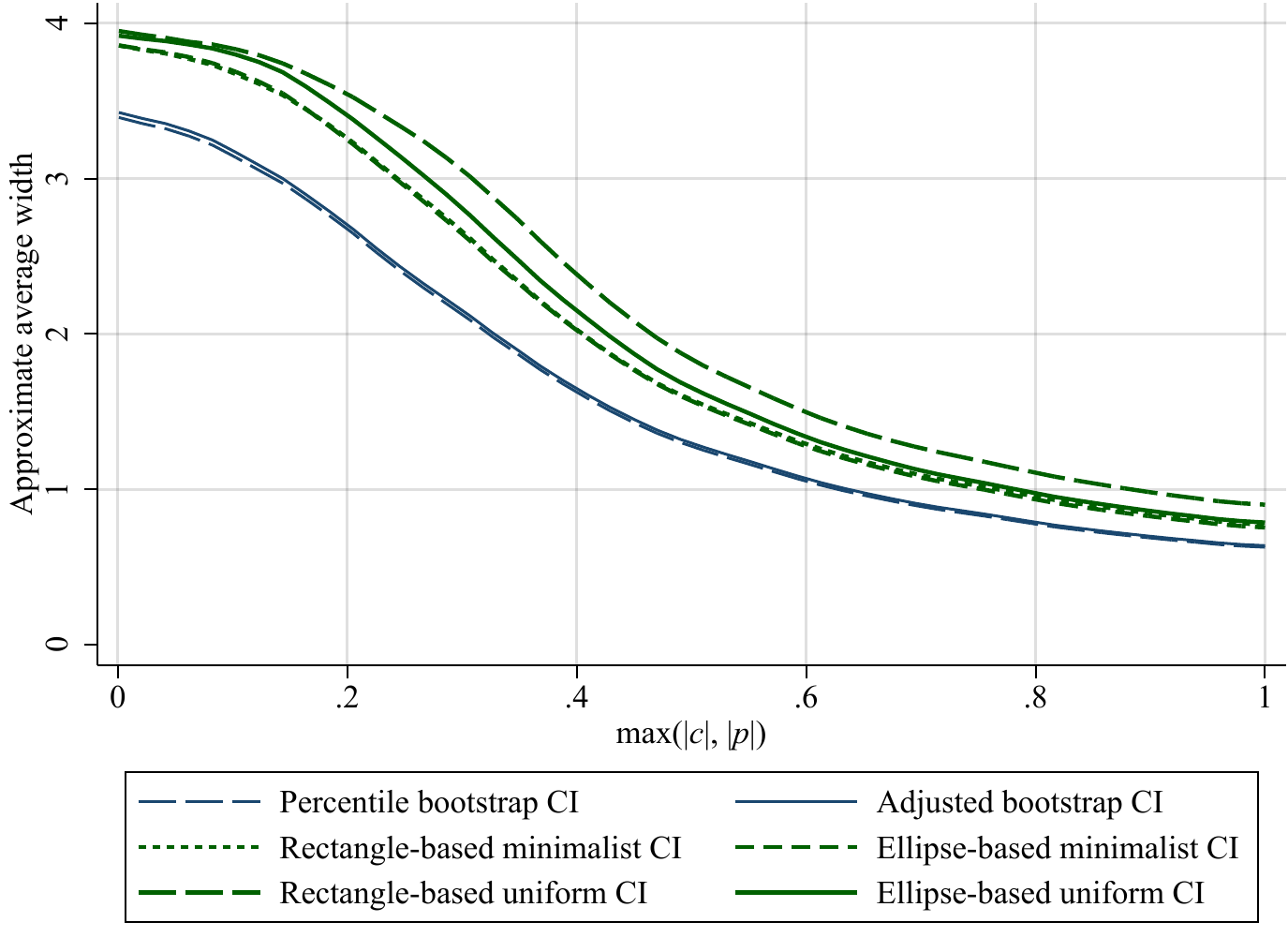}
         \caption{$n = 1000$}
         \label{fig:width_amax_1000}
    \end{subfigure}
\end{center}
\footnotesize \textit{Note}: The above figures plot the approximate average width (of various confidence intervals for the RPV) for two sample sizes ($n = 100$ and $n = 1000$) as a function of $\mathrm{max}\{|\,c\,|,\,|\,p\,|\}$, the Chebyshev distance from the origin. These graphs are produced by carrying out the following procedure a quarter million times: (i) randomly draw $(c,p)$ from $\mathrm{Unif}([-1,1]^2)$; (ii) use $\mathcal{N}((c,p),\bigl( \begin{smallmatrix} \hphantom{-}20 & -10\\ -10 & \hphantom{-}20\end{smallmatrix}\bigr))$ to obtain a sample, using which $(c,p)$ can be estimated and a thousand resampled estimates can be generated; and (iii) use the estimate and resampled estimates to form six confidence intervals (CIs) for $\phi(c,p)$ using a 95\% confidence level. The six CIs are the usual percentile bootstrap CI, the adjusted bootstrap CI (which is the union of the percentile bootstrap CI and the bias-corrected bootstrap CI), the minimalist CIs $\hat{C}^\odot_{n,\mathcal{B}}(\hat{E}_n(\hat{t}^\alpha_n))$ and $\hat{C}^\odot_{n,\mathcal{B}}(\hat{S}_n(\hat{q}^\alpha_n))$, and the uniform CIs $\hat{C}^{\dagger}_{n,K}(\hat{E}_n(\hat{t}^\alpha_n))$ and $\hat{C}^{\dagger}_{n,K}(\hat{S}_n(\hat{q}^\alpha_n))$.
\end{figure}

\begin{figure}
\begin{center}
\caption{Simulation Evidence on Average CI Width versus Magnitude of the RPV}
\label{fig:width_adsi}
	\begin{subfigure}[b]{1\textwidth}
         \centering
         \caption{$n = 100$}
         \includegraphics[scale=0.8]{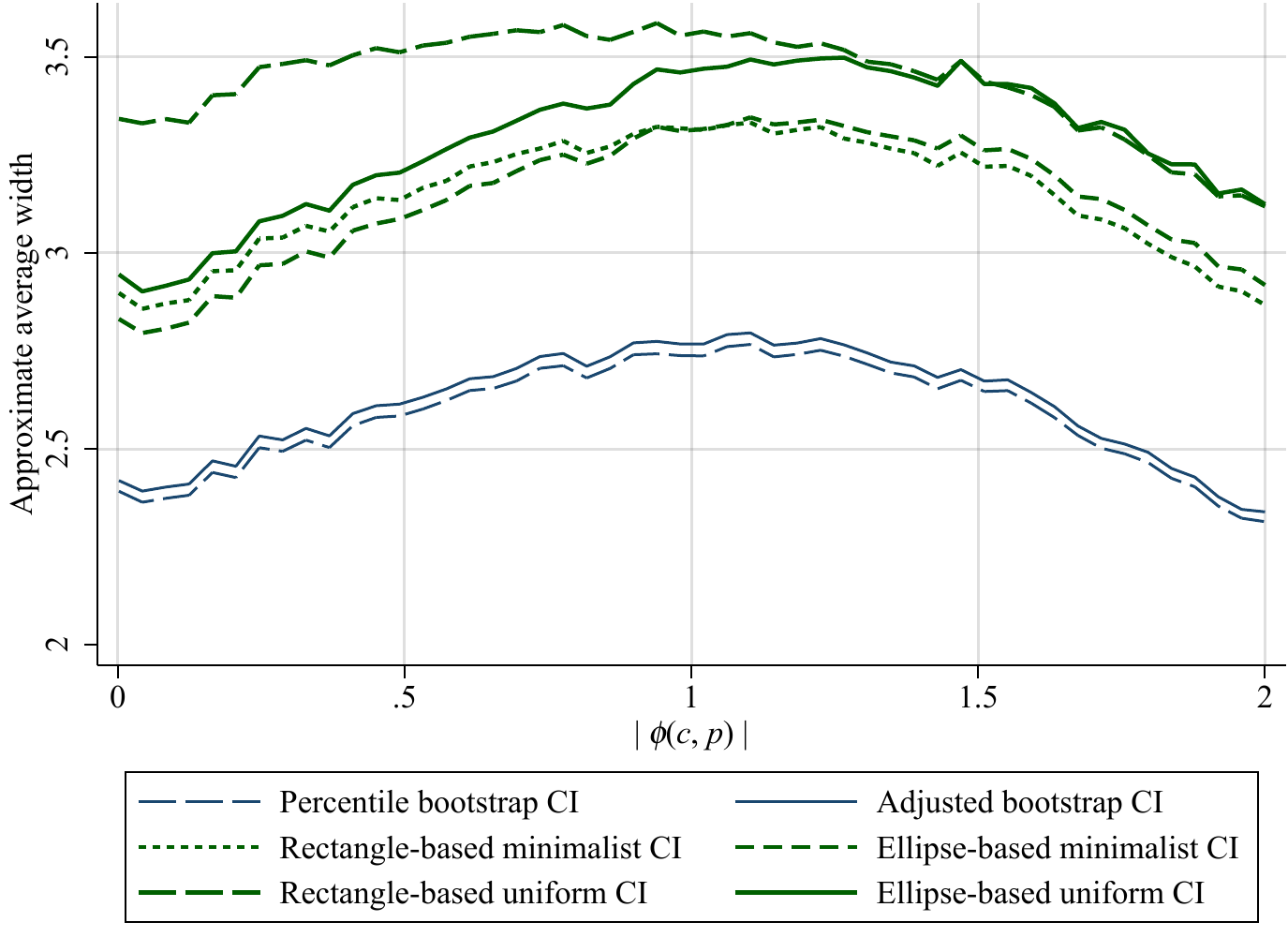}
         \label{fig:width_adsi_100}
     \end{subfigure}
    \begin{subfigure}[b]{1\textwidth}
         \centering
         \includegraphics[scale=0.8]{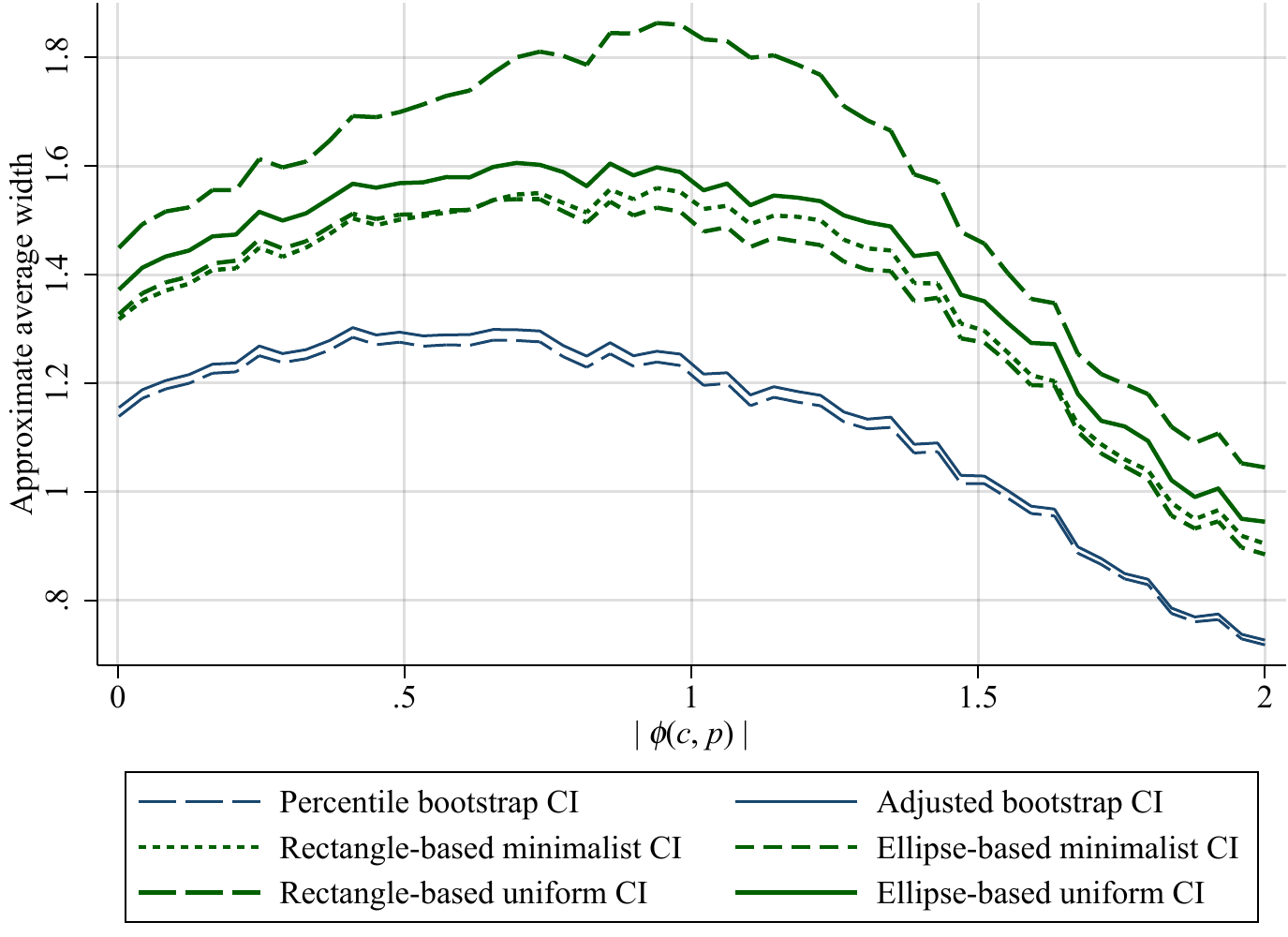}
         \caption{$n = 1000$}
         \label{fig:width_adsi_1000}
     \end{subfigure}
\end{center}
\footnotesize \textit{Note}: The above figures plot the approximate average width (of various confidence intervals for the RPV) for two sample sizes ($n = 100$ and $n = 1000$) as a function of $|\,\phi(c,p)\,|$, the magnitude of the true RPV parameter value. These graphs are produced by carrying out the following procedure a quarter million times: (i) randomly draw $(c,p)$ from $\mathrm{Unif}([-1,1]^2)$; (ii) use $\mathcal{N}((c,p),\bigl( \begin{smallmatrix} \hphantom{-}20 & -10\\ -10 & \hphantom{-}20\end{smallmatrix}\bigr))$ to obtain a sample, using which $(c,p)$ can be estimated and a thousand resampled estimates can be generated; and (iii) use the estimate and resampled estimates to form six confidence intervals (CIs) for $\phi(c,p)$ using a 95\% confidence level. The six CIs are the usual percentile bootstrap CI, the adjusted bootstrap CI (which is the union of the percentile bootstrap CI and the bias-corrected bootstrap CI), the minimalist CIs $\hat{C}^\odot_{n,\mathcal{B}}(\hat{E}_n(\hat{t}^\alpha_n))$ and $\hat{C}^\odot_{n,\mathcal{B}}(\hat{S}_n(\hat{q}^\alpha_n))$, and the uniform CIs $\hat{C}^{\dagger}_{n,K}(\hat{E}_n(\hat{t}^\alpha_n))$ and $\hat{C}^{\dagger}_{n,K}(\hat{S}_n(\hat{q}^\alpha_n))$.
\end{figure}

The above points are further corroborated by Figures \ref{fig:coverage_amax}, \ref{fig:coverage_adsi}, \ref{fig:width_amax}, and \ref{fig:width_adsi}, which present simulation evidence on the practical finite-sample performance of the aforementioned inferential procedures when $\cL = \{1\}$. To produce these graphs, the following procedure is first carried out a quarter million times: (i) randomly draw $(c,p)$, representing the true values of the net fiscal cost and WTP unknown to the analyst, from $\mathrm{Unif}([-1,1]^2)$; (ii) use the data-generating process $\mathcal{N}((c,p),\bigl( \begin{smallmatrix} \hphantom{-}20 & -10\\ -10 & \hphantom{-}20\end{smallmatrix}\bigr))$ to obtain a sample, using which $(c,p)$ can be estimated and a thousand resampled estimates can be generated; and (iii) use the estimate and resampled estimates to form six confidence intervals (CIs) for $\phi(c,p)$ using a $(1 - \alpha) = 0.95$ confidence level. The six CIs are the usual percentile bootstrap CI, the adjusted bootstrap CI (which is the union of the percentile bootstrap CI and the bias-corrected bootstrap CI), the minimalist CIs $\hat{C}^\odot_{n,\mathcal{B}}(\hat{E}_n(\hat{t}^\alpha_n))$ and $\hat{C}^\odot_{n,\mathcal{B}}(\hat{S}_n(\hat{q}^\alpha_n))$, and the uniform CIs $\hat{C}^{\dagger}_{n,K}(\hat{E}_n(\hat{t}^\alpha_n))$ and $\hat{C}^{\dagger}_{n,K}(\hat{S}_n(\hat{q}^\alpha_n))$. Figures \ref{fig:coverage_amax} and \ref{fig:coverage_adsi} graph the average coverage probabilities (up to simulation error) of these intervals as functions of $\mathrm{max}\{|\,c\,|,\,|\,p\,|\}$ and $|\,\phi(c,p)\,|$, respectively, for two different sample sizes: $n = 100$ and $n = 1000$. Figures \ref{fig:width_amax} and \ref{fig:width_adsi} are similar but show the average width (up to simulation error) of each CI rather than the average coverage probability.

Figures \ref{fig:coverage_amax} and \ref{fig:coverage_adsi} show\footnote{In the these figures, depending on whether the approximate average coverage probability is plotted against the true maximum norm $\mathrm{max}\{|\,c\,|,\,|\,p\,|\}$ or against the magnitude of the true RPV $|\,\phi(c,p)\,|$, the horizontal axis can be thought of as representing an index for the class of distributions ($\mathbb{F}$) mentioned in the previous theorems. In Figure \ref{fig:coverage_amax}, each element of $\mathbb{F}$ is a set of distributions with the same $\mathrm{max}\{|\,c\,|,\,|\,p\,|\}$. In Figure \ref{fig:coverage_adsi}, each element of $\mathbb{F}$ is a set of distributions with the same $|\,\phi(c,p)\,|$. In addition, the reported approximate average coverage probabilities are perhaps higher than the true coverage probabilities because I use a specific data generating process. The coverage probabilities may decrease if other data generating processes are also considered, strengthening the case for the proposed methods.} that the coverage of the percentile and bias-corrected bootstrap CIs can be well below the desired 95\% nominal level, especially when $(c,p)$ is near the origin or near the anti-diagonal axis where the lack of differentiability, Lipschitz continuity and convexity can be severe for RPV (see Remark \ref{remark:not_lipschitz_or_convex}), as expected. On the other hand, the uniform CIs have the coverage above the nominal level. The coverage of mininalist CIs can be slightly below 95\% for small sample sizes but then increases to at least the nominal level as the sample size increases. However, as Figures \ref{fig:width_amax} and \ref{fig:width_adsi} show, the uniform and minimalist CIs can be much wider than the conventional CIs. Another important takeaway from these figures is that the minimalist CIs are good alternatives to uniform CIs when the sample size is reasonably large and computational resources are limited.

\clearpage

\section{Reanalysis of a Selected Set of Government Policies}
\label{section:reanalysis}

This section uses the Relative Policy Value (RPV) to reanalyze \citeauthor{hendren2020unified}'s (\citeyear{hendren2020unified}) selected set of more than a hundred ``policy changes over the past half-century in the United States.'' Tables \ref{table:reanalysis_single_policies} and \ref{table:reanalysis_policy_categories} in \ref{appendix:tables} present a complete list of RPV-based counterparts of the main MVPF estimates and confidence intervals reported in Table II of \cite{hendren2020unified} for both individual policies and categories of policies. \cite{hendren2020unified} also report additional analyses and robustness checks, but these are not reanalyzed in this paper to avoid losing focus on the main comparisons between the MVPF and the RPV. In these tables, I do not separately report uniformly valid confidence intervals for the ``fixed'' MVPF (or the other comparative welfare indices) because of the sufficiency of RPV (formalized in Theorems \ref{theorem:rpv_mvpf_conversion} and \ref{theorem:rpv_Lp_conversion}).

Inferences based on RPV can substantially differ from those based on MVPF in several cases. For example, Table \ref{table:mvpf_infinity} shows examples of policies for which the 95\% bias-corrected bootstrap confidence interval (Efron CI) for MVPF is a degenerate singleton $\{\infty\}$ but the confidence intervals (CIs) for RPV include values below 1. In other words, the inferences reported by \cite{hendren2020unified} indicate that the policies listed in Table \ref{table:mvpf_infinity} are Pareto superior (using the Efron CIs), but even the minimalist CI for the RPV includes values below 1 (the RPV threshold for Pareto superiority), meaning that the data do not warrant strong conclusions about the Pareto superiority of these policies at the 95\% confidence level.

One of the policies that \cite{hendren2020unified} use prominently to explain the construction of the MVPF is the FIU GPA (Florida International University Admissions at GPA Threshold) policy. Its MVPF-based Efron CI is $\{\infty\}$, implying that the policy more than pays for itself, but the uniform and minimalist CIs for the RPV of the FIU GPA policy include values below $-1$ (indicating Pareto inferiority) and also values above $1$ (indicating Pareto superiority), and so the MVPF-based conclusion is misleading (at the 95\% confidence level). Note that Table \ref{table:mvpf_infinity} shows two confidence intervals (CIs) for the RPV: a uniformly valid CI, which could be potentially conservative, but also a minimalist CI, which is a less conservative subset of the uniform CI. Thus, even if we do not employ statistical methods that are a bit conservative, there is reason to be cautious about the strong MVPF-based conclusions of Pareto superiority of the policies listed in Table \ref{table:mvpf_infinity}.

Table \ref{table:mvpf_above_one} shows policies whose MVPF-based Efron CIs lie above 1 (indicating that the policies lie above the break-even line $p = c$). However, the 95\% CIs for the RPV include values below 0 (i.e., below the break-even line), even when the minimalist CIs (that are less conservative than the uniform CIs) are used. Similarly, Table \ref{table:mvpf_above_zero} shows policies for which the MVPF-based Efron CIs lie above 0 (ruling out Pareto inferiority) but the CIs for RPV include values below $-1$ (that do not rule out Pareto inferiority). These empirical examples show the practical importance of using the statistical methods proposed in this paper.

\clearpage

\begin{table}[ht]
\setstretch{1.5}
\caption{Cases Where MVPF CIs Indicate Pareto Superiority While RPV CIs Express Uncertainty}
\label{table:mvpf_infinity}
\vspace{-5mm}
    \begin{center} \footnotesize
    \begin{tabular}{l|ccc|cc}
    \hline\hline
\textit{Program} & \textit{RPV} & \textit{Uniform CI} & \textit{Minimalist CI} & \textit{MVPF} & \textit{Efron CI} \\\hline
CPC Preschool & $ 1.16 $ & $ [ 0.36 , 2.00 ] $ & $ [ 0.57 , 1.99 ] $ & $ \infty $ & $ [ \infty , \infty ]$ \\
FIU GPA & $ 1.22 $ & $ [ -2.00 , 1.90 ] $ & $ [ -1.37 , 1.27 ] $ & $ \infty $ & $ [ \infty , \infty ] $ \\
MC Pregnant \& Infants & $ 1.15 $ & $ [ -2.00 , 1.20 ] $ & $ [ 0.77 , 1.20 ] $ & $ \infty $ & $ [ \infty , \infty ] $ \\
\hline\hline
    \end{tabular}
    \end{center}
\end{table}
\begin{table}[ht]
\setstretch{1.5}
\caption{Cases Where MVPF CIs Lie Above Break-Even Point While RPV CIs Show Uncertainty}
\label{table:mvpf_above_one}
\vspace{-5mm}
    \begin{center} \footnotesize
    \begin{tabular}{l|ccc|cc}
    \hline\hline
\textit{Program} & \textit{RPV} & \textit{Uniform CI} & \textit{Minimalist CI} & \textit{MVPF} & \textit{Efron CI} \\\hline
CC Texas & $ 1.00 $ & $ [ -1.68 , 1.12 ] $ & $ [ -1.24 , 1.12 ] $ & $ 349.51 $ & $ [ 1.61 , \infty ] $ \\
DC Grant & $ 0.96 $ & $ [ -0.52 , 1.09 ] $ & $ [ -0.45 , 1.08 ] $ & $ 22.98 $ & $ [ 1.11 , \infty ] $ \\
Florida Grant & $ 0.87 $ & $ [ -1.93 , 1.05 ] $ & $ [ -0.70 , 1.04 ] $ & $ 7.42 $ & $ [ 1.09 , \infty ] $ \\
College Spend & $ 0.75 $ & $ [ -0.56 , 0.95 ] $ & $ [ -0.24 , 0.95 ] $ & $ 4.00 $ & $ [ 1.25 , 20.44 ] $ \\
WI Scholarship & $ 0.30 $ & $ [ -0.30 , 0.57 ] $ & $ [ -0.13 , 0.56 ] $ & $ 1.43 $ & $ [ 1.00 , 2.32 ] $ \\
\hline\hline
    \end{tabular}
    \end{center}
\end{table}
\begin{table}[ht]
\setstretch{1.5}
\caption{Cases Where MVPF CIs Rule Out Pareto Inferiority While RPV CIs Show Uncertainty}
\label{table:mvpf_above_zero}
\vspace{-5mm}
    \begin{center} \footnotesize
    \begin{tabular}{l|ccc|cc}
    \hline\hline
\textit{Program} & \textit{RPV} & \textit{Uniform CI} & \textit{Minimalist CI} & \textit{MVPF} & \textit{Efron CI} \\\hline
K12 Spend Mich. & $ -0.35 $ & $ [ -1.40 , 0.54 ] $ & $ [ -1.12 , 0.52 ] $ & $ 0.65 $ & $ [ 0.05 , 2.19 ] $ \\
Free FAFSA (Dep) & $ 0.75 $ & $ [ -2.00 , 2.00 ] $ & $ [ -1.68 , 0.92 ] $ & $ 4.03 $ & $ [ 0.65 , 10.75 ] $ \\
Soc Sec College & $ 0.79 $ & $ [ -2.00 , 1.00 ] $ & $ [ -1.72 , 0.99 ] $ & $ 4.86 $ & $ [ 0.98 , 52.39 ] $ \\
\hline\hline
    \end{tabular}
    \end{center}
\end{table}
    \vspace{-1mm}
\setstretch{1}\noindent \footnotesize
\textit{Note}: For each of the above policies, the above tables report the following: the Relative Policy Value (RPV) and its ellipse-based uniform and minimalist confidence intervals (CIs) as well as the marginal value of public funds (MVPF) and its modified bias-corrected bootstrap confidence interval (Efron CI). All the CIs use a confidence level of 95\%. In Table \ref{table:mvpf_infinity}, the Efron CI is degenerate singleton $\{\infty\}$ but the CIs for RPV include values below 1. In Table \ref{table:mvpf_above_one}, the Efron CI for MVPF lies above 1, but a 95\% CIs for RPV include 0. In Table \ref{table:mvpf_above_zero}, the Efron CI lies above 0 but the CIs for RPV include values below $-1$. The abbreviations used in the first column of this table for the names of programs are the same as those used in Table II of \cite{hendren2020unified}. See their Table I for the full forms and descriptions of these abbreviations. ``CPC Preschool'' refers to the ``Chicago Child-Parent Centers Preschool Program.'' ``FIU GPA'' refers to the ``Florida International University Admissions at GPA Threshold'' policy. ``MC Pregnant \& Infants'' refers to the ``Medicaid Expansions to Pregnant Women \& Infants.'' ``CC Texas'' refers to ``Community College Tuition Changes in Texas.'' ``DC Grant'' refers to ``District of Columbia Tuition Assistance Grant Program.'' ``Florida Grant'' refers to ``Florida Student Access Grant.'' ``College Spend'' refers to ``Spending at Colleges from State Appropriations.'' ``WI Scholarship'' refers to ``Wisconsin Scholar Grant to Low-Income College Students.'' ``K12 Spend Mich.'' refers to ``K--12 School Spending in Michigan.'' ``Free FAFSA (Dep)'' refers to ``Free Application for Federal Student Aid, Dependent Year Impact.'' ``Soc Sec College'' refers to ``Social Security Student Benefit Program.''

\onehalfspacing \normalsize

\clearpage

Of course, if we were to take into account multiple hypothesis testing considerations in the previous discussion, we would obtain even more conservative inferences. However, the purpose of this section is to show that there is a lot of statistical uncertainty in the welfare of many policies, even if we use an inference procedure (such as the minimalist CI) that is barely valid in accounting for the statistical irregularities of comparative welfare measures.

Tables \ref{table:mvpf_infinity}, \ref{table:mvpf_above_one}, and \ref{table:mvpf_above_zero} list policies where there is a conflict between RPV- and MVPF-based inferences, but of course the aforementioned policies are only a subset of a broad set of public policies analyzed by \cite{hendren2020unified}. For each of those policies, Table \ref{table:reanalysis_single_policies} in \ref{appendix:tables} reports the estimates and inferences on the Relative Policy Value (RPV), the Marginal Value of Public Funds (MVPF), and the Marginal Social Surplus (MSS) plus one, which equals the benefit-to-cost ratio (BCR) in their case (i.e., BCR $=$ MSS $+$ 1) because of the way $p$ and $c$ are operationalized, as discussed in Section \ref{sec:motivation}. (I report the BCR $=$ MSS $+$ 1 to enable easy cross-checks between this paper and \cite{hendren2020unified}, who report the BCR and not the MSS.) Because there is a lot of statistical uncertainty in the data for most of the programs studied by \cite{hendren2020unified}, both the RPV-based and MVPF-based inferences lead to inconclusive results in a large number of cases, as shown in the unshaded rows of Table \ref{table:reanalysis_single_policies}.

There are also several policies for which RPV-based and MVPF-based inferences result in similar conclusions. In Table \ref{table:reanalysis_single_policies} of \ref{appendix:tables}, the rows shaded in teal green include programs for which RPV-based 95\% CIs lie above 0 and the MVPF-based Efron CIs lie above 1. For example, all three cost-benefit measures (RPV, MVPF, and MSS) indicate that the following policies generated positive welfare: the Head Start program, the 1993 K--12 school finance reform, the Cal Grant program (based on GPA threshold), the Oregon health insurance program (for single adults), the special supplemental nutrition program for women, infants, and children (WIC), the top tax rate reductions in the Tax Reform Act of 1986,\footnote{As \cite{hendren2020unified} explain, the result on the 1986 top tax rate reductions may be due to Laffer effects (i.e., the previous top tax rate being on the ``wrong side of the Laffer curve'') and may not be relevant today.} and the top tax rate increase in the Omnibus Budget Reconciliation Act of 1993. There are also several policies (in the maroon-shaded rows of Table \ref{table:reanalysis_single_policies}) with a negative RPV and an MVPF below one: programs providing job training, disability insurance, supplemental security income, unemployment insurance, housing vouchers, application help and information for supplemental nutrition assistance, term limits on aid to families with dependent children, and the Alaska permanent fund dividend.

\begin{figure}
\begin{center}
\caption{Total and Joint Policy Values (TPVs and JPVs) for Policy Categories}
\label{fig:category_tsi_dsi}
\includegraphics[scale=1]{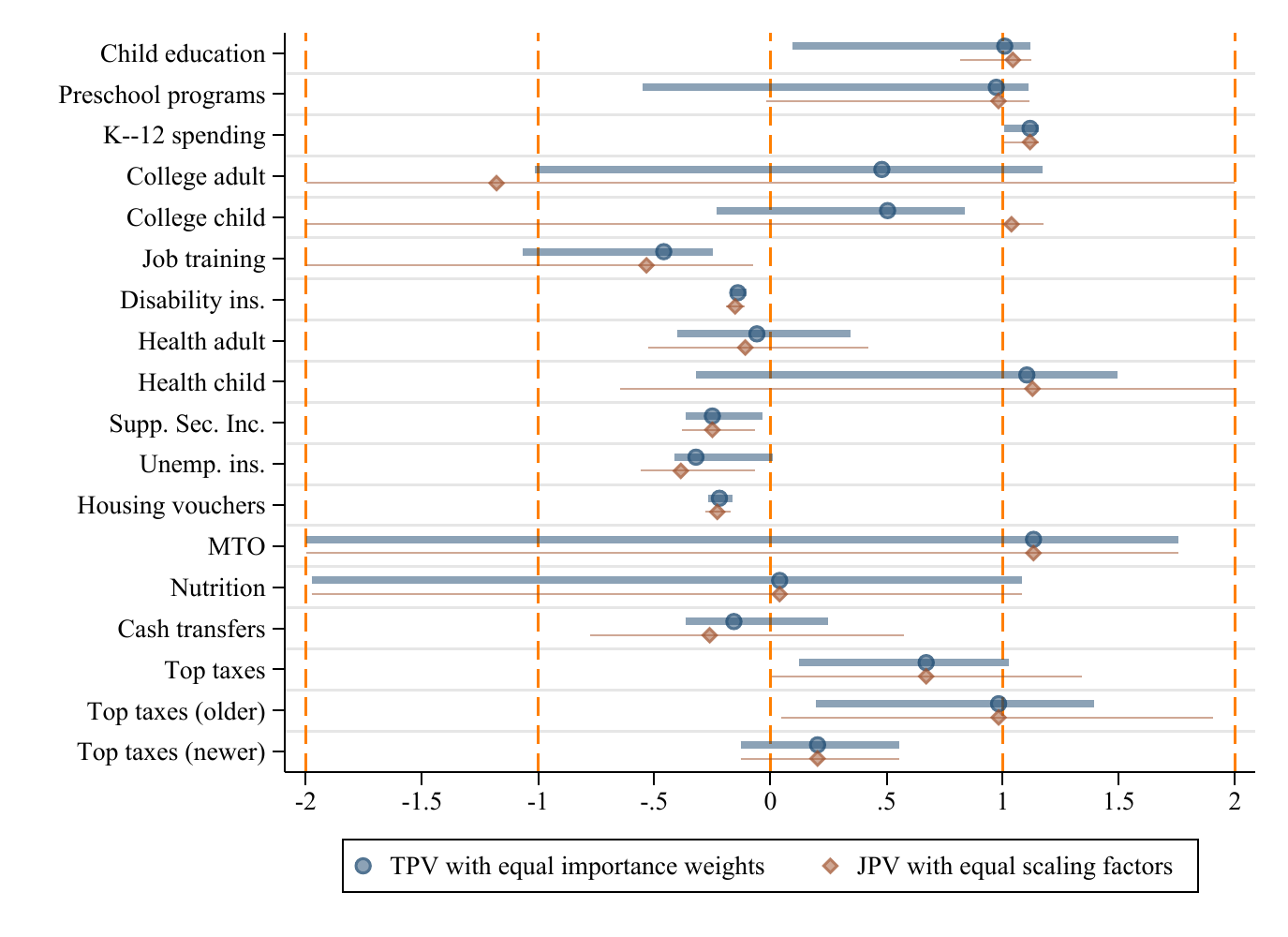}
\end{center}
\scriptsize\vspace{-5mm} \textit{Note}: The above figure graphically represents the TPVs and JPVs of policy categories as well as their 95\% minimalist confidence intervals, which are reported in Table \ref{table:reanalysis_policy_categories} of \ref{appendix:tables}. Each blue circle represents the Total Policy Value (TPV) of a policy category (using equal importance weights). In addition, each maroon diamond symbol represents the Joint Policy Value (JPV) of the policy category (using equal scaling factors).
\end{figure}

These conclusions are also reflected in Table \ref{table:reanalysis_policy_categories} presented in \ref{appendix:tables}, which shows the Total Policy Value (TPV) and the Joint Policy Value (JPV) for each policy category (using equal importance weights or scaling factors), along with the BCR and MVPF of the ``category average'' reported by \cite{hendren2020unified}. Figure \ref{fig:category_tsi_dsi} graphically displays the TPV and JPV for each category along with their minimalist confidence intervals. I do not bother with the full uniform confidence intervals for two reasons: the minimalist confidence intervals for many of the policy categories are already too wide; and the main point I wish to make is that some values of the TPVs and JPVs for a few policy categories cannot be rejected (as plausible values of the parameters), even if we use a minimalist inference procedure that is not too conservative.

The TPV and JPV differ substantially for the two college expenditure categories (namely, ``College Child'' and ``College Adult'' whose beneficiaries are 20 and 40 years old on average, respectively). According to the results of \cite{hendren2020unified}, the ``Health Child'' and ``College Child'' categories, representing child health insurance and college expenditure policies, have infinite estimate MVPFs and associated Efron CIs above 1, i.e., above the break-even line (see Table \ref{table:reanalysis_policy_categories} in \ref{appendix:tables}). However, such strong conclusions are not justified, since zero is included in the confidence intervals for the TPV and JPV for these categories (see Table \ref{table:reanalysis_policy_categories} in \ref{appendix:tables}). While the CIs for the TPV and JPV are above zero for the ``Child Education'' program category, this is largely a result of the inclusion of ``K12 Spend'' (i.e., the 1993 K--12 school finance reform), whose RPV and the associated confidence interval are above one. Finally, another clear advantage of the Total Policy Value is that the TPVs of various policy categories can themselves be averaged to get an even more aggregate TPV, which is not possible within an MVPF-based framework.

\clearpage

\section{Conclusion}
\label{section:conclusion}

There is a long tradition of axiomatization in decision theory and the economics of inequality, poverty, social choice, and consumer behavior \citep[see, e.g., ][]{marschak1950rational, debreu1954representation, vonneumann1944theory, anscombe1963definition, arrow1950difficulty, arrow2012social, sen1973economic, afriat2014index, ray2022measuring, ramsey2016truth, samuelson1938note, arrow1959rational}. ``Axiomatizations of general-purpose conceptual frameworks'' can be practically very useful; they ``can be powerful
rhetorical devices'' and ``can prove useful also for descriptive purposes'' \citep{gilboa2019axiomatizations}. In this paper, I develop an axiomatic framework for comparative welfare analysis, resulting in a unique econo-metric that I call the Relative Policy Value (RPV). I show how the RPV can be used to conduct general forms of cost--benefit analysis from both comparative and absolute perspectives (or a hybrid combination thereof). I also formalize two additional concepts called the Joint Policy Value (JPV) and the Total Policy Value (TPV), which incorporate the RPV, to conduct flexible forms of welfare aggregation across different policies or across population subgroups.

To better inform risk-averse policymakers, I also provide computationally convenient methods for making uniformly valid statistical inferences on the RPVs of policies as well as the JPVs and TPVs of policy collections. My empirical reanalysis shows that there is substantial economic and statistical uncertainty about welfare of some policies that were previously reported to have very high or even ``precisely estimated infinite'' MVPF values. For example, \cite{hendren2020unified} report that the confidence interval for the MVPF of the ``Medicaid Expansions to Pregnant Women \& Infants'' (or the ``MC Pregnant \& Infants'' policy) is a degenerate singleton $\{\infty\}$. Such results imply that ``expanding Medicaid coverage to pregnant women and their children, for instance, had a precisely estimated infinite payoff'' \citep{matthews2019government} and that the ``[child health insurance and college expenditure] policies pay for themselves'' \citep{hendren2020unified}. However, the RPV-based evidence in this paper weakens such conclusions. Thus, different types of policy preferences and econometric frameworks can lead to very different conclusive or inconclusive empirical statements about policies, even when the same data and the same resamples are utilized. However, there is conclusive evidence based on both the RPV and MVPF that a K--12 school finance reform, studied by \cite{jackson2016effects}, more than pays for itself.

Finally, this paper demonstrates that the recent replication policies adopted by several economics journals have value beyond replication: reuse of existing data for creation of new knowledge. \cite{hendren2020unified} took great care in ensuring that their code and data, which resulted from their big-budget project, are both fully replicable and easily accessible (i.e., clearly documented and publicly avaialable). Therefore, \cite{hendren2020unified} have actively reduced barriers to exploration and reuse of their data, effectively enabling new policy-relevant econometric perspectives at no additional data-related cost!

\clearpage

\setstretch{1}
\bibliographystyle{chicago}
\bibliography{references}

\clearpage

\appendix
\renewcommand{\thesection}{Appendix \Alph{section}}

\onehalfspacing

\section{}
\label{appendix:proofs}

\vspace{5mm}

\noindent\textbf{Lemma\,\ref{lemma:rpv_axiom}.}\textit{ The Relative Policy Value (RPV) satisfies the welfare-symmetry axioms.}

\begin{proof}
See the proofs of Lemmas \ref{lemma:rpv_axiom}.1, \ref{lemma:rpv_axiom}.2, and \ref{lemma:rpv_axiom}.3.
\end{proof}

\noindent\textbf{Lemma\,\ref{lemma:rpv_axiom}.1.}\textit{ The Relative Policy Value (RPV) $\phi$ satisfies Axiom \ref{axiom:1}.}

\begin{proof}
Let $(c, p) \in \R$. If $c = p \neq 0$, $\phi(c,p) = \phi(p, p) = (p - p)/|\,p\,| = 0$. Also, if $c = p = 0$, $\phi(0,0) = 0$, and so $c = p \implies \phi(c,p) = 0$. Suppose now that $\phi(c, p) = 0$. If $(c, p) = (0, 0)$, then $c = p$ trivially. Also, if $(c, p) \neq (0, 0)$, then $\phi(c, p) = 0 \implies (p - c)/\mathrm{max}\{|\,p\,|,|\,c\,|\} =~0 \implies p - c = 0 \implies c = p$. Thus, $\phi(c, p) = 0 \iff c = p$, proving that $\phi$ satisfies the first part of Axiom \ref{axiom:1}. To prove the second part, suppose $c > 0$ and $c \geq p \geq -c$. Then, $|\,p\,| \leq |\,c\,| = c \implies  \mathrm{max}\{|\,p\,|,|\,c\,|\} = |\,c\,| = c$. Then, $\phi(c, p) = (p - c)/\mathrm{max}\{|\,p\,|,|\,c\,|\}  = (p - c)/c = p/c - 1 = m(c,p) - 1$. Therefore, $\phi$ satisfies Axiom \ref{axiom:1}.
\end{proof}

\noindent\textbf{Lemma\,\ref{lemma:rpv_axiom}.2.}\textit{ The Relative Policy Value (RPV) $\phi$ satisfies Axiom \ref{axiom:2}.}

\begin{proof}
Let $(c, p) \in \R^2 \setminus \{(0,0)\}$. Then, $\mathrm{max}\{\left|\,p \, \right|,\left|\,c\, \right|\}\cdot[\phi(c,p) + \phi(p,c)] = (p - c) + (c - p) = 0$, implying that $\phi(c,p) + \phi(p,c) = 0\,/\,\mathrm{max}\{\left|\,p \, \right|,\left|\,c\, \right|\} = 0$. In addition, $\phi(c,p) + \phi(p,c) = 0$ also trivially holds when $(c,p) = (0,0)$. Therefore, $\phi$ satisfies Axiom \ref{axiom:2}.
\end{proof}

\noindent\textbf{Lemma\,\ref{lemma:rpv_axiom}.3.}\textit{ The Relative Policy Value (RPV) $\phi$ satisfies Axiom \ref{axiom:3}.}

\begin{proof}
Let $(c, p) \in \R^2 \setminus \{(0,0)\}$. Then, $\mathrm{max}\{\left|\,-p \, \right|,\left|\,-c\, \right|\} = \mathrm{max}\{\left|\,p \, \right|,\left|\,c\, \right|\}$, and so it follows that $\mathrm{max}\{\left|\,p \, \right|,\left|\,c\, \right|\}\cdot[\phi(c,p) + \phi(-c,-p)] = (p - c) + (-p -(-c)) = 0$, and so it follows that $\phi(c,p) + \phi(-c,-p) = 0$, which also holds at the origin.  Therefore, $\phi$ satisfies Axiom \ref{axiom:3}.
\end{proof}

\noindent\textbf{Theorem\,\ref{theorem:rpv}.}\textit{ The only function obeying the welfare-symmetry axioms is the Relative Policy Value.}

\begin{proof}
Axiom \ref{axiom:1} and Lemma \ref{lemma:rpv_axiom}.1 together provide that $\varphi(c, p) = \phi(c,p)$ for all $(c,p) \in \mathcal{H}_\triangleleft$, where $\mathcal{H}_\triangleleft \equiv \{(c',p'): c' > 0,\, c' \geq p' \geq -c'\} \cup \{(0,0)\}$. Now let $(c'', p'') \in \mathcal{H} \setminus \mathcal{H}_\triangleleft$, where $\mathcal{H} = \{(c',p') \in \R^2: p' \geq -c'\}$. It then follows that $(p'', c'') \in \mathcal{H}$ such that $p'' > 0$ and $p'' > c'' \geq -p''$. Thus, $(p'', c'') \in \mathcal{H}_\triangleleft$, and so Axiom \ref{axiom:1} implies that $\varphi(p'',c'') = \phi(p'', c'')$. By Axiom \ref{axiom:2}, $\varphi(c'',p'') + \varphi(p'',c'') = 0$, and so $\varphi(c'',p'') = -\phi(p'',c'') = \phi(c'',p'')$ by Lemma \ref{lemma:rpv_axiom}.2. Thus, $\varphi(c,p) = \phi(c,p)$ for all $(c,p) \in \mathcal{H}$. Now suppose $(c''',p''') \in \R^2 \setminus \mathcal{H}$. Then, $(-c''',-p''') \in \mathcal{H}$, so $\varphi(-c''',-p''') = \phi(-c''',-p''') = -\phi(c''',p''')$ by Lemma \ref{lemma:rpv_axiom}.3. By Axiom \ref{axiom:3}, $\varphi(c''',p''') = -\varphi(-c''',-p''') =$ $ -[-\phi(c''',p''')] = \phi(c''',p''')$. Thus, $\varphi(c,p) = \phi(c,p)$ for all $(c,p) \in \R^2 \setminus \mathcal{H}$. Hence, $\varphi(c,p) = \phi(c,p)$ for all $(c,p) \in \R^2$. Therefore, $\varphi \equiv \phi$.
\end{proof}

\clearpage

\noindent\textbf{Remark\,\ref{remark:homogeneous_degree_zero}.}\textit{ The Relative Policy Value (RPV) is homogeneous of degree zero.}

\begin{proof}
Let $\lambda \in \R$. Then, $\phi(\lambda(0,0)) = \phi(0,0)  = 0 = \sgn(\lambda) \cdot 0 = \sgn(\lambda)\,\phi(0,0)$ trivially. Suppose $(c,p) \in \R^2 \setminus \{(0,0)\}$. Then, note that $\phi(\lambda c, \lambda p) = \lambda(p - c)/\,\mathrm{max}\{\left|\,\lambda p \, \right|,\left|\,\lambda c\, \right|\} = (\lambda/|\,\lambda\,|)\cdot (p - c)/\,\mathrm{max}\{\left|\,p \, \right|,\left|\, c\, \right|\} = \sgn(\lambda)\,\phi(c,p)$ if $\lambda \neq 0$. If $\lambda = 0$, $\phi(\lambda c, \lambda p) = \phi(0, 0) = 0 = 0 \cdot \phi(c,p) = \sgn(\lambda)\,\phi(c,p)$. Therefore, $\phi(\lambda c, \lambda p) = \sgn(\lambda)\,\phi(c,p)$ for all $\lambda \in \R$. This is more general than degree-zero homogeneity, which states that $\phi(\lambda c, \lambda p) = \phi(c,p)$ for all $\lambda > 0$.
\end{proof}

\noindent\textbf{Theorem\,\ref{theorem:rpv_Linfinity}.}\textit{ The Relative Policy Value $\phi(c,p) = (p-c)/||(c,p)||_\infty = (p-c)/\mathrm{max}\{|\,p\,|, |\,c\,|\}$ is the only function satisfying both the welfare-symmetry axioms and surplus-normalization axioms.}

\begin{proof}
Let $\lambda_{c,\,p} \equiv ||(c,p)||_\infty = \mathrm{max}\{|\,p\,|,|\,c\,|\}$. Then, $(c/\lambda_{c,\,p}, \,p/\lambda_{c,\,p}) \in \mathcal{S}$ if $(c,p) \neq (0,0)$. For any $(c,p) \in \R^2 \setminus \{(0,0)\}$, Axiom \ref{axiom:6} implies that $\varphi(c,p) = \varphi(\lambda_{c,\,p}\,c/\lambda_{c,\,p}, \,\lambda_{c,\,p} \,p/\lambda_{c,\,p}) = \varphi(c/\lambda_{c,\,p}, \,p/\lambda_{c,\,p})$, which equals $p/\lambda_{c,\,p} - c/\lambda_{c,\,p}$ by Axiom \ref{axiom:5}. Thus, for any point that is not the origin, i.e., for $(c,p) \in \R^2 \setminus \{(0,0)\}$, $\varphi(c,p) = (p-c)/\lambda_{c,\,p} = (p - c)/\mathrm{max}\{|\,p\,|,|\,c\,|\}$. Finally, by Axiom \ref{axiom:4}, $\varphi(0,0) = 0$. Thus, $\varphi(c,p)$ is equivalent to the Relative Policy Value (RPV) $\phi(c,p)$ and is unique. It is also easy to check that the RPV $\phi(c,p)$ satisfies Axioms \ref{axiom:4}, \ref{axiom:5}, and \ref{axiom:6}.
\end{proof}

\noindent\textbf{Theorem\,\ref{theorem:rwi_Lq}.}\textit{ For each $q \in [1, \infty)$, there is a unique comparative welfare index $\varphi_q: \R^2 \to \R$ that satisfies the associated $L^q$-normalization axioms: $\varphi_q \equiv \phi_q$, where $\phi_q$ is the $L^q$-normalized welfare index that equals zero at the origin $(0,0)$ but otherwise equals $\phi_q(c,p) = 2^{1/q}\,(p-c)/||(c,p)||_q$.}

\begin{proof}
Let $\lambda_{c,\,p} \equiv ||(c,p)||_q = (|\,c\,|^q + |\,p\,|^q)^{1/q}$ for a given $q \in [1, \infty)$. If $(c,p) \in \R^2 \setminus \{(0,0)\}$, then $(c/\lambda_{c,\,p}, \,p/\lambda_{c,\,p}) \in \mathcal{C}_q \equiv \{(c',p') \in \R^2: \lambda_{c',\,p'} = 1\}$. For any $(c,p) \in \R^2 \setminus \{(0,0)\}$, Axiom \ref{axiom:9} implies that $\varphi_q(c,p) = \varphi_q(\lambda_{c,\,p}\,c/\lambda_{c,\,p}, \,\lambda_{c,\,p} \,p/\lambda_{c,\,p}) = \varphi_q(c/\lambda_{c,\,p}, \,p/\lambda_{c,\,p})$, which equals $2^{1/q}\,(p/\lambda_{c,\,p} - c/\lambda_{c,\,p})$ by Axiom \ref{axiom:8}. Thus, for any point that is not the origin, i.e., for $(c,p) \in \R^2 \setminus \{(0,0)\}$, $\varphi_q(c,p) = 2^{1/q}\,(p-c)/\lambda_{c,\,p} = 2^{1/q}\,(p - c)/||(c,p)||_q$. In addition, by Axiom \ref{axiom:7}, $\varphi_q(0,0) = 0$. Thus, $\varphi_q(c,p)$ is equivalent to the $L^q$-normalized welfare index $\phi_q(c,p)$ and is unique. It is also easy to check that the function $\phi_q(c,p)$ satisfies Axioms \ref{axiom:7}, \ref{axiom:8}, and \ref{axiom:9}.
\end{proof}

\noindent\textbf{Theorem\,\ref{theorem:fixed_mvpf}.}\textit{ The ``fixed'' MVPF $\tilde{m}: \R^2 \to [0, \infty]$ satisfying the above Axioms \ref{axiom:10}, \ref{axiom:11}, \ref{axiom:12} is given by $\tilde{m}(c,p) = p/c$ if $(c,p) \in \R^2_{>0}$, $\tilde{m}(c,p) = c/p$ if $(c,p) \in \R^2_{<0}$, $\tilde{m}(c,p) = \infty$ if $(c,p) \in \R^2_{\leq 0} \times \R^2_{\geq 0}$ (excluding the origin), $\tilde{m}(c,p) = 0$ if $(c,p) \in \R^2_{\geq 0} \times \R^2_{\leq 0}$ (excluding the origin), and $\tilde{m}(0,0) = 1$.}

\begin{proof}
Let $(c,p) \in \R^2_{>0}$. Then, by Axiom \ref{axiom:11}, $\tilde{m}(c,p) = m(c,p) = p/c$. Let $(c',p') \in \R^2_{<0}$. Then, by Axiom \ref{axiom:12}, $\tilde{m}(c',p')\,\tilde{m}(-c',-p') = 1$. Since $(-c',-p') \in \R^2_{>0}$, $\tilde{m}(-c',-p') = (-p')/(-c') = p'/c'$, and so $\tilde{m}(c',p')\,p'/c' = 1 \implies \tilde{m}(c',p') = c'/p'$. If $(c'',p'') \in \R_{\leq 0} \times \R_{\geq 0} \setminus \{(0,0)\}$, then $\tilde{m}(c'',p'') = m(c'',p'') = \infty$ by Axiom \ref{axiom:11}. If $(c''',p''') \in \R_{\geq 0} \times \R_{\leq 0} \setminus \{(0,0)\}$, then Axiom \ref{axiom:12} provides that $\tilde{m}(c''',p''')\,\tilde{m}(-c''',-p''') = 1 \implies \tilde{m}(c''',p''')\times\infty = 1 \implies \tilde{m}(c''',p''') = 0$, since $(-c''',-p''') \in \R_{\leq 0} \times \R_{\geq 0} \setminus \{(0,0)\}$. Finally, $\tilde{m}(0,0) = 1$ by Axiom \ref{axiom:10}.
\end{proof}

\noindent\textbf{Theorem\,\ref{theorem:rpv_mvpf_conversion}.}\textit{ Let $(c, p) \in \R^2$ and $\phi(c,p) = \tilde{\phi}$. Then, $\tilde{m}(c,p) = 0$ if $\tilde{\phi} \leq -1$, $\tilde{m}(c,p) = \infty$ if $\tilde{\phi} \geq 1$, and $\tilde{m}(c,p) = 1\{-1 < \tilde{\phi} < 0\}\,(\tilde{\phi} + 1) + 1\{0 \leq \tilde{\phi} < 1\}\,(1 - \tilde{\phi})^{-1}$ if $-1 < \tilde{\phi} < 1$.}

\begin{proof}
Note that $\tilde{\phi} \leq -1 \iff (c, p) \in \R_{\geq 0} \times \R_{\leq 0} \setminus \{(0,0)\} \iff \tilde{m}(c,p) = 0$. Similarly, note that $\tilde{\phi} \geq 1 \iff (c, p) \in \R_{\leq 0} \times \R_{\geq 0} \setminus \{(0,0)\} \iff \tilde{m}(c,p) = \infty$. If $-1 < \tilde{\phi} < 0$, then either $0 < p < c$ and $p/c - 1 = \tilde{\phi} \implies \tilde{m}(c,p) = \tilde{\phi} + 1$, or it is the case that $0 > c > p$ and $c/p - 1 = \tilde{\phi} \implies \tilde{m}(c,p) = \tilde{\phi} + 1$, and so in either case $\tilde{m}(c,p) = \tilde{\phi} + 1$ if $-1 < \tilde{\phi} < 0$. Using a similar logic, $\tilde{m}(c,p) = (1 - \tilde{\phi})^{-1}$ if $0 < \tilde{\phi} < 1$.  If $\tilde{\phi} = 0$, then $c = p \implies \tilde{m}(c,p) = 1$.
\end{proof}

\noindent\textbf{Theorem\,\ref{theorem:rpv_Lp_conversion}.}\textit{ Let $(c, p) \in \R^2$ and $\phi(c,p) = \tilde{\phi}$. Then, for any $q \geq 1$, $\varphi_q(c,p)$ can be expressed in terms of $\tilde{\phi}$. Specifically, $\varphi_q(c,p) = \varphi_q(1\{\tilde{\phi} < 0\}\,(1,\,\tilde{\phi} +1) + 1\{\tilde{\phi} \geq 0\}\,(1-\tilde{\phi},\,1))$.}

\begin{proof}
If $\tilde{\phi} = 0$, then $p = c$, and so $\varphi_q(c,p) = 0 = \varphi_q(1 - \tilde{\phi},1)$. Now suppose that $\tilde{\phi} \neq 0$ so that $c \neq p$. Let $(c',p') = 1\{p \geq -c\}\,(c,p) + 1\{p < -c\}\,(-p,-c)$. Then, Theorem \ref{theorem:rwi_Lq} provides that $\varphi_q(c,p) = \varphi_q(c',p') = \varphi_q((c',p')/||(c',p')||_\infty)$. Since $\phi(c',p') = \phi(c,p) = \tilde{\phi}$, $\varphi_q(c,p) = \varphi_q((c',p')/||(c',p')||_\infty) = \varphi_q(1\{\tilde{\phi} < 0\}\,(1,\,\tilde{\phi} +1) + 1\{\tilde{\phi} > 0\}\,(1-\tilde{\phi},\,1))$.
\end{proof}

\noindent\textbf{Remark\,\ref{remark:not_lipschitz_or_convex}.}\textit{ The Marginal Value of Public Funds $m(c,p)$ and the Relative Policy Value $\phi(c,p)$ are functions that satisfy neither Lipschitz continuity nor convexity nor full differentiability.}

\begin{proof}
The Marginal Value of Public Funds (MVPF) $m(c,p)$ is not differentiable on the vertical axis, and the Relative Policy Value (RPV) $\phi(c,p)$ is not differentiable on the anti-diagonal axis, among other places. Note that $m(0.5(1,2) + 0.5(2,10)) = m(1.5, 6) = 4 > 3.5 = (0.5)(2) + (0.5)(5) = 0.5 \,m(1, 2) + 0.5\,m(2, 10)$, and so the MVPF is not convex. Since $\phi(0.5(1,2) + 0.5(2,10)) = \phi(1.5, 6) = 0.75 > 0.65 = (0.5)(0.5) + (0.5)(0.8) = 0.5\,\phi(1, 2) + 0.5\,\phi(2, 10)$, convexity also fails for the RPV. Thus, the MVPF and RPV are neither differentiable nor convex.

Suppose for the sake of contradiction that $\phi$ is Lipschitz continuous, i.e., there exists some $L > 0$ such that $|\phi(c',p') - \phi(c,p)| \leq L\, \lVert (c',p') - (c,p) \rVert$ for all $(c',p'), (c,p) \in \R^2$. Consider two points $(c',p'), (c,p) \in \mathcal{H}$ such that $c \geq p$, $c' \geq p'$, $c = c' > 0$, $p = -c$, and $p' = p + h$ for some small $h > 0$. Then, the magnitude of the difference in the RPVs of the two points is $|\phi(c',p') - \phi(c,p)| = p'/c' - 1 - (p/c - 1) = p'/c' - p/c = (-c+h)/c - (-c)/c = -1 + h/c + 1 = h/c$. In addition, note that the distance between the two points is $\lVert (c',p') - (c,p) \rVert = \lVert (c, -c+h) - (c, -c) \rVert = \lVert (0, h) \rVert = h$. In the previous calculations, $c$ is left unspecified, but suppose $c < 1/L$. Then, $|\phi(c',p') - \phi(c,p)| = h/c > L\,h = L\,\lVert (c',p') - (c,p) \rVert$, contradicting the initial assumption that $|\phi(c',p') - \phi(c,p)| \leq L\, \lVert (c',p') - (c,p) \rVert$. Thus, $\phi$ is not Lipschitz continuous. This argument can be applied without loss of generality to the MVPF as well, since $|m(c',p') - m(c,p)| = p'/c' - p/c = h/c$. Therefore, both the MVPF $m(c,p)$ and RPV $\phi(c, p)$ are not Lipschitz continuous.
\end{proof}

\noindent\textbf{Lemma\,\ref{lemma:cp_uniform_cs}.}\textit{ Suppose $F_l$ satisfies the standardized uniform integrability condition for all $l \in \cL$. Let $H_n(\cdot, F)$ be the distribution of the root $R_n$. Let $\hat{d}^\alpha_n = H^{-1}_n(1 - \alpha, \hat{F}_n)$, where $\hat{F}_n = \{\hat{F}_{l,n_l}\}_{l \in \mathcal{L}}$ and $\hat{F}_{l,n_l}$ is the empirical distribution function based on the sample $x_l^{(n_l)}$ for all $l \in \cL$. Then, the hyperrectangle $\hat{S}_n(\hat{d}^\alpha_n)$ is a uniformly valid asymptotic $(1 - \alpha)$-confidence set for $\{(c_l, p_l)\}_{l \in \cL}$, i.e., $\mathrm{lim}_{n \to \infty} \,\mathrm{inf}_{F \in \mathbb{F}} \,\mathbb{P}_F\{\{(c_l, p_l)\}_{l \in \cL} \in \hat{S}_n(\hat{d}^\alpha_n)\} \geq 1 - \alpha$ for any $\alpha \in (0, 1)$. If $H^\circ_n(\cdot, F)$ is the distribution of the root $R^\circ_n$, and $\hat{t}^\alpha_n = H^{\circ-1}_n(1 - \alpha, \hat{F}_n)$, then $\hat{E}_n(\hat{t}^\alpha_n)$ is also a uniform $(1 - \alpha)$-confidence set for $\{(c_l, p_l)\}_{l \in \cL}$, i.e., $\mathrm{lim\,inf}_{n \to \infty} \,\mathrm{inf}_{F \in \mathbb{F}} \,\mathbb{P}_F\{\{(c_l, p_l)\}_{l \in \cL} \in \hat{E}_n(\hat{t}^\alpha_n)\} \geq 1 - \alpha.$}

\begin{proof}
Let $z_{l,n_l}(F_l) = (z^c_{l,n_l}(F_l), z^p_{l,n_l}(F_l)) = \sqrt{n_l}\,([\overline{c}_{l,n_l} - \mu_c(F_l)]/s_{c,l,n_l}, \,[\overline{p}_{l,n_l} - \mu_p(F_l)]/s_{p,l,n_l})$ for all $l \in \mathcal{L}$. Then, it follows that $R^\circ_n = \mathrm{max}_{l \in \mathcal{L}}\, (z^c_{l,n_l}(F_l), z^p_{l,n_l}(F_l))'\,\Omega^{-1}(F_l)\,(z^c_{l,n_l}(F_l), z^p_{l,n_l}(F_l))$ and $R_n = \mathrm{max}_{l \in \mathcal{L}}\,\mathrm{max}\{|\,z^c_{l,n_l}(F_l)\,|,\,|\,z^p_{l,n_l}(F_l)\,|\}$, since $\mathcal{L}$ is assumed to be discrete. Then, Theorem~3.8 of \cite{romano2012uniform} applies to both the roots $R_n$ and $R^\circ_n$ as $n = \mathrm{min}_{l \in \cL}\,n_l \to \infty$ because $\{F_l\}_{l \in \mathcal{L}}$ satisfy standardized uniform integrability. As a consequence, $\hat{S}_n(\hat{d}^\alpha_n)$ and $\hat{E}_n(\hat{t}^\alpha_n)$ are uniformly valid asymptotic $(1 - \alpha)$-confidence sets for $\{(c_l, p_l)\}_{l \in \cL}$. In other words, for the confidence set $\hat{S}_n(\hat{d}^\alpha_n)$, $\mathrm{lim}_{n \to \infty} \,\mathrm{inf}_{F \in \mathbb{F}} \,\mathbb{P}_F\{\{(c_l, p_l)\}_{l \in \cL} \in \hat{S}_n(\hat{d}^\alpha_n)\} \geq 1 - \alpha$. In addition, for the confidence set $\hat{E}_n(\hat{t}^\alpha_n)$, $\mathrm{lim\,inf}_{n \to \infty} \,\mathrm{inf}_{F \in \mathbb{F}} \,\mathbb{P}_F\{\{(c_l, p_l)\}_{l \in \cL} \in \hat{E}_n(\hat{t}^\alpha_n)\} \geq 1 - \alpha$.
\end{proof}

\noindent\textbf{Theorem\,\ref{theorem:param_uniform_cs}.}\textit{ Suppose $F_l$ satisfies the standardized uniform integrability condition for all $l \in \cL$, and suppose $\hat{B}^\alpha_n$ is chosen to be either $\hat{S}_n(\hat{d}^\alpha_n)$ or $\hat{E}_n(\hat{t}^\alpha_n)$, as defined in Lemma \ref{lemma:cp_uniform_cs}. Then, $\hat{C}_n(\hat{B}^\alpha_n)$ is a uniform $(1 - \alpha)$-confidence set, i.e., $\mathrm{lim\,inf}_{n \to \infty} \,\mathrm{inf}_{F \in \mathbb{F}} \,\mathbb{P}_F\{\{\Theta_k(X_\mathcal{L})\}_{k \in \mathcal{K}} \in \hat{C}_n(\hat{B}^\alpha_n)\} \geq 1 - \alpha.$}

\begin{proof}
If $X_\cL \equiv \{(c_l,p_l)\}_{l \in \cL} \in \hat{B}^\alpha_n$, then $\mathrm{inf}_{\tilde{X}_\mathcal{L} \,\in\, \hat{B}^\alpha_n}\, \Theta_k(\tilde{X}_\mathcal{L}) \leq \Theta_k(X_\mathcal{L})\leq \mathrm{sup}_{\tilde{X}_\mathcal{L} \,\in\, \hat{B}^\alpha_n}\, \Theta_k(\tilde{X}_\mathcal{L})$ for all $k\in \mathcal{K}$. Since $X_\cL \equiv \{(c_l,p_l)\}_{l \in \cL} \in \hat{B}^\alpha_n$ implies $\Theta^*_\mathcal{K} \equiv \{\Theta_k(X_\mathcal{L})\}_{k \in \mathcal{K}} \in \hat{C}_n(\hat{B}^\alpha_n)$, it then follows that $\mathbb{P}_F\{\{\Theta^*_\mathcal{K} \equiv \Theta_k(X_\mathcal{L})\}_{k \in \mathcal{K}} \in \hat{C}_n(\hat{B}^\alpha_n)\} \geq \mathbb{P}_F\{X_\mathcal{L} \equiv \{(c_l, p_l)\}_{l \in \cL} \in \hat{B}^\alpha_n\}$. Then, by Lemma~\ref{lemma:cp_uniform_cs}, $\mathrm{lim\,inf}_{n \to \infty} \,\mathrm{inf}_{F \in \mathbb{F}} \,\mathbb{P}_F\{\{\Theta^*_\mathcal{K} \in \hat{C}_n(\hat{B}^\alpha_n)\} \geq  \mathrm{lim\,inf}_{n \to \infty} \,\mathrm{inf}_{F \in \mathbb{F}} \,\mathbb{P}_F\{X_\mathcal{L} \in \hat{B}^\alpha_n\} \geq {1 - \alpha}$.
\end{proof}

\noindent\textbf{Theorem\,\ref{theorem:param_rpv}.}\textit{ Suppose $\mathcal{L}$ and $\mathcal{K}$ are singletons and $\phi(c,p)$ is the parameter of interest for a single policy with $X_\mathcal{L} \equiv (c,p)$, and suppose $\hat{B}^\alpha_n$ is chosen to be either $\hat{S}_n(\hat{d}^\alpha_n)$ or $\hat{E}_n(\hat{t}^\alpha_n)$ defined in Lemma~\ref{lemma:cp_uniform_cs}. Then, $\hat{C}_n(\hat{B}^\alpha_n) = \phi(\partial \hat{B}^\alpha_n)$, where $\partial \hat{B}^\alpha_n$ represents the boundary of the region $\hat{B}^\alpha_n$, is a uniform $(1 - \alpha)$-confidence set for $\phi(c,p)$, i.e., $\mathrm{lim\,inf}_{n \to \infty} \,\mathrm{inf}_{F \in \mathbb{F}} \,\mathbb{P}_F\{\phi(c,p) \in \phi(\partial \hat{B}^\alpha_n)\} \geq 1 - \alpha.$}

\begin{proof}
If $(c,p) \in \hat{B}^\alpha_n$, then $\mathrm{inf}_{(\tilde{c}, \tilde{p}) \,\in\, \hat{B}^\alpha_n}\, \phi(\tilde{c}, \tilde{p}) \leq \phi(c,p) \leq \mathrm{sup}_{(\tilde{c}, \tilde{p}) \,\in\, \hat{B}^\alpha_n}\, \phi(\tilde{c}, \tilde{p})$. Since $(c,p) \in \hat{B}^\alpha_n$ implies that $\phi(c,p) \in \hat{C}_n(\hat{B}^\alpha_n)$, it then follows that $\mathbb{P}_F\{\{\phi(c,p) \in \hat{C}_n(\hat{B}^\alpha_n)\} \geq \mathbb{P}_F\{(c,p) \in \hat{B}^\alpha_n\}$, and so $\mathrm{lim\,inf}_{n \to \infty} \,\mathrm{inf}_{F \in \mathbb{F}} \,\mathbb{P}_F\{\{\phi(c,p) \in \hat{C}_n(\hat{B}^\alpha_n)\} \geq  \mathrm{lim\,inf}_{n \to \infty} \,\mathrm{inf}_{F \in \mathbb{F}} \,\mathbb{P}_F\{(c,p) \in \hat{B}^\alpha_n\} \geq {1 - \alpha}$ by Lemma~\ref{lemma:cp_uniform_cs}. Since $\mathcal{L}$ is a singleton, $\hat{B}^\alpha_n$ is either a rectangular region if $\hat{B}^\alpha_n = \hat{S}_n(\hat{d}^\alpha_n)$ or an ellipsoidal region if $\hat{B}^\alpha_n = \hat{E}_n(\hat{t}^\alpha_n)$. In either case, $\hat{B}^\alpha_n$ is a closed convex set, and so the degree-zero homogeneity of the Relative Policy Value (RPV), as stated in Remark \ref{remark:homogeneous_degree_zero}, implies that $\hat{C}_n(\hat{B}^\alpha_n) = \phi(\hat{B}^\alpha_n) = \phi(\partial \hat{B}^\alpha_n)$. Therefore, $\mathrm{lim\,inf}_{n \to \infty} \,\mathrm{inf}_{F \in \mathbb{F}} \,\mathbb{P}_F\{\phi(c,p) \in \phi(\partial \hat{B}^\alpha_n)\} \geq 1 - \alpha$.
\end{proof}

\clearpage

\noindent\textbf{Theorem\,\ref{theorem:rpv_rectangle_ci}.}\textit{ Suppose $\hat{B}^\alpha_n = \hat{S}_n(\hat{d}^\alpha_n)$ such that its center is $\overline{x}_n = (\overline{c}_n, \overline{p}_n)$, its length equals $2\hat{r}^c_n$, and its width equals $2\hat{r}^p_n$. Let $\{(b^1_k,b^2_k,u_k)\}_{k = 1}^K$ be $K$ random vectors such that $(b^1_k,b^2_k,u_k)$ are mutually independent, $b^1_k, b^2_k \sim \mathrm{Bernoulli}(0.5)$, and $u_k \sim \mathrm{Uniform}[-1,1]$ for $k \in \{ 1, \dots, K\}$. Let $\upsilon_k \equiv \overline{x}_n + b^1_k((-1)^{b^2_k}\hat{r}^c_n,u_k\hat{r}^p_n) + (1 - b^1_k)(u_k\hat{r}^c_n,(-1)^{b^2_k}\hat{r}^p_n)$ for all $k \in \{ 1, \dots, K\}$. Then, $\hat{C}^{\dagger}_{n,K}(\hat{B}^\alpha_n) = [\mathrm{min}\{\phi(\upsilon_k)\}_{k = 1}^K,\,\mathrm{max}\{\phi(\upsilon_k)\}_{k = 1}^K]$ is a rectangle-based approximate asymptotically valid uniform confidence interval such that $\hat{C}^{\dagger}_{n,K}(\hat{B}^\alpha_n)\to \phi(\partial \hat{B}^\alpha_n)$ almost surely as $K \to \infty$.}

\begin{proof}
Since $\upsilon_k \equiv \overline{x}_n + b^1_k((-1)^{b^2_k}\hat{r}^c_n,u_k\hat{r}^p_n) + (1 - b^1_k)(u_k\hat{r}^c_n,(-1)^{b^2_k}\hat{r}^p_n)$, the random vectors $\{\upsilon_k\}_{k = 1}^K$ are independent and identically distributed uniformly over the rectangular boundary $\partial \hat{S}_n(\hat{d}^\alpha_n)$. Then, $\hat{C}^{\dagger}_{n,K}(\hat{B}^\alpha_n) = \mathrm{conv}\{\phi(\upsilon_k): K \geq k \in \mathbb{N}\} = [\mathrm{min}\{\phi(\upsilon_k)\}_{k = 1}^K,\,\mathrm{max}\{\phi(\upsilon_k)\}_{k = 1}^K]$, whose end points are order statistics, converges almost surely to the interval $\phi(\partial \hat{S}_n(\hat{d}^\alpha_n)) = \phi(\partial \hat{B}^\alpha_n)$ as $K \to \infty$.
\end{proof}

\noindent\textbf{Theorem\,\ref{theorem:rpv_ellipse_ci}.}\textit{ Suppose $\hat{B}^\alpha_n = \hat{E}_n(\hat{t}^\alpha_n) = \{\tilde{x} \in \R^2: (\overline{x}_n - \tilde{x})'[\Sigma^{-1}(\hat{F}_n)](\overline{x}_n - \tilde{x}) \leq \hat{t}^\alpha_n\}$. Let $u_1, \dots, u_K$ be $K$ independent $\mathrm{Uniform}[0, 2\pi]$ random variables,  and let $v_k = \sqrt{\hat{t}^\alpha_n}(\mathrm{cos}(u_k), \mathrm{sin}(u_k))$ for all $k \in \{1,\dots, K\}$. Let $\tilde{x}_k = \overline{x}_n + [\Sigma^{1/2}(\hat{F}_n)] v_k$ for all random draws $k \in \{1,\dots, K\}$. Then, $\hat{C}^{\dagger}_{n,K}(\hat{B}^\alpha_n) = [\mathrm{min}\{\phi(\tilde{x}_k)\}_{k = 1}^K,\,\mathrm{max}\{\phi(\tilde{x}_k)\}_{k = 1}^K]$ is an ellipse-based approximate asymptotically valid uniform confidence interval such that $\hat{C}^{\dagger}_{n,K}(\hat{B}^\alpha_n) \to \phi(\partial \hat{B}^\alpha_n)$ almost surely as $K \to \infty$.}

\begin{proof}
Let $\tilde{w} =\Sigma^{-1/2}(\hat{F}_n)(\tilde{x} - \overline{x}_n)/\sqrt{\hat{t}^\alpha_n}$. Then, $(\overline{x}_n - \tilde{x})'[\Sigma^{-1}(\hat{F}_n)](\overline{x}_n - \tilde{x}) \leq \hat{t}^\alpha_n \iff \tilde{w}'\tilde{w} \leq 1$, and so $\tilde{w}$ is a point on the unit circle, i.e., $\tilde{w} = (\mathrm{cos}(\tilde{u}), \mathrm{sin}(\tilde{u}))$ for some $\tilde{u} \in [0, 2\pi]$. Note that $\tilde{x} = \overline{x}_n + \Sigma^{1/2}(\hat{F}_n)(\sqrt{\hat{t}^\alpha_n}\,\tilde{w})$. Thus, $\hat{B}^\alpha_n = \{\overline{x}_n + [\Sigma^{1/2}(\hat{F}_n)]\sqrt{\hat{t}^\alpha_n}\,(\mathrm{cos}(\tilde{u}), \mathrm{sin}(\tilde{u})): \tilde{u} \in [0, 2\pi]\}$. Since $\{u_k\}_{k = 1}^K$ are $K$ independent $\mathrm{Uniform}[0, 2\pi]$ random variables, it follows that $\{\tilde{x}_k\}_{k = 1}^K$, where $\tilde{x}_k = \overline{x}_n + [\Sigma^{1/2}(\hat{F}_n)] \sqrt{\hat{t}^\alpha_n}(\mathrm{cos}(u_k), \mathrm{sin}(u_k))$ for all $k \in \{1, \dots, K\}$, are random vectors that are uniformly distributed on the elliptical boundary $\partial \hat{B}^\alpha_n  = \partial \hat{E}_n(\hat{t}^\alpha_n)$. Then, it follows that $\hat{C}^{\dagger}_{n,K}(\hat{B}^\alpha_n) = \mathrm{conv}\{\phi(\tilde{x}_k): K \geq k \in \mathbb{N}\} = [\mathrm{min}\{\phi(\tilde{x}_k)\}_{k = 1}^K,\,\mathrm{max}\{\phi(\tilde{x}_k)\}_{k = 1}^K]$, whose end points are order statistics that converge almost surely to those of the interval $\phi(\partial \hat{E}_n(\hat{t}^\alpha_n)) = \phi(\partial \hat{B}^\alpha_n)$ as $K \to \infty$.
\end{proof}

\clearpage

\section{}
\label{appendix:tables}

\begin{center}
\scriptsize
\begin{longtable}{l|rrr|rr|rr}
\caption{Reanalysis of \citeauthor{hendren2020unified}'s (\citeyear{hendren2020unified}) Selected Set of Policies} \label{table:reanalysis_single_policies} \\

\hhline{========} \multicolumn{1}{l|}{\textit{Program}} &  \multicolumn{1}{c}{\textit{RPV}} & \multicolumn{1}{c}{\textit{Uniform CI}}  & \multicolumn{1}{c|}{\textit{Minimalist CI}}  & \multicolumn{1}{c}{\textit{MVPF}}  & \multicolumn{1}{c|}{\textit{Efron CI}}  & \multicolumn{1}{c}{\textit{MSS+$1$}} & \multicolumn{1}{r}{\textit{Minimalist CI}} \\ \hline
\endfirsthead

\multicolumn{8}{c}%
{\normalsize {\bfseries \tablename\ \thetable{}:} Reanalysis of \citeauthor{hendren2020unified}'s (\citeyear{hendren2020unified}) Selected Set of Policies} \\[4mm]
\hhline{========} \multicolumn{1}{l|}{\textit{Program}} &  \multicolumn{1}{c}{\textit{RPV}} & \multicolumn{1}{c}{\textit{Uniform CI}}  & \multicolumn{1}{c|}{\textit{Minimalist CI}}  & \multicolumn{1}{c}{\textit{MVPF}}  & \multicolumn{1}{c|}{\textit{Efron CI}}  & \multicolumn{1}{c}{\textit{MSS+$1$}} & \multicolumn{1}{r}{\textit{Minimalist CI}} \\ \hline
\endhead

\hhline{========}
\endfoot

\hhline{========}
\multicolumn{8}{l}{\parbox{6in}{\singlespacing \textit{Note}:  The abbreviations used in the first column of this table for the names of programs are the same as those used in Table II of \cite{hendren2020unified}. See their Table I for the full forms and descriptions of these abbreviations. For each of the policies, the above table reports the following: the Relative Policy Value (RPV) and its ellipse-based uniform and minimalist confidence intervals (CIs); the Marginal Value of Public Funds (MVPF) and its modified bias-corrected bootstrap confidence interval (Efron CI); the Marginal Social Surplus (MSS) plus one, which in this case equals the benefit-to-cost ratio (BCR) with a zero discount rate, and the associated ellipse-based minimalist CI. All the CIs use a confidence level of 95\%. Rows that are shaded in light gray indicate that the Efron CI for MVPF leads to a stronger conclusion than warranted by the uniform or minimalist CIs for the RPV, such as in the following cases: (i) the 95\% bias-corrected bootstrap confidence interval (Efron CI) for MVPF does not include 1, but a 95\% CI for RPV includes 0; (ii) the Efron CI lies above 0 but the CI for RPV includes values below $-1$; and (iii) the Efron CI is degenerate singleton $\{\infty\}$ but the CI for RPV includes values below 1. For policies with RPV CIs above zero and the MVPF CI above one, the associated rows are shaded in teal green. For policies with RPV CIs below zero and the MVPF CI below one, the associated rows are shaded in light maroon.}}
\endlastfoot

Abecedarian & $ 0.92 $ & $ [ -1.27 , 1.13 ] $ & $ [ -1.25 , 1.13 ] $ & $ 11.89 $ & $ [ -0.18 , \infty ] $ & $ 3.40 $ & $ [  -0.5 ,   7.3 ] $ \\
CPC Extended & $ 1.29 $ & $ [ -2.00 , 2.00 ] $ & $ [ -1.22 , 1.97 ] $ & $ \infty $ & $ [ -\infty , \infty ] $ & $ 6.37 $ & $ [ -28.1 ,  40.9 ] $ \\
\rowcolor{lightgray!70}  CPC Preschool & $ 1.16 $ & $ [ 0.36 , 2.00 ] $ & $ [ 0.57 , 1.99 ] $ & $ \infty $ & $ [ \infty , \infty ] $ & $ 3.58 $ & $ [   1.4 ,   5.6 ] $ \\
CPC School & $ -0.24 $ & $ [ -2.00 , 2.00 ] $ & $ [ -1.99 , 1.99 ] $ & $ 1.32 $ & $ [ -\infty , \infty ] $ & $ 0.96 $ & $ [   0.0 ,   1.9 ] $ \\
\rowcolor{teal!20}  Head Start & $ 1.02 $ & $ [ 0.90 , 1.08 ] $ & $ [ 0.90 , 1.08 ] $ & $ \infty $ & $ [ 10.58 , \infty ] $ & $ 5.53 $ & $ [   3.6 ,   7.5 ] $ \\
Head Start RD & $ -0.28 $ & $ [ -2.00 , 1.05 ] $ & $ [ -1.96 , 1.04 ] $ & $ 0.72 $ & $ [ -0.02 , \infty ] $ & $ 0.73 $ & $ [ -27.3 ,  38.7 ] $ \\
\rowcolor{teal!20}  Head Start RCT & $ 0.58 $ & $ [ 0.46 , 0.67 ] $ & $ [ 0.47 , 0.67 ] $ & $ 2.41 $ & $ [ 1.90 , 3.15 ] $ & $ 1.76 $ & $ [   1.5 ,   2.0 ] $ \\
\rowcolor{teal!20}  K12 Spend & $ 1.12 $ & $ [ 1.01 , 1.15 ] $ & $ [ 1.01 , 1.15 ] $ & $ \infty $ & $ [ \infty , \infty ] $ & $ 10.81 $ & $ [   5.6 ,  15.9 ] $ \\
\rowcolor{lightgray!70}  K12 Spend Mich. & $ -0.35 $ & $ [ -1.40 , 0.54 ] $ & $ [ -1.12 , 0.52 ] $ & $ 0.65 $ & $ [ 0.05 , 2.19 ] $ & $ 0.67 $ & $ [  -0.3 ,   1.9 ] $ \\
Perry Preschool & $ 0.98 $ & $ [ 0.44 , 1.09 ] $ & $ [ 0.46 , 1.09 ] $ & $ 43.61 $ & $ [ 1.85 , \infty ] $ & $ 4.37 $ & $ [   1.6 ,   7.2 ] $ \\ \hline
AOTC (IS) & $ 0.85 $ & $ [ -2.00 , 2.00 ] $ & $ [ -2.00 , 1.97 ] $ & $ 6.75 $ & $ [ -1.61 , \infty ] $ & $ 3.08 $ & $ [ -27.2 ,  28.9 ] $ \\
AOTC (JE) & $ -1.57 $ & $ [ -2.00 , 2.00 ] $ & $ [ -2.00 , 1.98 ] $ & $ -1.77 $ & $ [ -17.06 , \infty ] $ & $ -10.94 $ & $ [ -109.2 ,  76.1 ] $ \\
AOTC (JS) & $ 1.14 $ & $ [ -2.00 , 2.00 ] $ & $ [ -1.99 , 1.66 ] $ & $ \infty $ & $ [ -5.96 , \infty ] $ & $ 12.33 $ & $ [ -140.8 , 120.5 ] $ \\
AOTC (SI) & $ 0.90 $ & $ [ -2.00 , 2.00 ] $ & $ [ -1.99 , 1.97 ] $ & $ 10.05 $ & $ [ -18.36 , \infty ] $ & $ 5.83 $ & $ [ -135.4 , 148.3 ] $ \\
AOTC (SE) & $ -1.02 $ & $ [ -2.00 , 2.00 ] $ & $ [ -2.00 , 1.99 ] $ & $ -0.02 $ & $ [ -2.25 , \infty ] $ & $ -0.14 $ & $ [ -16.3 ,  17.5 ] $ \\
AOTC (SS) & $ 1.07 $ & $ [ -2.00 , 2.00 ] $ & $ [ -1.99 , 1.30 ] $ & $ \infty $ & $ [ -8.00 , \infty ] $ & $ 26.00 $ & $ [ -344.6 , 360.4 ] $ \\
HOPE Cred. & $ 0.92 $ & $ [ -2.00 , 2.00 ] $ & $ [ -1.68 , 1.83 ] $ & $ 12.58 $ & $ [ -24.72 , \infty ] $ & $ 5.85 $ & $ [ -1424 , 890.6 ] $ \\
HTC (IS) & $ 0.95 $ & $ [ -2.00 , 2.00 ] $ & $ [ -1.99 , 1.89 ] $ & $ 18.86 $ & $ [ -2.87 , \infty ] $ & $ 6.60 $ & $ [ -86.0 ,  83.2 ] $ \\
HTC (JE) & $ 0.58 $ & $ [ -2.00 , 2.00 ] $ & $ [ -1.98 , 1.93 ] $ & $ 2.37 $ & $ [ -2.22 , \infty ] $ & $ 5.74 $ & $ [ -122.7 , 169.6 ] $ \\
HTC (JS) & $ 1.17 $ & $ [ -2.00 , 2.00 ] $ & $ [ -2.00 , 2.00 ] $ & $ \infty $ & $ [ -3.51 , \infty ] $ & $ 22.56 $ & $ [ -294.4 , 378.8 ] $ \\
HTC (SE) & $ 0.92 $ & $ [ -2.00 , 2.00 ] $ & $ [ -2.00 , 1.23 ] $ & $ 11.83 $ & $ [ -4.48 , \infty ] $ & $ 5.20 $ & $ [ -54.8 ,  64.6 ] $ \\
HTC (SS) & $ -1.61 $ & $ [ -2.00 , 2.00 ] $ & $ [ -1.99 , 1.99 ] $ & $ -1.64 $ & $ [ -13.38 , \infty ] $ & $ -2.07 $ & $ [ -27.4 ,  22.3 ] $ \\
HOPE/LLC & $ -1.11 $ & $ [ -2.00 , 2.00 ] $ & $ [ -2.00 , 1.94 ] $ & $ -8.81 $ & $ [ -\infty , \infty ] $ & $ -46.68 $ & $ [ -284.1 ,  36.3 ] $ \\
Adult Pell & $ 0.54 $ & $ [ -0.97 , 0.86 ] $ & $ [ -0.67 , 0.86 ] $ & $ 2.18 $ & $ [ 0.71 , 6.11 ] $ & $ 2.85 $ & $ [  -0.4 ,   6.5 ] $ \\
Tuition deduc (JE) & $ -0.23 $ & $ [ -2.00 , 1.03 ] $ & $ [ -2.00 , 1.03 ] $ & $ 0.77 $ & $ [ -1.92 , 38.88 ] $ & $ 0.71 $ & $ [  -8.2 ,   9.6 ] $ \\
Tuition deduc (JS) & $ -1.02 $ & $ [ -2.00 , 0.99 ] $ & $ [ -2.00 , 0.99 ] $ & $ -0.02 $ & $ [ -2.50 , 5.62 ] $ & $ -0.41 $ & $ [  -8.3 ,   7.6 ] $ \\
Tuition deduc (SE) & $ 1.89 $ & $ [ -1.00 , 2.00 ] $ & $ [ -0.99 , 2.00 ] $ & $ \infty $ & $ [ -\infty , \infty ] $ & $ 3.13 $ & $ [  -8.5 ,  15.0 ] $ \\
Tuition deduc (SS) & $ 1.95 $ & $ [ -0.05 , 2.00 ] $ & $ [ 0.23 , 2.00 ] $ & $ \infty $ & $ [ -\infty , \infty ] $ & $ 11.47 $ & $ [   2.4 ,  22.0 ] $ \\ \hline
\rowcolor{teal!20}  Cal Grant GPA & $ 1.06 $ & $ [ 0.45 , 1.10 ] $ & $ [ 0.75 , 1.10 ] $ & $ \infty $ & $ [ 10.72 , \infty ] $ & $ 10.98 $ & $ [   2.6 ,  19.9 ] $ \\
Cal Grant Inc & $ -1.69 $ & $ [ -2.00 , 0.90 ] $ & $ [ -2.00 , 0.89 ] $ & $ -0.69 $ & $ [ -2.36 , 7.41 ] $ & $ -1.54 $ & $ [  -8.3 ,   5.3 ] $ \\
CUNY Pell & $ 0.28 $ & $ [ -2.00 , 0.91 ] $ & $ [ -1.99 , 0.90 ] $ & $ 1.39 $ & $ [ -2.95 , 12.88 ] $ & $ 1.40 $ & $ [  -5.1 ,   8.0 ] $ \\
CC Mich & $ 0.97 $ & $ [ -2.00 , 1.15 ] $ & $ [ -2.00 , 1.13 ] $ & $ 29.46 $ & $ [ -2.33 , \infty ] $ & $ 8.53 $ & $ [ -20.9 ,  39.2 ] $ \\
\rowcolor{lightgray!70}  CC Texas & $ 1.00 $ & $ [ -1.68 , 1.12 ] $ & $ [ -1.24 , 1.12 ] $ & $ 349.51 $ & $ [ 1.61 , \infty ] $ & $ 11.66 $ & $ [  -2.1 ,  26.3 ] $ \\
\rowcolor{lightgray!70}  DC Grant & $ 0.96 $ & $ [ -0.52 , 1.09 ] $ & $ [ -0.45 , 1.08 ] $ & $ 22.98 $ & $ [ 1.11 , \infty ] $ & $ 8.28 $ & $ [   0.2 ,  16.4 ] $ \\
\rowcolor{lightgray!70}  FIU GPA & $ 1.22 $ & $ [ -2.00 , 1.90 ] $ & $ [ -1.37 , 1.27 ] $ & $ \infty $ & $ [ \infty , \infty ] $ & $ 17.71 $ & $ [ -66.7 , 133.7 ] $ \\
\rowcolor{lightgray!70}  Florida Grant & $ 0.87 $ & $ [ -1.93 , 1.05 ] $ & $ [ -0.70 , 1.04 ] $ & $ 7.42 $ & $ [ 1.09 , \infty ] $ & $ 7.40 $ & $ [  -0.8 ,  17.6 ] $ \\
\rowcolor{lightgray!70}  Free FAFSA (dep) & $ 0.75 $ & $ [ -2.00 , 2.00 ] $ & $ [ -1.68 , 0.92 ] $ & $ 4.03 $ & $ [ 0.65 , 10.75 ] $ & $ 26.32 $ & $ [  -8.8 ,  85.0 ] $ \\
Free FAFSA (indep) & $ 0.53 $ & $ [ -2.00 , 2.00 ] $ & $ [ -1.84 , 1.87 ] $ & $ 2.12 $ & $ [ -0.06 , 9.71 ] $ & $ 3.81 $ & $ [  -6.0 ,  13.3 ] $ \\
Georgia Hope & $ 0.75 $ & $ [ -0.94 , 0.98 ] $ & $ [ -0.84 , 0.97 ] $ & $ 4.00 $ & $ [ 0.37 , 20.63 ] $ & $ 3.70 $ & $ [  -0.4 ,   7.7 ] $ \\
HAIL Aid & $ 0.23 $ & $ [ -1.06 , 0.76 ] $ & $ [ -0.97 , 0.75 ] $ & $ 1.30 $ & $ [ 0.24 , 3.65 ] $ & $ 1.22 $ & $ [   0.0 ,   2.5 ] $ \\
Kalamazoo & $ 0.48 $ & $ [ -1.02 , 0.80 ] $ & $ [ -0.55 , 0.79 ] $ & $ 1.93 $ & $ [ 0.97 , 5.61 ] $ & $ 1.93 $ & $ [   0.4 ,   4.0 ] $ \\
MA scholarship & $ -0.28 $ & $ [ -2.00 , 0.81 ] $ & $ [ -2.00 , 0.78 ] $ & $ 0.72 $ & $ [ -0.92 , 3.05 ] $ & $ 0.53 $ & $ [  -3.9 ,   4.9 ] $ \\
Ohio Pell & $ 0.60 $ & $ [ -0.38 , 0.86 ] $ & $ [ -0.29 , 0.85 ] $ & $ 2.49 $ & $ [ 0.80 , 5.40 ] $ & $ 2.72 $ & $ [   0.5 ,   4.9 ] $ \\
TN Pell & $ -0.16 $ & $ [ -2.00 , 0.83 ] $ & $ [ -1.99 , 0.82 ] $ & $ 0.84 $ & $ [ -1.59 , 3.57 ] $ & $ 0.85 $ & $ [  -2.1 ,   3.7 ] $ \\
Texas Pell & $ 1.20 $ & $ [ -0.92 , 2.00 ] $ & $ [ 1.19 , 1.81 ] $ & $ \infty $ & $ [ \infty , \infty ] $ & $ 104.12 $ & $ [   6.2 , 211.4 ] $ \\
\rowcolor{lightgray!70}  Soc Sec College & $ 0.79 $ & $ [ -2.00 , 1.00 ] $ & $ [ -1.72 , 0.99 ] $ & $ 4.86 $ & $ [ 0.98 , 52.39 ] $ & $ 4.99 $ & $ [  -1.0 ,  11.7 ] $ \\
\rowcolor{lightgray!70}  College spend & $ 0.75 $ & $ [ -0.56 , 0.95 ] $ & $ [ -0.24 , 0.95 ] $ & $ 4.00 $ & $ [ 1.25 , 20.44 ] $ & $ 3.38 $ & $ [   0.7 ,   6.4 ] $ \\
TN Hope & $ 0.46 $ & $ [ -0.90 , 0.78 ] $ & $ [ -0.51 , 0.78 ] $ & $ 1.86 $ & $ [ 0.92 , 5.08 ] $ & $ 1.90 $ & $ [   0.5 ,   3.8 ] $ \\
College tuition & $ 0.02 $ & $ [ -2.00 , 0.87 ] $ & $ [ -1.99 , 0.87 ] $ & $ 1.02 $ & $ [ -1.06 , 5.47 ] $ & $ 1.02 $ & $ [  -2.6 ,   4.8 ] $ \\
\rowcolor{lightgray!70}  WI scholarship & $ 0.30 $ & $ [ -0.30 , 0.57 ] $ & $ [ -0.13 , 0.56 ] $ & $ 1.43 $ & $ [ 1.00 , 2.32 ] $ & $ 1.44 $ & $ [   0.9 ,   2.2 ] $ \\ \hline
\rowcolor{maroon!20}  Job Corps & $ -0.85 $ & $ [ -1.24 , -0.42 ] $ & $ [ -1.23 , -0.42 ] $ & $ 0.15 $ & $ [ -0.23 , 0.58 ] $ & $ 0.16 $ & $ [  -0.3 ,   0.6 ] $ \\
JTPA adult & $ 0.27 $ & $ [ -2.00 , 2.00 ] $ & $ [ -1.99 , 0.60 ] $ & $ 1.38 $ & $ [ -0.21 , 2.13 ] $ & $ 1.32 $ & $ [   0.2 ,   2.4 ] $ \\
JTPA youth & $ -1.23 $ & $ [ -2.00 , 0.50 ] $ & $ [ -1.99 , 0.29 ] $ & $ -0.23 $ & $ [ -3.43 , 1.27 ] $ & $ -0.12 $ & $ [  -1.7 ,   1.4 ] $ \\
\rowcolor{maroon!20}  JobStart & $ -0.80 $ & $ [ -0.95 , -0.57 ] $ & $ [ -0.95 , -0.57 ] $ & $ 0.20 $ & $ [ 0.04 , 0.42 ] $ & $ 0.18 $ & $ [  -0.2 ,   0.6 ] $ \\
NSW Women & $ 0.33 $ & $ [ -2.00 , 0.47 ] $ & $ [ -1.99 , 0.47 ] $ & $ 1.48 $ & $ [ -\infty , \infty ] $ & $ 1.19 $ & $ [   0.6 ,   1.8 ] $ \\
NSW Ex-Addict & $ -0.56 $ & $ [ -0.56 , -0.56 ] $ & $ [ -0.56 , -0.56 ] $ & $ 0.44 $ & $ [ 0.44 , 0.44 ] $ & $ 0.56 $ & $ [   0.6 ,   0.6 ] $ \\
NSW Ex-Offender & $ -0.36 $ & $ [ -0.36 , -0.36 ] $ & $ [ -0.36 , -0.36 ] $ & $ 0.64 $ & $ [ 0.64 , 0.64 ] $ & $ 0.71 $ & $ [   0.7 ,   0.7 ] $ \\
NSW Youth & $ -0.40 $ & $ [ -2.00 , 0.05 ] $ & $ [ -1.99 , 0.05 ] $ & $ 0.60 $ & $ [ -\infty , \infty ] $ & $ 0.69 $ & $ [   0.1 ,   1.3 ] $ \\
Work Advance & $ -0.22 $ & $ [ -0.78 , 0.25 ] $ & $ [ -0.78 , 0.24 ] $ & $ 0.78 $ & $ [ 0.26 , 1.34 ] $ & $ 0.81 $ & $ [   0.4 ,   1.3 ] $ \\
\rowcolor{maroon!20}  Year Up & $ -0.57 $ & $ [ -0.62 , -0.52 ] $ & $ [ -0.62 , -0.52 ] $ & $ 0.43 $ & $ [ 0.37 , 0.48 ] $ & $ 0.45 $ & $ [   0.4 ,   0.5 ] $ \\ \hline
\rowcolor{maroon!20}  DI generosity & $ -0.04 $ & $ [ -0.05 , -0.03 ] $ & $ [ -0.05 , -0.03 ] $ & $ 0.96 $ & $ [ 0.95 , 0.97 ] $ & $ 0.96 $ & $ [   1.0 ,   1.0 ] $ \\
\rowcolor{maroon!20}  DI judge & $ -0.26 $ & $ [ -0.29 , -0.22 ] $ & $ [ -0.29 , -0.22 ] $ & $ 0.74 $ & $ [ 0.71 , 0.78 ] $ & $ 0.66 $ & $ [   0.6 ,   0.7 ] $ \\
\rowcolor{maroon!20}  DI examiner & $ -0.22 $ & $ [ -0.28 , -0.15 ] $ & $ [ -0.28 , -0.15 ] $ & $ 0.78 $ & $ [ 0.72 , 0.85 ] $ & $ 0.72 $ & $ [   0.6 ,   0.8 ] $ \\
\rowcolor{maroon!20}  DI veterans & $ -0.05 $ & $ [ -0.08 , -0.02 ] $ & $ [ -0.07 , -0.02 ] $ & $ 0.95 $ & $ [ 0.92 , 0.98 ] $ & $ 0.95 $ & $ [   0.9 ,   1.0 ] $ \\ \hline
Mass HI (150\%FPL) & $ -0.20 $ & $ [ -0.20 , -0.20 ] $ & $ [ -0.20 , -0.20 ] $ & $ 0.80 $ & $ [ 0.80 , 0.80 ] $ & $ 0.75 $ & $ [   0.8 ,   0.8 ] $ \\
Mass HI (200\%FPL) & $ -0.15 $ & $ [ -0.15 , -0.15 ] $ & $ [ -0.15 , -0.15 ] $ & $ 0.85 $ & $ [ 0.85 , 0.85 ] $ & $ 0.82 $ & $ [   0.8 ,   0.8 ] $ \\
Mass HI (250\%FPL) & $ 0.08 $ & $ [ 0.08 , 0.08 ] $ & $ [ 0.08 , 0.08 ] $ & $ 1.09 $ & $ [ 1.09 , 1.09 ] $ & $ 1.08 $ & $ [   1.1 ,   1.1 ] $ \\
Medicare intro & $ 0.38 $ & $ [ -0.84 , 0.78 ] $ & $ [ -0.79 , 0.77 ] $ & $ 1.63 $ & $ [ 0.52 , 3.83 ] $ & $ 1.77 $ & $ [   0.0 ,   3.5 ] $ \\
\rowcolor{teal!20}  Oregon Health & $ 0.14 $ & $ [ 0.05 , 0.21 ] $ & $ [ 0.06 , 0.21 ] $ & $ 1.16 $ & $ [ 1.08 , 1.25 ] $ & $ 1.20 $ & $ [   1.1 ,   1.4 ] $ \\
Medigap tax & $ -0.60 $ & $ [ -0.77 , 0.37 ] $ & $ [ -0.77 , 0.36 ] $ & $ 0.40 $ & $ [ 0.22 , 1.54 ] $ & $ -0.53 $ & $ [  -2.4 ,   1.4 ] $ \\ \hline
MC child 83+ & $ 1.23 $ & $ [ -0.81 , 2.00 ] $ & $ [ -0.75 , 1.71 ] $ & $ \infty $ & $ [ 0.26 , \infty ] $ & $ 2.06 $ & $ [  -5.7 ,   8.2 ] $ \\
\rowcolor{lightgray!70}  MC pregnant \& infants & $ 1.15 $ & $ [ -2.00 , 1.20 ] $ & $ [ 0.77 , 1.20 ] $ & $ \infty $ & $ [ \infty , \infty ] $ & $ 16.67 $ & $ [   2.9 ,  52.4 ] $ \\
MC child (state exp) & $ 1.13 $ & $ [ -0.02 , 1.17 ] $ & $ [ 0.05 , 1.16 ] $ & $ \infty $ & $ [ -0.37 , \infty ] $ & $ 10.21 $ & $ [   1.0 ,  19.1 ] $ \\
MC intro & $ 0.90 $ & $ [ -0.08 , 1.89 ] $ & $ [ -0.08 , 1.88 ] $ & $ 10.24 $ & $ [ 0.93 , \infty ] $ & $ 2.61 $ & $ [   0.8 ,   4.3 ] $ \\ \hline
SSI review & $ -0.24 $ & $ [ -0.42 , 0.07 ] $ & $ [ -0.42 , 0.07 ] $ & $ 0.76 $ & $ [ 0.56 , 1.00 ] $ & $ 0.68 $ & $ [   0.3 ,   1.1 ] $ \\
\rowcolor{maroon!20}  SSI judge & $ -0.26 $ & $ [ -0.28 , -0.23 ] $ & $ [ -0.28 , -0.23 ] $ & $ 0.74 $ & $ [ 0.72 , 0.77 ] $ & $ 0.66 $ & $ [   0.6 ,   0.7 ] $ \\ \hline
UI ben (state max) & $ -0.32 $ & $ [ -0.55 , 0.30 ] $ & $ [ -0.55 , 0.27 ] $ & $ 0.68 $ & $ [ 0.48 , 1.13 ] $ & $ 0.46 $ & $ [  -0.4 ,   1.3 ] $ \\
UI ben (DD) & $ -0.57 $ & $ [ -0.73 , 0.02 ] $ & $ [ -0.73 , -0.05 ] $ & $ 0.43 $ & $ [ 0.28 , 0.78 ] $ & $ -0.57 $ & $ [  -2.1 ,   0.9 ] $ \\
UI ben (DD w UR) & $ -0.52 $ & $ [ -0.73 , 0.60 ] $ & $ [ -0.72 , 0.59 ] $ & $ 0.48 $ & $ [ 0.30 , 1.69 ] $ & $ -0.26 $ & $ [  -2.1 ,   1.7 ] $ \\
UI ben (GA) & $ 0.03 $ & $ [ -0.05 , 0.09 ] $ & $ [ -0.04 , 0.09 ] $ & $ 1.03 $ & $ [ 0.97 , 1.09 ] $ & $ 1.03 $ & $ [   1.0 ,   1.1 ] $ \\
\rowcolor{maroon!20}  UI ben (MO Exp.) & $ -0.26 $ & $ [ -0.34 , -0.17 ] $ & $ [ -0.34 , -0.17 ] $ & $ 0.74 $ & $ [ 0.67 , 0.81 ] $ & $ 0.58 $ & $ [   0.4 ,   0.8 ] $ \\
\rowcolor{maroon!20}  UI ben (MO Rec.) & $ -0.56 $ & $ [ -0.62 , -0.48 ] $ & $ [ -0.62 , -0.49 ] $ & $ 0.44 $ & $ [ 0.39 , 0.50 ] $ & $ -0.51 $ & $ [  -0.9 ,  -0.1 ] $ \\
UI ben (NY) & $ -0.11 $ & $ [ -0.20 , 0.00 ] $ & $ [ -0.20 , -0.01 ] $ & $ 0.89 $ & $ [ 0.82 , 0.97 ] $ & $ 0.86 $ & $ [   0.7 ,   1.0 ] $ \\
\rowcolor{maroon!20}  UI ben (RK) & $ -0.16 $ & $ [ -0.26 , -0.05 ] $ & $ [ -0.25 , -0.06 ] $ & $ 0.84 $ & $ [ 0.76 , 0.92 ] $ & $ 0.77 $ & $ [   0.6 ,   0.9 ] $ \\
UI dur (DD) & $ -0.55 $ & $ [ -0.78 , 1.01 ] $ & $ [ -0.78 , 0.94 ] $ & $ 0.45 $ & $ [ 0.25 , 2.12 ] $ & $ -0.59 $ & $ [  -3.5 ,   2.2 ] $ \\
\rowcolor{maroon!20}  UI dur (MO) & $ -0.17 $ & $ [ -0.26 , -0.07 ] $ & $ [ -0.25 , -0.08 ] $ & $ 0.83 $ & $ [ 0.76 , 0.90 ] $ & $ 0.73 $ & $ [   0.6 ,   0.9 ] $ \\ \hline
\rowcolor{maroon!20}  HCV RCT to welfare & $ -0.09 $ & $ [ -0.14 , -0.04 ] $ & $ [ -0.14 , -0.04 ] $ & $ 0.91 $ & $ [ 0.86 , 0.96 ] $ & $ 0.90 $ & $ [   0.8 ,   1.0 ] $ \\
\rowcolor{maroon!20}  HCV Chicago lottery & $ -0.35 $ & $ [ -0.39 , -0.30 ] $ & $ [ -0.39 , -0.30 ] $ & $ 0.65 $ & $ [ 0.61 , 0.70 ] $ & $ 0.56 $ & $ [   0.5 ,   0.7 ] $ \\
Jobs+ & $ 0.29 $ & $ [ -0.59 , 0.64 ] $ & $ [ -0.58 , 0.64 ] $ & $ 1.42 $ & $ [ 0.45 , 2.83 ] $ & $ 1.34 $ & $ [   0.5 ,   2.2 ] $ \\ \hline
MTO & $ 1.13 $ & $ [ -2.00 , 2.00 ] $ & $ [ -2.00 , 1.75 ] $ & $ \infty $ & $ [ -2.80 , \infty ] $ & $ 21.83 $ & $ [ -32.8 ,  74.9 ] $ \\ \hline
\rowcolor{teal!20}  WIC & $ 0.27 $ & $ [ 0.10 , 0.40 ] $ & $ [ 0.10 , 0.40 ] $ & $ 1.38 $ & $ [ 1.10 , 1.66 ] $ & $ 1.35 $ & $ [   1.1 ,   1.6 ] $ \\
\rowcolor{maroon!20}  SNAP assist & $ -0.08 $ & $ [ -0.11 , -0.05 ] $ & $ [ -0.09 , -0.05 ] $ & $ 0.92 $ & $ [ 0.91 , 0.96 ] $ & $ 0.92 $ & $ [   0.9 ,   1.0 ] $ \\
\rowcolor{maroon!20}  SNAP info & $ -0.11 $ & $ [ -0.11 , -0.11 ] $ & $ [ -0.11 , -0.11 ] $ & $ 0.89 $ & $ [ 0.89 , 0.89 ] $ & $ 0.89 $ & $ [   0.9 ,   0.9 ] $ \\
SNAP intro & $ 0.04 $ & $ [ -1.97 , 1.09 ] $ & $ [ -1.97 , 1.08 ] $ & $ 1.04 $ & $ [ -0.97 , \infty ] $ & $ 1.04 $ & $ [  -3.9 ,   6.0 ] $ \\ \hline
\rowcolor{teal!20}  EITC 1986 & $ 0.16 $ & $ [ 0.05 , 0.28 ] $ & $ [ 0.05 , 0.28 ] $ & $ 1.20 $ & $ [ 1.05 , 1.38 ] $ & $ 1.16 $ & $ [   1.1 ,   1.3 ] $ \\
EITC 1993 & $ 0.11 $ & $ [ -0.08 , 0.30 ] $ & $ [ -0.07 , 0.29 ] $ & $ 1.12 $ & $ [ 0.82 , 1.21 ] $ & $ 1.11 $ & $ [   0.9 ,   1.3 ] $ \\
AFDC generosity & $ -0.09 $ & $ [ -0.19 , 0.02 ] $ & $ [ -0.19 , 0.01 ] $ & $ 0.91 $ & $ [ 0.83 , 1.00 ] $ & $ 0.90 $ & $ [   0.8 ,   1.0 ] $ \\
\rowcolor{maroon!20}  AFDC term limits & $ -0.19 $ & $ [ -0.27 , -0.09 ] $ & $ [ -0.27 , -0.09 ] $ & $ 0.81 $ & $ [ 0.73 , 0.90 ] $ & $ 0.77 $ & $ [   0.6 ,   0.9 ] $ \\
\rowcolor{maroon!20}  Alaska UBI & $ -0.08 $ & $ [ -0.11 , -0.05 ] $ & $ [ -0.11 , -0.05 ] $ & $ 0.92 $ & $ [ 0.89 , 0.96 ] $ & $ 0.91 $ & $ [   0.9 ,   1.0 ] $ \\
Paycheck+ & $ 0.00 $ & $ [ -0.14 , 0.15 ] $ & $ [ -0.13 , 0.15 ] $ & $ 1.00 $ & $ [ 0.87 , 1.19 ] $ & $ 1.00 $ & $ [   0.8 ,   1.2 ] $ \\
Neg. inc tax & $ -1.01 $ & $ [ -1.88 , 0.86 ] $ & $ [ -1.88 , 0.85 ] $ & $ -0.01 $ & $ [ -0.82 , 9.83 ] $ & $ -0.98 $ & $ [  -5.5 ,   3.5 ] $ \\ \hline
Top tax 2013 & $ 0.14 $ & $ [ -0.14 , 0.44 ] $ & $ [ -0.14 , 0.44 ] $ & $ 1.16 $ & $ [ 0.87 , 1.92 ] $ & $ 1.14 $ & $ [   0.8 ,   1.4 ] $ \\
\rowcolor{teal!20}  Top tax 1993 & $ 0.46 $ & $ [ 0.16 , 0.75 ] $ & $ [ 0.17 , 0.75 ] $ & $ 1.85 $ & $ [ 1.19 , 4.07 ] $ & $ 1.46 $ & $ [   1.2 ,   1.8 ] $ \\
\rowcolor{teal!20}  Top tax 1986 & $ 0.98 $ & $ [ 0.58 , 1.38 ] $ & $ [ 0.58 , 1.37 ] $ & $ 44.27 $ & $ [ 2.37 , \infty ] $ & $ 1.98 $ & $ [   1.6 ,   2.4 ] $ \\
Top tax 2001 & $ 0.27 $ & $ [ -0.09 , 0.64 ] $ & $ [ -0.09 , 0.64 ] $ & $ 1.37 $ & $ [ 0.92 , 2.86 ] $ & $ 1.27 $ & $ [   0.9 ,   1.6 ] $ \\
Top tax 1981 & $ 1.51 $ & $ [ -0.10 , 2.00 ] $ & $ [ -0.09 , 2.00 ] $ & $ \infty $ & $ [ 0.94 , \infty ] $ & $ 2.51 $ & $ [   0.9 ,   4.1 ] $ \\
\end{longtable}
\end{center}

\vspace{-15mm}

\begin{center}
\scriptsize
\begin{longtable}{l|rr|rr|rr|rr}
\caption{Reanalysis of \citeauthor{hendren2020unified}'s (\citeyear{hendren2020unified}) Policy Categories} \label{table:reanalysis_policy_categories} \\

\hhline{=========} \multicolumn{1}{l|}{\textit{Program}} &  \multicolumn{1}{c}{\textit{TPV}} & \multicolumn{1}{c|}{\textit{95\% MCI}}  & \multicolumn{1}{c}{\textit{JPV}} & \multicolumn{1}{c|}{\textit{95\% MCI}}  & \multicolumn{1}{c}{\textit{MVPF}}  & \multicolumn{1}{c|}{\textit{Efron CI}}  & \multicolumn{1}{c}{\textit{MSS+$1$}} & \multicolumn{1}{c}{\textit{95\% MCI}} \\ \hline
\endfirsthead

\endhead

\endfoot

\hhline{=========}
\multicolumn{9}{l}{\parbox{6in}{\singlespacing \textit{Note}: The above table reports the following welfare measures and associated confidence intervals (CIs) for each policy category: the Total Policy Value (TPV) using equal importance weights, and its 95\% minimalist CI (MCI); the Joint Policy Value (JPV) with equal scaling factors, and its 95\% MCI; the Marginal Value of Public Funds (MVPF) of the ``category average'' and the associated modified bias-corrected bootstrap 95\% CI (Efron CI); the Marginal Social Surplus (MSS) plus one, i.e., MSS+$1$ (which in this case equals the undiscounted benefit-to-cost ratio) of the policy category and the associated 95\% MCI. ``N/R'' indicates that the values for the sub-categories under consideration are not reported in Table II of \cite{hendren2020unified}. The main categories considered above are the same as those defined in Table II of \cite{hendren2020unified}. Not every policy considered in Table~\ref{table:reanalysis_single_policies} (or Table II of \cite{hendren2020unified}) is included in the categories that \cite{hendren2020unified} define. The ``top taxes (older)'' subcategory consists of the ``top tax 1981, 1986, and 1993'' policies. The ``top taxes (newer)'' subcategory consists of the ``top tax 2001 and 2013'' policies. Since the goal of the above table is to show that the RPV-based CIs have much more uncertainty than the MVPF-based CIs, the above table only reports the minimalist confidence intervals (MCIs), which are strictly contained in the uniform confidence intervals (UCIs). Using UCIs rather than MCIs only strengthens the points made in Sections \ref{section:reanalysis} and \ref{section:conclusion}.}}

\endlastfoot
\rowcolor{teal!20}  Child education & $ 1.01 $ & $ [ 0.21 , 1.11 ] $ & $ 1.04 $ & $ [ 0.84 , 1.12 ] $ &  $ \infty $ & $ [17.83,\infty] $  & $ 6.03 $ & $ [ 3.11 , 8.92 ] $ \\
Preschool programs & $ 0.97 $ & $ [ -0.03 , 1.09 ] $ & $ 0.98 $ & $ [ 0.57 , 1.10 ] $ &  N/R & [N/R, N/R]  & $ 4.43 $ & $ [ 1.87 , 7.00 ] $ \\
\rowcolor{teal!20}  K--12 spending & $ 1.12 $ & $ [ 1.01 , 1.15 ] $ & $ 1.12 $ & $ [ 1.01 , 1.15 ] $ &  $ \infty $ & $ [\infty,\infty] $  & $ 10.81 $ & $ [ 5.63 , 16.00 ] $ \\
College adult & $ 0.48 $ & $ [ -1.02 , 1.17 ] $ & $ -1.18 $ & $ [ -2.00 , 2.00 ] $ &  $ -5.59 $ & $ [-\infty, \infty] $  & $ -2.16 $ & $ [ -969 , 599 ] $ \\
\rowcolor{lightgray!70}  College child & $ 0.50 $ & $ [ -0.23 , 0.84 ] $ & $ 1.04 $ & $ [ -2.00 , 1.17 ] $ &  $ \infty $ & $ [4.18, \infty] $  & $ 10.05 $ & $ [ -6.30 , 32.80 ] $ \\
\rowcolor{maroon!20}  Job training & $ -0.46 $ & $ [ -1.07 , -0.26 ] $ & $ -0.53 $ & $ [ -2.00 , -0.08 ] $ &  $ 0.44 $ & $ [-19.6, 0.91] $  & $ 0.56 $ & $ [ 0.27 , 0.86 ] $ \\
\rowcolor{maroon!20}  Disability ins. & $ -0.14 $ & $ [ -0.17 , -0.10 ] $ & $ -0.15 $ & $ [ -0.19 , -0.11 ] $ &  $ 0.85 $ & $ [0.82, 0.88] $  & $ 0.82 $ & $ [ 0.76 , 0.88 ] $ \\
Health adult & $ -0.06 $ & $ [ -0.40 , 0.35 ] $ & $ -0.11 $ & $ [ -0.53 , 0.42 ] $ &  $ 0.89 $ & $ [0.56, 1.57] $  & $ 0.85 $ & $ [ 0.07 , 1.64 ] $ \\
\rowcolor{lightgray!70}  Health child & $ 1.10 $ & $ [ -0.32 , 1.49 ] $ & $ 1.13 $ & $ [ -0.65 , 2.00 ] $ &  $ \infty $ & $ [24.82, \infty] $  & $ 7.89 $ & $ [ -4.22 , 26.26 ] $ \\
\rowcolor{maroon!20}   Supp. Sec. Inc. & $ -0.25 $ & $ [ -0.37 , -0.04 ] $ & $ -0.25 $ & $ [ -0.38 , -0.07 ] $ &  $ 0.75 $ & $ [0.64, 0.85] $  & $ 0.67 $ & $ [ 0.38 , 0.93 ] $ \\
Unemp. ins. & $ -0.32 $ & $ [ -0.41 , 0.01 ] $ & $ -0.39 $ & $ [ -0.56 , -0.07 ] $ &  $ 0.61 $ & $ [0.53, 0.74] $  & $ 0.25 $ & $ [ -0.51 , 0.91 ] $ \\
\rowcolor{maroon!20}   Housing vouchers & $ -0.22 $ & $ [ -0.27 , -0.16 ] $ & $ -0.23 $ & $ [ -0.28 , -0.17 ] $ &  $ 0.77 $ & $ [0.74, 0.81] $  & $ 0.73 $ & $ [ 0.65 , 0.81 ] $ \\
MTO & $ 1.13 $ & $ [ -2.00 , 1.75 ] $ & $ 1.13 $ & $ [ -2.00 , 1.75 ] $ &  $ \infty $ & $ [-2.80, \infty] $  & $ 21.83 $ & $ [ -32.8 , 74.9 ] $ \\
Nutrition & $ 0.04 $ & $ [ -1.97 , 1.08 ] $ & $ 0.04 $ & $ [ -1.97 , 1.08 ] $ &  $ 1.04 $ & $ [-0.97, \infty] $  & $ 1.04 $ & $ [ -3.97 , 5.90 ] $ \\
Cash transfers & $ -0.16 $ & $ [ -0.37 , 0.25 ] $ & $ -0.26 $ & $ [ -0.78 , 0.57 ] $ &  $ 0.74 $ & $ [0.36, 1.47] $  & $ 0.70 $ & $ [ -0.18 , 1.95 ] $ \\
\rowcolor{teal!20}  Top taxes & $ 0.67 $ & $ [ 0.12 , 1.02 ] $ & $ 0.67 $ & $ [ 0.01 , 1.34 ] $ &  $ 3.03 $ & $ [1.35, \infty] $  & $ 1.67 $ & $ [ 1.01 , 2.34 ] $ \\
\rowcolor{teal!20} Top taxes (older) & $ 0.98 $ & $ [ 0.20 , 1.39 ] $ & $ 0.98 $ & $ [ 0.05 , 1.90 ] $ &  N/R & [N/R, N/R]  & $ 1.98 $ & $ [ 1.05 , 2.90 ] $ \\
Top taxes (newer) & $ 0.20 $ & $ [ -0.13 , 0.55 ] $ & $ 0.20 $ & $ [ -0.13 , 0.55 ] $ &  N/R & [N/R, N/R]  & $ 1.20 $ & $ [ 0.85 , 1.55 ] $ 
\end{longtable}
\end{center}

\end{document}